# Atomic Clocks for Geodesy

**Tanja E. Mehlstäubler[1], Gesine Grosche[1], Christian Lisdat[1], Piet O. Schmidt[1,2], and Heiner Denker[3]**

[1] Physikalisch-Technische Bundesanstalt, Bundesallee 100, 38116 Braunschweig, Germany

[2] Institut für Quantenoptik, Leibniz Universität Hannover, Welfengarten 1, 30167 Hannover, Germany

[3] Institut für Erdmessung, Leibniz Universität Hannover, Schneiderberg 50, 30167 Hannover, Germany

**Abstract.** We review experimental progress on optical atomic clocks and frequency transfer, and consider the prospects of using these technologies for geodetic measurements. Today, optical atomic frequency standards have reached relative frequency inaccuracies below $10^{-17}$, opening new fields of fundamental and applied research. The dependence of atomic frequencies on the gravitational potential makes atomic clocks ideal candidates for the search for deviations in the predictions of Einstein's general relativity, tests of modern unifying theories and the development of new gravity field sensors. In this review, we introduce the concepts of optical atomic clocks and present the status of international clock development and comparison. Besides further improvement in stability and accuracy of today's best clocks, a large effort is put into increasing the reliability and technological readiness for applications outside of specialized laboratories with compact, portable devices. With relative frequency uncertainties of $10^{-18}$, comparisons of optical frequency standards are foreseen to contribute together with satellite and terrestrial data to the precise determination of fundamental height reference systems in geodesy with a resolution at the cm-level. The long-term stability of atomic standards will deliver excellent long-term height references for geodetic measurements and for the modelling and understanding of our Earth.

## 1. Introduction

The measurement of time has always been closely related to the observation of our Earth, which in its daily rotation and yearly orbit around the sun defines the rhythm of our day and time periods of our life. Thus, the definition of time was based over centuries on our Earth's celestial motion, serving as a stable long-term frequency reference. Short time intervals were measured with man-made frequency standards such as hour glasses, pendulum clocks, and later spring based clocks (Landes 2000). In the early twentieth century, with the invention of quartz crystal oscillators (Walls and Vig 1995), a major leap in frequency resolution and accuracy was realized, translating to an uncertainty of time measurement of only 200 µs per day. This led to the discovery that Earth's rotation frequency varies in time (Scheibe and Adelsberger 1936). Besides tidal friction that is slowing down Earth's rotation over billions of years, a multitude of periodic and non-periodic events changing the mass distribution and angular momentum of our planet affect its rotation: tropospheric storms – following the seasonal warming of the hemispheres; the exchange of angular momentum between Earth's core and mantle – following an approximate ten-year cycle; or unforeseen events such as earthquakes and changes in ocean currents (McCarthy and Seidelmann 2009). From this a new field of Earth observation emerged. Today, the celestial motion of our planet is monitored in a global network including radio-astronomy ground stations, linked by stable hydrogen maser-based frequency standards (IERS 2010, Drewes 2007).

As a consequence of the 1930s discovery, the definition of the second was changed to the ephemeris second, defined by the tropical year, and since 1967, has been based on an atomic hyperfine transition in the Cs atom (Terrien 1968). With the ability to precisely measure frequencies associated with the energy difference between discrete levels in atoms, the definition of time was for the first time based on a non-astronomical time scale. Over many decades Cs clocks remained unsurpassed, providing microwave frequency standards with superior long-term stability and accuracy, which today are approaching relative uncertainties of $10^{-16}$ (Guena *et al.* 2017, BIPM), corresponding to only some ten ps per day.

However, in the last decade, the tremendous development of both optical atomic clocks and new methods to compare them over large distances has opened up a new era of time and frequency measurements and corresponding applications. As a consequence of the relativistic redshift effect in gravitational fields, precise frequency comparisons between optical atomic clocks with a fractional frequency resolution to the 18th digit are strongly related to the (otherwise not directly observable) gravitational potential (difference) at an entirely new level of sensitivity. In this context, a fractional frequency shift of $10^{-18}$ corresponds to about 0.1 $m^2s^{-2}$ in terms of the potential difference, which is equivalent to 1 cm in height difference on Earth's surface. Besides fundamental contributions to basic research (Ludlow *et al.* 2015, Delva *et al.* 2017), this makes clock comparisons competitive for new applications in astronomy and geodesy, contributing to studies on the shape and mass of our planet. The method of deriving potential differences from clock frequency comparisons has variously been termed "chronometric levelling", "relativistic geodesy", or "chronometric geodesy" (Bjerhammar 1975, Vermeer 1983, Bjerhammar 1985, Shen *et al.* 2011, Delva and Lodewyck 2013), and throughout this paper we will use the term "chronometric levelling", as it describes quite accurately the idea of clock-based levelling. This technique is especially useful for the establishment of a unified International Height Reference System (IHRS), the connection of tide gauges, and the measurement of time variations of Earth's gravity field. However, to contribute to the international height reference system, new developments around clocks and their comparison are required to achieve the cm level during field work.

This review reports on the state-of-the-art in atomic optical clock development and comparison and details on the implementation of clock measurements for geodetic applications. A particular focus is put on the need of stable and reproducible frequency measurements, the development of portable standards and long-distance clock comparisons. Section 2 introduces the foundations of classical geodesy and geodetic methods, today's state-of-the-art and challenges in defining height reference systems, together with the possible inclusion of clock measurements for geodetic applications. Section 3 details on the validity of evaluating relativistic time dilation shifts and discusses fundamental tests of gravitational and motional redshift measurements using precision spectroscopy. Section 4 introduces the terminology and ingredients of time and frequency measurements. Section 5 presents an overview of the current state-of-the art of optical clock performance in view of the requirements for chronometric levelling. Finally, the current capabilities and developments to compare clocks at remote places are discussed in section 6.

Since this review is intended for physicists interested in geodetic applications of optical atomic clocks as well as for geodesists curious about the capabilities of clocks and their comparison, it provides a basic introduction and description of the state-of-the art aimed for readers of the respective other field. Further details can be found in recent reviews on optical clocks (Margolis 2010, Poli *et al.* 2013, Ludlow *et al.* 2015, Hong 2017) and on clocks in the context of geodesy (Delva and Lodewyck 2013, Denker *et al.* 2017). The broader physics background of frequency standards and physical geodesy can be found in text books e.g. (Audoin and Guinot 2001, Riehle 2004) and (Heiskanen and Moritz 1967, Torge and Müller 2012), respectively.

## 2 Geodesy and Clocks

According to the classical definition of F. R. Helmert (1880), "geodesy is the science of the measurement and mapping of the Earth's surface". This definition includes the determination of the Earth's external gravity field, since almost every geodetic measurement depends on it, as well as reference systems and the Earth's orientation in space. These three areas (positioning, gravity field,

Earth rotation) are also considered the three pillars of geodesy. The classical geodesy definition of Helmert has been extended to ocean and space research, the study of other celestial bodies, and the determination of temporal variations in all three pillars of geodesy, for further details see Herring (2009), Vanicek (2003), and Torge and Müller (2012).

On this basis, geodesy is a combination of an observational and a theoretical science, and thus new theories or new observations determined the direction of geodetic science in the past. However, during recent decades improved observational accuracies have often been the main driver for new theoretical developments to explain these measurements. On large scales (over the whole Earth), modern geodetic measurements are precise to better than 1 part per billion (1 ppb) in many cases, e.g. corresponding to better than 6 mm (1 ppb) uncertainty for global height measurements. Especially the development of distance measurements based on propagating electromagnetic signals and the launch of Earth-orbiting satellites allowed global measurements of positions, Earth rotation changes, and gravity field parameters at the 1 ppb level of uncertainty. These modern geodetic measurement systems can be divided into three basic classes: ground based positioning systems (e.g. GNSS – Global Navigation Satellite Systems) that provide geometric positioning and tracking of external bodies; satellite systems that sense the Earth's gravity field and/or make measurements directly from space (e.g. the satellite missions Jason, GRACE, and GOCE); and ground-based instrumentation that measure the gravity field on or near the Earth's surface, especially gravimeters (Herring 2009). Furthermore, in the future, optical clocks could be useful as ground-based instrumentation and as satellite systems.

Coordinate systems and the associated reference frames form a core theme and provide the foundation for all three pillars of geodesy. With the advent of space-based methods that allow global measurements to be made (with GNSS 24 hours a day, seven days a week), it became possible to define a purely geometrical global coordinate system with position accuracies at the level of one centimetre (or below), allowing the detection of plate tectonic and other deformations. These coordinate systems are primarily Cartesian and have their origin at the centre of mass of the Earth, with the axes aligned in a well-defined manner to the outer surface of the Earth (approximately along the rotation axis and the Greenwich meridian). The Cartesian coordinates can also be transformed into equivalent ellipsoidal coordinates based on a given reference ellipsoid with ellipsoidal latitude, longitude, and height, all being again purely geometrical quantities. However, historically, before the era of GNSS, coordinate systems in geodesy were divided into (geometry-based) horizontal coordinates (latitude and longitude) and a (potential-based or physical) vertical coordinate called height, which was a consequence of the methods available for making measurements. These physical heights are based on equipotential surfaces within the Earth's gravity field, along which fluid will not flow. Consequently, equipotential or level surfaces are of paramount importance for geodesy, oceanography, geophysics, and other disciplines to define (physical) heights on the continents and for the dynamic ocean surface in order to answer questions on the direction of fluid (water) flow. One special equipotential surface is called the geoid and is associated with the surface representing mean sea level (MSL).

The following sections are devoted to some foundations of classical geodesy and the possible application of atomic clocks in this field. Section 2.1 outlines some fundamentals of physical geodesy related to the gravity field in general, coordinate reference systems, the gravity potential, equipotential or level surfaces, and the definition of the geoid. Then, sections 2.2 and 2.3 describe two geodetic methods for determining the gravity potential, considering the geometric levelling approach and the so-called GNSS/geoid approach. In section 2.4, some aspects of general relativity and appropriate spacetime reference systems are given, which provide the background for all applications of atomic clocks in geodesy. In general, two ideal clocks run at different rates with respect to a common (coordinate) timescale, if they move or are under the influence of a gravitational field, which is known as the relativistic redshift effect. For the usual case of two earthbound clocks at rest, the resulting frequency shift is directly proportional to the difference of the gravity (gravitational plus centrifugal) potential associated with corresponding equipotential or level surfaces. This offers completely new perspectives for the measurement of potential-based (physical) heights in geodesy and other disciplines. Such geodetic applications as well as geodynamic investigations are outlined in sections 2.5 to 2.7, respectively, especially regarding further improved optical clocks in the $10^{-18}$ to $10^{-19}$ regime (and below).

## 2.1 Some fundamentals of physical geodesy

Following the classical textbook from Heiskanen and Moritz (1967), "the study of the physical properties of the gravity field and their geodetic application are the subject of physical geodesy". This includes the dominant static (spatially variable) and the (small) time-variable parts of the gravity field, where the description, acquisition, analysis, and interpretation of changes of the Earth's body and its gravity field are treated in the field of geodynamics. The largest temporal variations are due to solid Earth tide effects, which lead to predominantly vertical movements of the Earth's surface with a global maximum amplitude of about 0.4 m (equivalent to about 4 $m^2 s^{-2}$ in potential); this corresponds to a magnitude of about $10^{-7}$ with respect to quantities related to the entire Earth, such as the potential, gravity, or radius. The next largest contribution is the ocean tide effect with a magnitude of roughly 10 – 15 % of the solid Earth tides, but with significantly increased values towards the coast. All other time-variable effects are a further order of magnitude smaller and originate from atmospheric mass movements (on a global scale, ranging from hourly to seasonal variations), hydro-geophysical mass changes (on regional and continental scales, seasonal variations), and polar motion (pole tides). For details regarding the computation, magnitudes, and main time periods of all relevant time-variable gravity potential components, see Voigt *et al.* (2016). In this contribution, however, the focus is on the determination of the static (spatially variable) part of the potential field, while temporal variations in the potential quantities as well as in the station coordinates are assumed to be taken into account through appropriate reductions or by using sufficiently long averaging times. This is common geodetic practice and leads to a quasi-static state, such that the Earth can be considered as a rigid and non-deformable body, uniformly rotating about a body-fixed axis. Hence, all gravity field quantities including the level surfaces are considered as static quantities in the following.

In the context of physical geodesy and many other terrestrial geodetic applications, it is most convenient to work with an Earth-fixed coordinate system (co-rotating with the Earth) with its origin located at the geocentre and orientation based on the rotation axis and the Greenwich meridian. Such a coordinate system is the International Terrestrial Reference System (ITRS) and corresponding ITRF (International Terrestrial Reference Frame), both being maintained by the International Earth Rotation Service (IERS). With regard to the geodetic terminology, it is fundamental to distinguish between a "reference system", which is based on theoretical considerations or conventions, and its realization, the "reference frame", to which users have access, e.g. in the form of position catalogues. ITRF station coordinates are primarily given as Cartesian coordinates, but can also be transformed into ellipsoidal coordinates (ellipsoidal latitude, longitude, and height) based on a given reference ellipsoid (such as the Geodetic Reference System 1980, GRS80; Moritz 2000); Cartesian and ellipsoidal coordinates are both purely geometric quantities. The most recent realization of the ITRS is the ITRF2014, with GNSS being a major contributor (Altamimi *et al*. 2016). The ITRS and its frames are today based on a relativistic framework, which also provides the foundation for the use of atomic clocks in geodesy; for further details, see subsection 2.4.

Classical physical geodesy is largely based on the Newtonian theory with Newton's law of gravitation, giving the gravitational force between two point masses, to which a gravitational acceleration (also termed gravitation) can be ascribed by setting the mass at the attracted point *P* to unity. Then, by the law of superposition, the gravitational acceleration of an extended body like the Earth can be computed as the vector sum of the accelerations generated by the individual point masses (or mass elements), yielding

$$\mathbf{b} = \mathbf{b}(\mathbf{r}) = -G \iiint_{\text{Earth}} \frac{\mathbf{r} - \mathbf{r}'}{|\mathbf{r} - \mathbf{r}'|^3} dm \; , \quad dm = \rho dv \; , \quad \rho = \rho(\mathbf{r}') \; , \tag{2.1}$$

where **r** and **r'** are the position vectors of the attracted point *P* and the source point *Q*, respectively, *dm* is the differential mass element, $\rho$ is the volume density, *dv* is the volume element, and *G* is the gravitational constant. The SI unit of acceleration is $m\,s^{-2}$, but the non-SI unit Gal is still used frequently in geodesy and geophysics (1 Gal = 0.01 $m\,s^{-2}$, 1 mGal = $10^{-5}\,m\,s^{-2}$), see also BIPM (2006). In addition to this, a body rotating with the Earth also experiences a centrifugal force and a corresponding centrifugal acceleration **z**, which is directed outwards and perpendicular to the rotation axis:

$$\mathbf{z} = \mathbf{z}(\mathbf{p}) = \omega^2 \mathbf{p} \ . \tag{2.2}$$

In the above equation, $\omega$ is the angular velocity, and $\mathbf{p}$ is the distance vector from the rotation axis. Finally, the gravity acceleration (or gravity) vector $\mathbf{g}$ is the resultant of the gravitation $\mathbf{b}$ and the centrifugal acceleration $\mathbf{z}$:

$$\mathbf{g} = \mathbf{b} + \mathbf{z} \ , \tag{2.3}$$

The direction of $\mathbf{g}$ is the direction of the plumb line (vertical), the magnitude $g$ is called the gravity intensity (or often just gravity), see Torge and Müller (2012). As the gravitational and centrifugal acceleration vectors $\mathbf{b}$ and $\mathbf{z}$ both form conservative vector fields or potential fields, these can be represented as the gradient of corresponding potential functions by

$$\mathbf{g} = \operatorname{grad} W = \mathbf{b} + \mathbf{z} = \operatorname{grad} V_E + \operatorname{grad} Z_E = \operatorname{grad}(V_E + Z_E) \ , \tag{2.4}$$

where $W$ is the gravity potential, consisting of the gravitational potential $V_E$ and the centrifugal potential $Z_E$. Based on equations (2.1) to (2.4), the gravity potential $W$ can be expressed as

$$W = W(\mathbf{r}) = V_E + Z_E = G \iiint_{\text{Earth}} \frac{\rho dv}{l} + \frac{\omega^2}{2} p^2 \ , \tag{2.5}$$

where $l$ is the length of the vector $\mathbf{r} - \mathbf{r}'$ and $p$ is the length of the vector $\mathbf{p}$. All potentials are defined with a positive sign, which is common geodetic practice, but opposite to most physics applications. The gravitational potential $V_E$ is assumed to be regular (i.e. zero) at infinity and has the important property that it fulfils the Laplace equation outside the masses; hence it can be represented by harmonic functions in free space, with the spherical harmonic expansion playing a very important role.

The surfaces of constant gravity potential $W = W(\mathbf{r}) = $ const. are designated as equipotential or level surfaces (also geopotential surfaces) of gravity. The gravity vector $\mathbf{g}$ is everywhere normal to the corresponding equipotential surface and points in the direction of greatest change of the potential function $W$, while in general $g$ is not constant along an equipotential surface. The determination of the gravity potential $W$ as a function of position is one of the primary goals of physical geodesy; if $W(\mathbf{r})$ were known, then all other parameters of interest could be derived from it, including the gravity vector $\mathbf{g}$ according to equation (2.4) as well as the form of the equipotential surfaces (by solving the equation $W(\mathbf{r}) = $ const.). Furthermore, the gravity potential is also the ideal quantity for describing the direction of water flow, i.e., water flows from points with lower gravity potential to points with higher values. However, although the above equation (2.5) is fundamental in geodesy, it cannot be used directly to compute the gravity potential $W$ due to insufficient knowledge about the density structure of the entire Earth; this is evident from the fact that densities are at best known with two to three significant digits, while geodesy generally strives for a relative uncertainty of at least $10^{-9}$ for all relevant quantities (including the potential $W$). Therefore, the determination of the exterior potential field must be solved indirectly based on measurements performed at or above the Earth's surface, e.g. by gravity measurements, which leads to the area of geodetic boundary value problems (GBVPs; see Sect. 2.3).

The gravity potential is closely related to the question of heights, level or equipotential surfaces, and the geoid. The geoid is classically defined as a selected level surface with constant gravity potential $W_0$, conceptually chosen to approximate (in a mathematical sense) the mean ocean surface or mean sea level (MSL). However, MSL changes with time (e.g., because of global sea level rise) and does not coincide with a level surface due to the forcing of the oceans by winds, atmospheric pressure, and buoyancy in combination with gravity and the Earth's rotation. The deviation of MSL from a best fitting equipotential surface (geoid) is denoted as the (mean) dynamic ocean topography (DOT); it reaches maximum values of about ±2 m (Rapp and Balasubramania 1992; Bosch and Savcenko 2010) and is of vital importance for oceanographers for deriving ocean circulation models (Wunsch and Gaposchkin 1980; Condi and Wunsch 2004).

On the other hand, a substantially different approach was chosen by the International Association of Geodesy (IAG) in 2015 for the International Height Reference System (IHRS; see IAG 2016), where a numerical value $W_{0 \text{ (IHRS 2015)}} = 62,636,853.4$ m$^2$s$^{-2}$ is defined for the realization of the vertical reference level surface. Similarly, the International Astronomical Union (IAU) decided within Resolution B1.9

(2000) on the re-definition of Terrestrial Time (TT) to turn the constant $L_G$ (related to the rate between TT and Geocentric Coordinate Time TCG) into a defining constant with a fixed value and zero uncertainty, which implies a value of $W_{0\ (IAU\ 2000)} = 62{,}636{,}856.00$ m$^2$s$^{-2}$ (see section 2.4 for further details). Therefore, due to the significant differences between these two and other existing $W_0$ values, it is important to clearly document which zero level surface and gravity potential have been employed in any particular situation. Petit *et al.* (2014) denote these two definitions as "classical geoid" and "chronometric geoid", respectively. It is clear that the definition of the zero level surface ($W_0$ issue) is largely a matter of convention, where a good option is probably to select a conventional value for $W_0$ (referring to a certain epoch, without a strict relationship to MSL), accompanied by static modelling of the corresponding zero level surface, and to describe then the potential of the time-variable mean ocean surface for any given point in time as the deviation from this reference value.

## 2.2 The geometric levelling approach and heights

Level surfaces or equipotential surfaces are surfaces of constant gravity potential $W$. Based on equation (2.4), the gravity potential differential (associated with differentially separated level surfaces) can be expressed as

$$dW = \frac{\partial W}{\partial x}dx + \frac{\partial W}{\partial y}dy + \frac{\partial W}{\partial z}dz = \text{grad}\,W\,\mathbf{ds} = \mathbf{g}\,\mathbf{ds} = -g\,dn\ ,\qquad(2.6)$$

where $\mathbf{ds}$ is the vectorial line element, $g$ is the magnitude of the gravity vector, and $dn$ is the distance along the outer normal of the level surface (zenith or vertical). Since only the projection of $\mathbf{ds}$ along the plumb lime (direction of $\mathbf{g}$) enters, $dW$ is independent of the path, and hence no work is necessary for a displacement along the level surface $W$ = const. This means that the level surfaces are equilibrium surfaces, which may be described by the surface of freely-moving homogeneous water masses that are only affected by gravity. The geoid, defined as a selected equipotential surface $W = W_0$ related to MSL, may be considered as the surface of such an idealized ocean, being extended under the continents (e.g. by a system of conducting tubes). Furthermore, if $\mathbf{ds}$ is taken along the level surface, then it follows from $dW = 0$ that the gravity vector $\mathbf{g}$ is normal to the level surface, or, in other words, that the level surfaces are intersected at right angles by the plumb lines.

Equation (2.6) provides the link between the potential difference $dW$ (a physical quantity) and the difference in height $dn$ (a geometric quantity) of neighbouring level surfaces. Hence, a combination of (quasi)-differential height determinations, as provided by geometric levelling (see Fig. 2.1), and gravity measurements deliver potential differences. Because gravity $g$ is changing along the level surfaces, the distance *dn* to a neighbouring level surface also changes, which means that the level surfaces are not parallel and that the plumb lines are space curves. As a consequence of the gravity increase of about 0.05 m s$^{-2}$ (roughly 0.5 % of *g*) from the equator to the poles, the level surfaces of the Earth also converge towards the poles by about 0.5 % in a relative sense; for instance, two level surfaces that are 100.0 m apart at the equator are separated by only 99.5 m at the poles. This also implies that the raw levelling is path dependent, i.e. $\oint dn \neq 0$, which cannot be neglected over longer distances, while corresponding gravity potential differences are path independent ($\oint dW = 0$) because the gravity field is conservative.

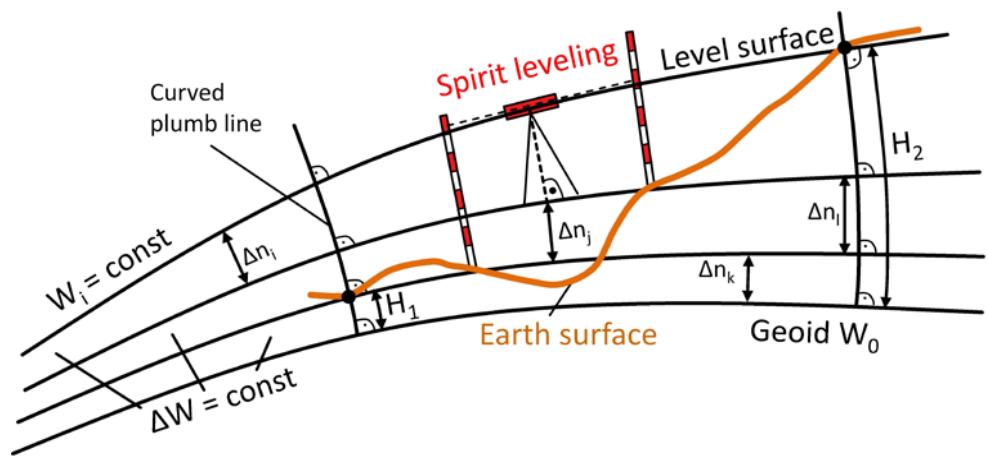

*Figure 2.1: Principle of geometric (or spirit) levelling along with equipotential (or level) surfaces, plumb lines and orthometric heights.*

The principle of geometric levelling (also denoted as spirit levelling) is further detailed in Fig. 2.1, showing also that in general the level surfaces are not parallel and hence the corresponding vertical separations are varying. Levelling is a quasi-differential technique and conducted with a levelling instrument (level) and two vertically posted levelling rods. The instrument is oriented along the plumb line to obtain horizontal lines of sight between points in close proximity to each other (target distances 30 to 50 m), providing height differences *δn* (backsight minus foresight reading) between the rods. In this way height differences can be obtained with an uncertainty of 1 mm or below over 1 km distance, depending on the equipment used.

Integrating equation (2.6) based on geometric levelling and gravity observations is the classical and most direct way to obtain gravity potential differences. It is denoted here as the geometric levelling approach and leads to the geopotential number *C* in the form

$$C = W_0 - W_P = -\int_{P_0}^{P} dW = \int_{P_0}^{P} g \, dn \ , \tag{2.7}$$

where *P* is a point at the Earth's surface, and $P_0$ is an arbitrary point on the selected height reference surface, which is typically related to a fundamental national tide gauge. Because MSL deviates from a level surface within the Earth's gravity field due to the DOT, and as tide gauges (connected to MSL) usually build the starting point for national and regional height reference systems, this leads to inconsistencies between these systems, known as the vertical datum problem, which reach more than 0.5 m across Europe. Consequently, for each national height system (vertical datum) based on a selected tide gauge, an index *i* has to be associated with each system and the corresponding zero potential value, but this is ignored here for reasons of simplicity. Furthermore, the corresponding numerical value of the zero potential is typically unknown, and hence geometric levelling (*dn*) together with gravity observations (*g*) gives only potential differences. The geopotential number *C* is defined such that it is positive for points *P* above the zero level surface, similar to heights (see Figure 2.1).

Although the geopotential numbers in the unit m$^2$s$^{-2}$ (or in the geopotential unit 1 gpu = 10 m$^2$s$^{-2}$ = 1 kGal m) are ideal quantities for describing the direction of water flow, they are somewhat inconvenient in disciplines such as civil engineering, etc. Therefore, a conversion to metric heights is desirable, which can be achieved by dividing the *C* values by an appropriate gravity value. Widely used are the orthometric heights (e.g. in the USA, Canada, Austria, and Switzerland) and normal heights (e.g. in Germany and many other European countries). The orthometric height *H* is defined as the distance between the surface point *P* and the zero level surface (geoid), measured along the curved plumb line (see Fig. 2.1), which explains the common understanding of this term as "height above sea level" (Torge and Müller 2012). The orthometric height can be derived from equation (2.7) by integrating along the plumb line, giving

$$H = \frac{C}{\bar{g}} \ , \quad \bar{g} = \frac{1}{H} \int_0^H g \, dH \ , \tag{2.8}$$

where $\bar{g}$ is the mean gravity along the plumb line (inside the Earth). As $\bar{g}$ cannot be observed directly, hypotheses about the interior gravity field are necessary, which is one of the main drawbacks of the orthometric heights. Therefore, in order to avoid these hypotheses, the normal heights $H^N$ were introduced by Molodensky (e.g., Molodenskii *et al.* 1962) in the form

$$H^N = \frac{C}{\bar{\gamma}} \ , \quad \bar{\gamma} = \frac{1}{H^N} \int_0^{H^N} \gamma \, dH^N \ , \tag{2.9}$$

where $\bar{\gamma}$ is a mean normal gravity value along the normal plumb line (within the normal gravity field, associated with the level ellipsoid, see also section 2.3), and $\gamma$ is the normal gravity acceleration along this line. Consequently, the normal height $H^N$ is measured along the slightly curved normal plumb line. Furthermore, as $\bar{g}$ and $\bar{\gamma}$ are both location dependent, points on the same level surface may have slightly different heights $H$ and $H^N$, which can be avoided only by using a constant conventional gravity value, leading to the so-called dynamic heights (for further details, see line Torge and Müller 2012).

Precise levelling in fundamental networks is generally carried out in closed loops as double-run levelling (in opposing direction). The loops are composed of levelling lines, connecting the nodal points of a whole network. Prior to the (least-squares) adjustment of the levelling network, usually potential differences are formed by taking gravity observations into account (according to equation (2.7)). The adjustment uses the loop misclosure condition of zero ($\oint dW = 0$), while the zero level reference surface (vertical datum) is generally defined by MSL derived from a fundamental tide gauge. The adjustment then delivers the geopotential numbers $C$ of all nodal points, which are usually converted to metric heights, e.g. by using equations (2.8) or (2.9). However, it is also worth mentioning that the raw levelling results along the lines ($\Delta n$) can also be converted directly into corresponding height differences (e.g. $\Delta H$, $\Delta H^N$) by the orthometric and normal corrections, respectively (Torge and Müller 2012). Hence the network adjustment may also be performed based on heights instead of geopotential numbers (as also $\oint dH = \oint dH^N = 0$ holds).

The uncertainty of geometric levelling is rather low over shorter distances, where it can reach the sub-millimetre level, but it is a differential technique and hence susceptible to systematic errors that may exceed the decimetre level over 1000 km distance. Examples include the differences between the second and third geodetic levelling in Great Britain (about 0.2 m in the north–south direction over about 1000 km distance; Kelsey 1972), corresponding differences between an old and new levelling in France (about 0.25 m from the Mediterranean Sea to the North Sea, also mainly in north–south direction, distance about 900 km; Rebischung et al. 2008), and inconsistencies of more than ±1 m across Canada and the USA (differences between different levelling and with respect to an accurate geoid; Véronneau et al. 2006 and Smith et al. 2010 and 2013). A further complication with geometric levelling in different countries is that the results are usually based on different tide gauges with offsets between the corresponding zero level surfaces and that in some countries the levelling observations may be about 100 years old and thus not represent the actual situation due to possibly occurring recent vertical crustal movements.

While the orthometric and normal heights are related to the Earth's gravity field (physical heights), the ellipsoidal heights $h$, as derived from GNSS observations, are purely geometric quantities, describing the distance (along the ellipsoid normal) of a point $P$ from a conventional reference ellipsoid (see Fig. 2.2). As the geoid and quasigeoid serve as the zero height reference surface (vertical datum) for the orthometric and normal heights, respectively, the following relation holds

$$h = H + N = H^N + \zeta \ , \tag{2.10}$$

where $N$ is the geoid undulation, and $\zeta$ is the quasigeoid height or height anomaly, quantifying the distance from the geoid and quasigeoid to the level ellipsoid, respectively; for further details see, e.g., Torge and Müller (2012). Equation (2.10) and Figure 2.2 neglect the fact that strictly the relevant quantities are measured along slightly different lines in space, but the maximum effect is only at the sub-millimetre level (for further details cf. Denker 2013). The foregoing shows that the geopotential numbers and heights are fully equivalent, and hence the relations between various quantities, including equation (2.10), can be expressed in terms of height and potential

The deviations of the geoid and quasigeoid from a best-fitting geocentric reference ellipsoid reach about ±100 m (RMS about 30 m), and both surfaces show significant tilts (vertical deflections) with respect to the reference ellipsoid. Even in low mountain ranges, such as the Harz mountains in Germany (maximum elevation about 1000 m), these tilts lead to changes in the geoid and quasigeoid of up to about 10 cm over 1 km distance (see also Fig. 2.2). Hence, for describing the direction of water flow, one always needs some kind of physical heights and geopotential information based on equations (2.8), (2.9), and (2.7), respectively, while the purely geometric ellipsoidal heights cannot cope with this task. Moreover, equations (2.8) to (2.10) can be combined to derive the difference $H^N$–$H$ or $N$–$\zeta$, which can reach several centimetres to about 1 dm in low mountain ranges, about 3–5 dm (or even more) in the high mountains such as the European Alps or the Rocky Mountains, and about 3 m in the Himalayan Mountains (Rapp 1997; Marti and Schlatter 2001), while on the oceans, the geoid and quasigeoid practically coincide (Torge and Müller 2012).

As geometric levelling is an extremely costly and time-consuming procedure, and because it is a differential technique that is susceptible to significant systematic errors (see above), several countries have seriously considered to replace it by GNSS and a high-resolution gravimetric geoid/quasigeoid model, i.e. to derive physical heights from GNSS ellipsoidal heights by using equation (2.10) in the form $H = h - N$ or $H^N = h - \zeta$. This approach, abandoning geometric levelling completely, is also known as the "geoid-based vertical datum". It was first introduced in Great Britain 2002 (see, e.g. Iliffe *et al.* 2003), then Canada followed in 2013 (Véronneau and Huang 2016), and the U.S.A. plan to introduce it in 2022 (NGS/NOAA 2017).

Lastly, the geometric levelling approach gives only gravity potential differences, but the associated constant zero potential $W_0^{(i)}$ can be determined by at least one (better several) GNSS and levelling points in combination with the (gravimetrically derived) disturbing potential, as described in the next section. Rearranging the above equations gives the desired gravity potential values in the form

$$W_P = W_0 - C = W_0 - \bar{g}H = W_0 - \bar{\gamma}H , \qquad (2.11)$$

and hence the geopotential numbers and the heights $H$ and $H^N$ are fully equivalent. For further details, see Denker *et al.* (2017).

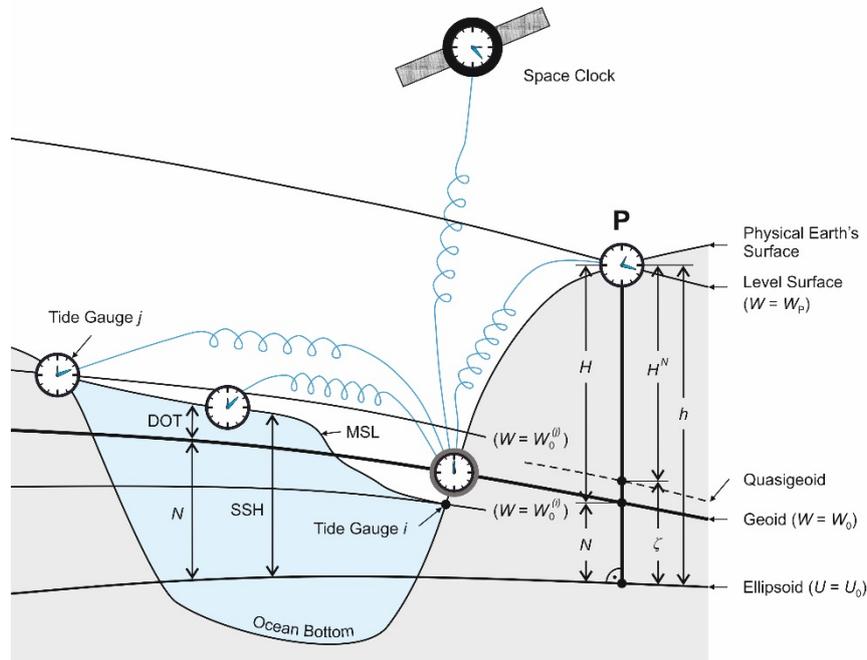

*Figure 2.2: Geoid and level surfaces, quasigeoid, ellipsoid, heights, continental topography, mean sea level (MSL), dynamic ocean topography (DOT), and sea surface height (SSH). The principle of chronometric levelling is indicated by the clock symbols and the spiral-shaped lines for the link technologies (for further details see Sect. 2.5).*

## 2.3 The GNSS/geoid approach

For the determination of the gravity potential $W$, one of the primary goals of geodesy, gravity measurements form one of the most important data sets. However, since gravity (represented as $g = |\mathbf{g}|$ = length of the gravity vector $\mathbf{g}$) and other relevant gravity field quantities depend in general in a nonlinear way on the potential $W$, which is to be determined, the observation equations must be linearized. This is done by introducing an a priori known (conventional) reference potential and by corresponding reference positions. The linearization process leads to the disturbing (or anomalous) potential $T$ defined for a point $P$ as

$$T = W_P - U_P , \tag{2.12}$$

where $U_P$ is the normal gravity potential associated with the normal gravity field, which is usually related to the level ellipsoid (a rotational ellipsoid, with mass and rotational velocity). The level ellipsoid is chosen as a conventional system because it is easy to compute (from just four fundamental parameters; e.g., two geometrical parameters for the ellipsoid plus the total mass $M$ and the angular velocity $\omega$), useful for other disciplines (like geophysics), and also utilized for describing station positions (e.g. in connection with GNSS coordinates). Furthermore, the normal gravity field is defined such that the ellipsoid surface is a level surface of its own gravity field, for further details see Torge and Müller (2012).

Accordingly, the gravity vector and other gravity field observationals can be approximated by corresponding reference quantities based on the level ellipsoid, leading to the gravity anomalies $\Delta g$, height anomalies $\zeta$, geoid undulations $N$, etc. In this context, Bruns formula is of fundamental importance, giving the height anomaly or quasigeoid height (see also figure 2.2) as a function of $T$ as

$$\zeta = \frac{T}{\gamma} - \frac{W_0 - U_0}{\gamma} = \frac{T}{\gamma} - \frac{\delta W_0}{\gamma} , \tag{2.13}$$

where $W_0$ is the potential for a selected zero level surface (geoid; see above), $U_0$ is the normal potential for the surface of the level ellipsoid, $\delta W_0$ equals $W_0 - U_0$, and $\gamma$ is the normal gravity acceleration; the second term on the right side of the above equation is also denoted as height system bias and frequently omitted in the literature, implicitly assuming that $W_0$ equals $U_0$ (can be reached by defining the level ellipsoid accordingly). A similar formula can be obtained for the geoid undulation $N$, involving the disturbing potential for the geoid. The main advantage of the linearization process is that the residual quantities (with respect to the known ellipsoidal reference field) are in general four to five orders of magnitude smaller than the original ones, and in addition, they are less position dependent.

Hence, the disturbing potential $T$ takes over the role of $W$ as the new fundamental target quantity, to which all other gravity field quantities of interest are related. As $T$ has the important property of being harmonic outside the Earth's surface and regular at infinity, solutions of $T$ are developed in the framework of potential theory and GBVPs, i.e. solutions of the Laplace equation are sought that fulfil certain boundary conditions. In this context, gravity observations $g$ at or above the Earth's surface form one of the most important data sets, providing the input data for corresponding boundary conditions. The discussion and solution of GBVPs is far beyond the scope of this contribution, but two important results are the well-known spherical harmonic expansion of $T$ and the Molodensky solution that provides $T$ from a series of surface integrals, involving gravity anomalies and heights over the entire Earth's surface. In the first instance, the Molodensky solution can be written as

$$T = \mathbf{M}(\Delta g) , \tag{2.14}$$

where $\mathbf{M}$ is the Molodensky operator and $\Delta g$ are the gravity anomalies over the entire Earth's surface. The main term of the Molodensky solution is the Stokes integral, given by

$$T_0 = \frac{R}{4\pi} \iint_\sigma \Delta g \ S(\psi) d\sigma , \tag{2.15}$$

where $\psi$ is the spherical distance between the computation and data points, $S(\psi)$ is Stokes's function, $R$ is a mean Earth radius, and $\sigma$ is the unit sphere. In addition to the main Stokes term, the so-called Molodensky correction terms $T_i$ (for $i > 0$), considering that the data are referring to the Earth's surface and not to a level surface, have to be added.

The above integral formulae require a global gravity anomaly data set, but in practice, only local and regional discrete gravity field data sets are typically available for the area of interest and the immediate surroundings. This problem is remedied by the remove-compute-restore (RCR) technique, where the short and long wavelength gravity field structures are obtained separately from digital elevation models and a global satellite geopotential model, while the medium wavelength field structures are derived from regional gravity field observations, with the additional side effect that the collection of observational data can be restricted to the region of interest plus a narrow edge zone around it. In addition to this, the very long wavelength gravity field structures can be determined much more accurately from satellite data than from terrestrial data, but due to the signal attenuation with satellite height, the spatial resolution of the satellite models is restricted to a few 100 km, and hence the omitted short-wavelength signals are still at the several decimetre level for the geoid. Consequently, satellite measurements alone will never be able to supply the complete geopotential field with sufficient accuracy, but only a combination with high-resolution terrestrial data, mainly gravity and topography data with a resolution down to a few kilometres and below, can cope with this task. In this respect, the satellite and terrestrial data complement each other in an ideal way, with the satellite data accurately providing the long wavelength field structures, while the terrestrial data sets mainly contribute the short wavelength features.

Based on the law of error propagation, the uncertainty (standard deviation) for a regionally computed height anomaly $\zeta$, based on the combination of a global geopotential model with terrestrial data, is about 2 cm (Denker et al. 2017); this estimate represents an optimistic scenario and it is only valid for the case that a state-of-the-art global geopotential model (e.g. 5th generation GOCE model; Brockmann et al. 2014) and sufficient high resolution and quality terrestrial data sets (especially gravity measurements with a spacing of a few km and an uncertainty better than 1 mGal) are available around the point of interest (e.g., within a distance of 50 – 100 km). In view of future improved satellite gravity field missions and better terrestrial data, the perspective exists to improve the uncertainty to the level of 1 cm and even below.

Now, once the disturbing potential values $T$ are computed, either from a global spherical harmonic geopotential model or from a regional solution by equation (2.14) based on Molodensky's theory, the gravity potential $W$, needed for the relativistic redshift corrections, can be computed most straightforwardly from equation (2.12) as

$$W_P = U_P + T_P \ , \tag{2.16}$$

where the basic requirement is that the position of the given point P in space must be known accurately (e.g. from GNSS observations), as the normal potential $U$ is strongly height-dependent, while $T$ is only weakly height dependent with a maximum vertical gradient of a few parts in $10^{-3}$ m$^2$s$^{-2}$ per metre. The above equation also makes clear that the predicted potential values $W_P$ are in the end independent of the choice of $W_0$ and $U_0$ used for the linearization. Furthermore, by combining equation (2.16) with (2.13), and representing $U$ as a function of $U_0$ and the ellipsoidal height $h$, the following alternative expressions for $W$ (at point P) can be derived as

$$W_P = U_0 - \bar{\gamma}(h - \zeta) + \delta W_0 \ , \tag{2.17}$$

which demonstrates that ellipsoidal heights $h$ (e.g. from GNSS) and the results from gravity field modelling in the form of the quasigeoid heights (height anomalies) $\zeta$ or the disturbing potential $T$ are required, whereby a similar equation can be derived for the geoid undulations $N$. Consequently, the above approach (equations (2.16) and (2.17) is denoted here somewhat loosely as the GNSS/geoid approach, which is also known in the literature as the GNSS/GBVP approach (the geodetic boundary value problem is the basis for computing the disturbing potential $T$; see, e.g., Rummel and Teunissen 1988, Heck and Rummel 1990).

The GNSS/geoid approach depends strongly on precise gravity field modelling (disturbing potential $T$, metric height anomalies $\zeta$ or geoid undulations $N$) and precise GNSS positions (ellipsoidal heights $h$) for the points of interest, with the advantage that it delivers the absolute gravity potential $W$, which is not directly observable and is therefore always based on the assumption that the gravitational potential is regular (zero) at infinity (see Sect. 2.1). In addition, the GNSS/geoid approach allows the derivation of the height system bias term $\delta W_0^{(i)}$ based on equations (2.13) and (2.10) together with at least one (better several) common GNSS and levelling stations in combination with the gravimetrically determined disturbing potential *T*.

With respect to the uncertainty of the GNSS/geoid approach based on equations (2.16) or (2.17), the uncertainty contributions from the station coordinates and the (gravimetric) height anomalies have to be considered. The uncertainty of the GNSS positions in the relevant international reference frames is today more or less independent of the interstation distance, reaching (in the vertical) about 5 – 10 mm for the ellipsoidal heights (Altamimi *et al.* 2016 or Seitz *et al.* 2013). Assuming an uncertainty of 1 cm for the GNSS ellipsoidal heights and 2 cm for the quasigeoid heights without correlations between both quantities, a combined uncertainty of 2.2 cm (in terms of heights) is finally obtained for the absolute potential values, where metre units are preferred, but corresponding potential values can easily be obtained by multiplying these figures with an average gravity value (e.g., 9.81 ms$^{-2}$ or roughly 10 ms$^{-2}$).

On the other hand, for potential differences over larger distances of a few 100 km, the statistical correlations of the quasigeoid values virtually vanish, which then leads to a standard deviation for the potential difference of 3.2 cm in terms of height, i.e. √2 times the figure given above for the absolute potential (according to the law of error propagation), which again has to be considered as a best-case scenario. This would also hold for intercontinental connections (e.g. between metrology institutes), provided again that sufficient regional high-resolution terrestrial data exist around these places. According to this, over long distances across national borders, the GNSS/geoid approach should be a better approach than geometric levelling. For further details, see Denker *et al.* (2017).

## 2.4  Spacetime reference systems and international timescales

Due to the demonstrated performance of atomic clocks and time transfer techniques, the definition of timescales and clock comparison procedures must be handled within the framework of general relativity. Einstein's general relativity theory (GRT) predicts that ideal clocks will in general run at different rates with respect to a common (coordinate) timescale if they move or are under the influence of a gravitational field, which is associated with the relativistic redshift effect (one of the classical general relativity tests). The international timescales TAI and UTC are of primary importance and probably represent the most important application of general relativity in worldwide metrology today. In general relativity, it is important to distinguish between proper quantities that are locally measurable and coordinate quantities that depend on conventions. An ideal clock can only measure local time and hence it realizes its own timescale that is only valid in the vicinity of the clock, i.e., proper time. On the other hand, coordinate time is the time defined for a larger region of space with associated conventional spacetime coordinates. In this context, the SI second, as defined in 1967, has to be considered as an ideal realization of the unit of proper time (e.g., Soffel and Langhans 2013).

The relation between proper and coordinate quantities can be derived in general from the relativistic line element $ds$ and the (coordinate-dependent) metric tensor $g_{\alpha\beta}$. Considering spacetime coordinates $x^\gamma = (x^0, x^1, x^2, x^3)$ with $x^0 = ct$, where $c$ is the speed of light in vacuum, and $t$ is the coordinate time, the line element along a time-like world line is given by

$$ds^2 = g_{\alpha\beta}(x^\gamma)dx^\alpha dx^\beta = -c^2 d\tau^2 \ , \tag{2.18}$$

where $\tau$ is the proper time along that world line. In this context, Einstein's summation convention over repeated indices is employed, with Greek indices ranging from 0 to 3 and Latin indices taking values from 1 to 3. The relation between proper and coordinate time is obtained by rearranging the above equation, resulting in

$$\left(\frac{d\tau}{dt}\right)^2 = -g_{00} - 2g_{0i}\frac{1}{c}\frac{dx^i}{dt} - g_{ij}\frac{1}{c^2}\frac{dx^i}{dt}\frac{dx^j}{dt} = -g_{00} - 2g_{0i}\frac{v^i}{c} - g_{ij}\frac{v^i v^j}{c^2} \ , \tag{2.19}$$

where $v^i(t)$ is the coordinate velocity along the path $x^i(t)$.

In this context, the International Astronomical Union (IAU) introduced in Resolution B1.3 (2000) the Geocentric Celestial Reference System (GCRS) and the associated Geocentric Coordinate Time (TCG) for the modelling of all processes in the near-Earth environment. The GCRS has its origin at the geocentre of the Earth and is a non-rotating system: for further detail see (Soffel *et al.* 2003). Inserting the GCRS metric, as recommended by IAU Resolution B1.3 (2000), and using a binomial series expansion leads to

$$\frac{d\tau}{dt} = 1 - \frac{1}{c^2}\left(V + \frac{v^2}{2}\right) + O(c^{-4}) \ , \tag{2.20}$$

where $V$ is the gravitational (scalar) potential (denoted as "$W$" in the IAU 2000 resolutions), which contains parts arising from the gravitational action of the Earth itself ($V_E$; approximately given by $V_E \approx GM_E/r$, where $G \approx 6.67384 \times 10^{-11}$ m$^3$ kg$^{-1}$ s$^{-2}$ is the gravitational constant and $M_E \approx 5.97219 \times 10^{24}$ kg is the mass of the Earth) and external parts due to tidal and inertial effects ($V_{ext}$), $v$ is the coordinate velocity of the observer in the GCRS, and $t$ is the coordinate time TCG, while terms of the order $c^{-4}$ are omitted. In this context, it should be noted again that all potentials are defined with a positive sign. The above equation corresponds to the first post-Newtonian approximation and is accurate to a few parts in $10^{19}$ for locations near the Earth's surface, which is fully sufficient, as in practice the limiting factor is the uncertainty with which $V_E$ can be determined; for further details see (Soffel *et al.* 2003, Petit *et al.* 2014, or Denker *et al.* 2017).

The above fundamental relationship between proper time and coordinate time refers to the (non-rotating) GCRS and therefore all quantities depend on the coordinate time $t$ (TCG) due to Earth rotation. However, for many practical applications it is more convenient to work with an Earth-fixed system (e.g., the International Terrestrial Reference System, ITRS, co-rotating with the Earth), which can be considered as static in the first instance. Then for an observer (clock) at rest in an Earth-fixed system, the velocity $v$ in the above equation is simply given by $v = \omega\, p$, where $\omega$ is the angular velocity about the Earth's rotation axis, and $p$ is the distance from the rotation axis. Taking all this into account, equation (2.20) may be rearranged and expressed in the Earth-fixed system (for an observer at rest) as

$$\frac{d\tau}{dt} = 1 - \frac{1}{c^2}W(t) \ + \ O(c^{-4}) \ , \tag{2.21}$$

with $W(t)$ being the slightly time-dependent (Newtonian) gravity potential related to the Earth-fixed system, as employed in classical geodesy (see the above equation (2.5)). $W(t)$ may be decomposed into

$$W(t) = W^{static}(t_0) + W^{temp}(t - t_0) \ , \tag{2.22}$$

where $W^{static}(t_0)$ is the dominant static (spatially variable) part of the gravity potential at a certain reference epoch $t_0$ (denoted simply as $W$ in the previous sections), while $W^{temp}(t)$ incorporates all temporal components of the gravity potential (inclusively tidal effects and other temporal variations of all terms in equation (2.20)); for a discussion of the relevant time-variable terms, including the magnitudes and accuracies with which these can be computed, see, e.g. Wolf and Petit (1995), Petit *et al.* (2014) Voigt *et al.* (2016), as well as section 2.1. Based on this, the temporal and static components can be added according to Eq. (2.22) to obtain the actual gravity potential value $W(t)$ at time $t$, as needed, e.g., for the evaluation of clock comparison experiments.

The above equations (2.20) and (2.21) are based on the coordinate timescale TCG (related to the GCRS). For an earthbound clock (at rest) near sea level (realizing proper time) the sum of the gravitational and centrifugal potential generates a relative frequency shift (see equation (2.21)) of approximately $7 \times 10^{-10}$ (corresponding to about 22 ms / year) with respect to TCG, because the latter is related to the non-rotating GCRS. In order to avoid this inconvenience for all practical timing issues at or near the Earth's surface, Terrestrial Time (TT) was introduced as another coordinate time associated with the GCRS. TT differs from TCG just by a constant rate, which is specified through a conversion constant $L_G$ in the form

$$\frac{dt_{(TT)}}{dt_{(TCG)}} = 1 - L_G \ . \tag{2.23}$$

Previously, the constant $L_G$ was based on the "SI second on the rotating geoid" (IAU Resolution A4, 1991), but due to the intricacy and problems associated with the definition, realization and changes of the geoid, the International Astronomical Union (IAU) decided in Resolution B1.9 (2000) to turn $L_G$ into a defining constant,

$$L_G = 6.969290134 \times 10^{-10} \ , \tag{2.24}$$

with a fixed value and zero uncertainty, where the numerical value of $L_G$ was chosen to maintain continuity with the previous definition (Soffel *et al.* 2003). From the above equations (2.21) and (2.23) it is clear that the relation $L_G = W_0 / c^2$ holds, which means that $L_G$ is directly linked to a corresponding (zero) reference gravity potential value, usually denoted as $W_0$, with

$$W_{0 \ (IAU2000)} = 62,636,856.00 \ \text{m}^2\text{s}^{-2} \ . \tag{2.25}$$

As the speed of light $c$ is also fixed and has no uncertainty, the parameters $L_G$ and $W_0$ can be considered as equivalent, both having zero uncertainty.

The relativistic time dilation according to the above equations (2.20) and (2.21) is closely related to the relativistic redshift effect. Since the (proper) frequency is inversely proportional to the proper time interval with $f = 1/d\tau$, the aforementioned equations can be used to derive the relativistic redshift correction for a clock at rest on the Earth's surface as

$$\frac{\Delta f}{f_P} = \frac{f_P - f_0}{f_P} = 1 - \frac{d\tau_P}{d\tau_0} = \frac{W_P - W_0}{c^2} + O(c^{-4}) \approx \frac{-gH}{c^2} \ , \tag{2.26}$$

where $f_P$ and $f_0$ are the proper frequencies of an electromagnetic wave as observed at points $P$ at the Earth's surface and $P_0$ on the zero level surface (by corresponding ideal clocks, showing the same time under the same conditions), respectively, while $W_P$ and $W_0$ are the associated gravity potential values. $H$ is the vertical distance of point $P$ relative to point $P_0$, measured positive upwards, and $g$ is the gravity acceleration. An exact relation for the potential difference is given by $W_P - W_0 = -C_P = -\bar{g}H = -\bar{\gamma}H^N$ (see equations (2.8) and (2.9)).

The above equation is the classic formula relating frequency differences and gravity potential differences, where a fractional frequency shift of 1 part in $10^{18}$ corresponds to about 0.1 m²s⁻² in terms of the gravity potential difference, which is equivalent to about 0.01 m in height. Equation (2.26) assumes that the two stations $P$ and $P_0$ are earthbound and at rest within the (rotating) Earth-fixed system, i.e., both points are affected by the Earth's gravity (gravitational plus centrifugal) field and relative velocities between them are non-existent. Therefore, the term "relativistic redshift effect" is preferred over "gravitational redshift effect" in this context, as not just the gravitational, but also the centrifugal potential is involved; similar conclusions are drawn by Delva and Lodewyck (2013), Pavlis and Weiss (2003) and Petit and Wolf (1997). Regarding the sign of the relativistic redshift correction, assuming that $P$ is located above $P_0$, $W_P - W_0$ is negative and so is $\Delta f$; this means that the frequency or rate of an individual clock at $P$ above the zero level surface must be reduced by an amount $\Delta f$, given by equation (2.26), in order to produce the desired signal corresponding to a hypothetical clock at the zero level surface; hence, the relativistic redshift correction establishes a link to the frequency that an ideal clock would generate on the zero level reference surface. As all existing primary frequency standards are situated above the zero level surface, in practice, the relativistic redshift correction is always negative and can become quite significant, e.g., Pavlis and Weiss (2003) estimated a correction $\Delta f / f$ of about $-1800 \times 10^{-16}$ for the clocks at NIST in Boulder, Colorado, USA, with an altitude of about 1650 m.

TCG and TT are equivalent timescales related to the GCRS and hence they are both (conventional) coordinate timescale, which can have different realizations, such as TAI (International Atomic Time), UTC (Coordinated Universal Time), GNSS time, etc. Consequently, for contributions to international timescales, a relativistic redshift correction according to equation (2.26) is needed to transform the proper time observed by a local clock into TT, where the (conventional) zero potential value $W_{0 \ (IAU2000)}$ according to equation (2.25) should be employed in order to be consistent with the IAU definition of TT; for this task, only the uncertainty of $W$ matters (as $W_{0 \ (IAU2000)}$ has zero uncertainty), while potential differences suffice for local and remote clock comparisons, with the proviso that the actual potential values and the clock frequency measurements must refer to the same epochs. The latter requirement means that the magnitude of time-variable effects in the gravity potential due to solid Earth and ocean tides as well as other effects (see above) must be taken into account for all clock measurements at a performance level below roughly 5 parts in $10^{17}$ (see equation (2.22)). This is especially important for contributions to international timescales and remote clock comparisons over large distances in cases where relatively short averaging times are used, since in such situations the time-variable gravity potential components may not average out sufficiently. Moreover, the tidal peak-to-peak signal, which can be computed with an uncertainty better than $10^{-18}$ in terms of clock frequency (Voigt *et al.* 2016), could also prove useful for evaluating the performance of optical clocks, by providing a detectability test.

## 2.5 Clocks for establishing physical height reference systems

An important consequence of equation (2.26) is that geodetic knowledge of the Earth's gravity potential and heights is required to predict frequency shifts between local and remote (optical) clocks, and vice versa, frequency standards can be used to determine gravity potential differences. The idea of using clocks to determine potential (and corresponding height) differences in geodesy dates back to Bjerhammar and Vermeer (Bjerhammar 1975 and 1985; Vermeer, 1983). This "chronometric levelling" offers the great advantage of being independent of any other geodetic data and infrastructure, with the perspective to overcome some of the limitations inherent in the classical geodetic approaches. Furthermore, Bjerhammar (1985) also gave the following remarkable definition for the geoid, stating that "The relativistic geoid is the surface where precise clocks run with the same speed and the surface is nearest to mean sea level". In this context, a comparison of equations (2.26) and (2.7) shows that the (classical) definition of the geoid as an equipotential surface is, to the level of approximation used above (roughly a few parts in $10^{19}$), identical with a surface on which clocks tick with the same rate. However, with the advent of optical clocks with below the level of $10^{-18}$ and corresponding links for their remote comparison, proposals for a new definition of a relativistic geoid based on optical clocks have become more and more realistic (Kopeikin *et al.* 2011; Müller *et al.* 2017; Phillip *et al.* 2017).

The most straightforward application of atomic clocks in geodesy is chronometric levelling. Based on equation (2.26), the frequency difference between two clocks, observed via an appropriate link, directly gives a gravity potential difference between the two sites, which then can be introduced easily as an additional observation into the levelling network adjustment (as described already in Sect. 2.2). Assuming that atomic clocks are approaching soon the $10^{-18}$ regime (or below), the corresponding height uncertainty would be 1 cm (or better). In this context, the main advantage of chronometric levelling is that its uncertainty is almost independent of the distance, in contrast to geometric levelling, which is very accurate over short distances, but susceptible to systematic errors over long distances (see above). Hence, both techniques complement each other in an optimal way, such that a few clock points per country integrated into a continental levelling network, for instance the EVRF (European Vertical Reference Frame), could greatly stabilize the entire network and improve the long-distance uncertainty. In addition, clocks at the $10^{-18}$ level and below could be used for long-time measurements of the variability of the gravity potential, e.g. associated with geodynamic effects (see below). Furthermore, once the optical clocks and link technologies become operationally available, for instance similar to GNSS receivers, clocks would be much more economic and faster than the time-consuming and costly geometric levelling.

The principle of chronometric levelling is also shown in Fig. 2.2, where a spiral-shaped line indicates the (fibre) link. In the figure it is assumed that one (reference) clock is located on the geoid, and a second clock is located at some distance at a point P on the Earth's surface, giving the potential difference between both sites as well as a physical height for P (either $H$ or $H^N$ based on equations (2.8) and (2.9), respectively). The same principle could also be applied to relate different tide gauges to a common reference level surface, and to derive the DOT on the oceans as direct input to ocean circulation models and studies, again provided that the clocks and link technologies can be realized, where the DOT case is a special challenge, as it requires continuous operation on ships. Also for the tide gauge and DOT cases, the links in Fig. 2.2 are indicated by a spiral-shaped line, but in the future only satellite links appear to be practical and feasible.

At the moment, only fibre links (see Sec. 6.2) have sufficient accuracy for clock comparisons at the $10^{-18}$ level and below, which means that only height and DOT differences can be determined with respect to a reference clock station. However, both cases would greatly benefit from a space-based master or reference clock, e.g. on a geostationary satellite, where, due to the relatively high altitudes, spatial and temporal variations of the Earth's gravitational field are significantly reduced and smoothed out. Then already with the presently existing satellite gravity models, the uncertainty of the gravitational potential at satellite level is primarily depending on the orbit uncertainty, such that an orbit uncertainty of about 10 cm (which seems to be achievable) would allow the computation of the (absolute) satellite reference gravitational potential (presupposing that it is regular or zero at infinity) with an uncertainty of about 0.01 $m^2s^{-2}$. Then, assuming that the link between the satellite and ground stations can be realized with appropriate uncertainty, this would allow the derivation of corresponding (absolute) gravity potential values for the ground stations, again with an uncertainty at the level of 0.01 $m^2s^{-2}$, corresponding to 1 mm in height. Consequently, with a space-master clock and corresponding link technologies, a global height reference system based on atomic clocks could be realized with respect to a conventional reference potential value $W_0$, as used for instance for the definition of the IHRS and the IAU zero reference level surfaces, denoted by Petit *et al.* (2014) as "classical geoid" and "chronometric geoid", respectively (see above). Such a clock-based geodetic height reference would also provide a long-term stability, needed for the direct monitoring of the geopotential field.

However, at present, geodetic results serve mainly for the evaluation of the optical clocks and their uncertainty budgets in order to gain confidence in the new generation of clocks within the international metrology community and beyond, but in the future, this may change as outlined above. The GNSS/geoid approach gives absolute gravity potential values, presently with an uncertainty of about 2 cm in terms of heights, but with the perspective for further improvements (Denker *et al.* 2017); this requires sufficient high resolution and quality terrestrial (gravity and terrain) data around the sites of interest, which may not exist in remote areas such as parts of Africa, South America, Asia, etc. On the other hand, the geometric levelling approach can deliver potential differences with millimetre uncertainties over shorter distances, but is susceptible to systematic errors at the decimetre level over large distances. Consequently, over long distances across national borders, the GNSS/geoid approach should be a better approach than geometric levelling. Practical results from both the geometric levelling and the GNSS/geoid approach are in line with these considerations, showing an agreement at the few centimetre level over a few 100 km distance, while the two approaches are presently inconsistent at the decimetre level across Europe (see Denker *et al.* 2017, Kenyeres *et al.* 2010, Gruber *et al.* 2011). For this reason, the more or less direct observation of gravity potential differences through optical clock comparisons (with targeted fractional accuracies of $10^{-18}$, corresponding to 1 cm in height) is eagerly awaited as a means for resolving the existing discrepancies between different geodetic techniques and remedying the geodetic height determination problem over large distances (see below).

**2.6 Clocks for gravity field modelling and geoid determination**

Closely related to height reference systems is the topic of gravity field modelling including geoid and quasigeoid determination, because the resulting high-resolution geoid and quasigeoid models may also be used together with GNSS measurements to define a so-called "geoid based vertical datum" (see above). In this context, geoid determination is understood as the determination of the shape and size of the geoid with respect to a well-defined coordinate reference system, which usually means the determination of the height of the geoid (geoid height) above a given reference ellipsoid. The problem is solved within the framework of potential theory and GBVPs, where the task is to find a harmonic function (i.e. the disturbing potential $T$) everywhere outside the Earth's masses (possibly after mass displacements and reductions), which fulfils certain boundary conditions. In principle, all measurements that can be mathematically linked to the disturbing potential $T$ (e.g. gravity anomalies, vertical deflections, gradiometer observations, and point-wise disturbing potential values itself), can contribute to the solution, but in practice gravity measurements play the main role in combination with topographic and global satellite gravity information – also denoted as the gravimetric method (see also above). A very flexible approach, with the capability to combine all the aforementioned (inhomogeneous) measurements of different kinds and the option to predict (output) heterogeneous quantities related to $T$, is the least-squares collocation (LSC) method (Moritz 1980).

Regarding the use of clocks for gravity field modelling and geoid determination, this always implies that also precise positions of the clock points with respect to a well-defined reference system are required. This concerns mainly the ellipsoidal height, which should be available with the same (or lower) uncertainty than the clock-based physical heights, such that gravity field related quantities $N = h - H$ or $\zeta = h - H^N$ (cf. equation (2.10)) can be obtained, establishing a direct link to the disturbing potential $T$ (e.g. through equation (2.13); this is exactly the same situation as a combination of GNSS and geometric levelling, so-called GNSS/levelling points). However, this point is frequently overlooked in publications and sometimes it is somewhat misleadingly mentioned that clocks can be used to determine the geoid (e.g. Bondarescu *et al.* 2012), but this is an incomplete statement, as always clock plus GNSS measurements are required for gravity field modelling. Furthermore, in view of further improved clocks at (or below) the $10^{-18}$ level, it should also be noted that a (ellipsoidal) height uncertainty of 5 to 10 mm is about the limit of what is achievable with GNSS today, requiring static and sufficiently long observation sessions and an appropriate post-processing. Clocks alone, if compared with a (space) reference clock with known potential value, can only help to realize the geoid, i.e. to find its position with respect to a given measurement point on the Earth's surface (usually at a certain distance below this measurement point), but this still does not mean that we know the coordinates of the corresponding geoid point (e.g. its ellipsoidal height or geoid height).

A first study in the direction of gravity field modelling and geoid determination with simulated clock and GNSS measurements is the publication of Lion *et al.* (2017), although the study never mentions that precise GNSS is an essential part of it. Presupposing that the GNSS and clock measurements can provide the disturbing potential $T$ according to equations (2.13) and (2.10), these data are utilized within an LSC approach together with gravity data in the Massif Central and the French Alps to predict $T$ at arbitrary points. The results suggest that a bias and trend in the recovered geoid or disturbing potential can be reduced significantly by adding about 1% of clock data to the existing gravimetry data, but the results may also be affected by the quite small data collection area and the covariance functions used in their approach. Furthermore, the stabilization of geoid and quasigeoid solutions by some additional GNSS/clock points is exactly the same situation as using GNSS plus geometric levelling (GNSS/levelling) points, as done since many years (e.g. Denker 1988 and 1998). Finally, with continuously operating clocks and GNSS positions, the GNSS/clock based approach could also be used in kinematic mode to do areal geoid/quasigeoid surveys, e.g. along survey lines (along roads) crossing each other. Another geodetic application in this direction would be spaceborne clock measurements of the redshift effect with respect to some reference clock (optimally a master clock in space as described above) for gravity field recovery missions, as discussed, for instance, in (Mayrhofer and Pail 2012) and (Müller *et al.* 2017).

## 2.7 Clocks for the monitoring of geodynamics

Facing the societal challenges ongoing with climatic change and increasing population, the IAG initiated the Global Geodetic Observing System (GGOS; Plag and Pearlman 2009) with the aim to integrate different geodetic techniques for providing the observational base to monitor phenomena and processes in the System Earth.

While geometric methods like GNSS and geometric levelling can observe position and height variations with time, ground based gravimetric techniques are sensitive not only to vertical movements of the measurement point, but also to near-surface mass variations within the crust. Therefore, geometric and gravimetric techniques can be employed as complementary methods to enable the discrimination between subsurface mass movements that are associated with or without surface deformations. However, gravimetry generally suffers from not well-known local variations, for instance, all gravimetric measurements in Europe are affected by irregular groundwater changes or other hydrological processes (e.g. soil moisture variations), and such local signals can also originate from human activities (mass changes due to the withdrawal of water, oil, or gas, large construction activities, etc.). These local effects are superimposed on the target signal and should be removed as much as possible. Nevertheless, the applied reductions have their inherent limitations in accuracy and completeness, which may cause misleading interpretations of the originally well-observed raw data. In this context, the prospect of optical clocks as a new geodetic tool may help to overcome to a large extent the drawback of superimposed and dominating local effects in gravimetry (measuring the gradient of the potential), because clocks (observing the potential) are mainly sensitive to large mass changes, while being quite insensitive to (small) local mass effects.

To further illustrate this point, it is assumed in a first scenario that mass shifts in the upper layer of the Earth's crust occur due to groundwater changes, which are not accompanied by any position change of the instrument site (no surface deformation, no distance change with respect to the geocenter). The example case assumes a vertical rise of the groundwater table by 1 m within a radius $s_0$ varying from 10 m to 1000 km, a water density of 1000 kgm$^{-3}$, and a porosity of 33 %, leading to a density contrast of 333 kgm$^{-3}$ for the calculation of the gravitational effect. Furthermore, both potential and gravity effects are calculated for a station located centrally 5 m above the circular disk using a simplified formula for a planar disk as an approximation (see Strang van Hees 1977). The set-up of the example, assuming that a gravimeter and a clock are installed side by side at the central calculation point, and the results are shown in Fig. 2.3, where the potential effects are given in metre units obtained by deviding the potential values by a corresponding gravity value – thus the given values in mm correspond to a vertical change of the equipotential surface (or the geoid). The figure clearly shows that local mass variations within the Earth's crust ($s_0$ = 10 … 1000 m) result in measurable signals for the gravimeter (77 to 138 nms$^{-2}$), but the optical clock (measuring the potential effect) is quite insensitive to such local phenomena (less than 0.02 mm). Furthermore, it is clear that the gravity effect is dominated by the local surrounding, while the potential effect requires a sufficiently large total mass change (over a large area), exceeding the 1 mm level in the present example for a disk size ($s_0$) of 100 km. Hence, gravimeters are more suitable for prospection and local groundwater studies, where local mass anomalies and variations close to the Earth's surface have to be detected, whereas clocks can only resolve very large mass anomalies, e.g. a 20% density anomaly concentrated in a sphere of 1.5 km radius in 2 km depth (Bondarescu *et al.* 2012).

Regarding the measurability of the aforementioned signals due to ground water changes, stationary gravimeters (recording time series) can significantly observe such a variation with a resolution of some 1 nms$^{-2}$ (spring meters) up to the sub 0.1 nms$^{-2}$ level (superconducting gravimeters) if it occurs in the vicinity of the gravimeter (within a radius of some 100 m to a few 1000 m). However, employing clocks for the measurement of such geodynamic effects requires first of all a sufficiently large total mass change (i.e. a 100 km circular disk in the present example) and an improved clock sensitivity in the 10$^{-19}$ regime, corresponding to 0.01 m²s$^{-2}$ in potential or 1 mm in height. In addition, a link is required to a second clock station that is unaffected by the local or regional mass changes, where the optimal case would be a reference clock in space at high altitude, as explained earlier in chapter 2.5. Consequently, clocks at this level of performance could also complement present and future satellite gravity missions for enhancing the spatial and temporal resolution of corresponding mass change maps. The clocks could provide results with a high temporal resolution (e.g. 1 hour or less) for understanding the daily to annual evolution of corresponding phenomena, which makes the clocks unique in their ability to continuously monitor regional variations of the gravity potential field, especially when using a well distributed clock network (e.g. clocks separated by 100 km or less). At present, the GRACE satellite mission provides monthly gravity field models (due to the span it takes to cover the whole Earth), e.g. in the form of geoid maps, with an uncertainty of about 1 mm at a resolution (half wavelength) of about 200 km and 1 cm at 150 km, which may be disturbed by regional signals of transient or episodic character. On the other hand, the GOCE mission aims mainly at the static gravity field, giving a geoid uncertainty of about 1 – 2 cm at a resolution of about 100 km, showing an agreement with external GNSS, levelling, and other data in Germany at the level of about 3 cm (Pail 2017). Hence, overall, clocks seem to be a good tool to complement satellite gravimetry missions in terms of resolution and amplitude uncertainty.

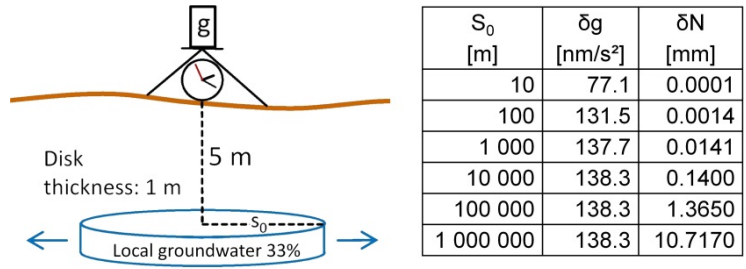

| $S_0$ [m] | $\delta g$ [nm/s²] | $\delta N$ [mm] |
|---|---|---|
| 10 | 77.1 | 0.0001 |
| 100 | 131.5 | 0.0014 |
| 1 000 | 137.7 | 0.0141 |
| 10 000 | 138.3 | 0.1400 |
| 100 000 | 138.3 | 1.3650 |
| 1 000 000 | 138.3 | 10.7170 |

*Figure 2.3: Impact of groundwater body on gravity acceleration δg and gravitational potential δN (given in metre units obtained by dividing the potential values by a corresponding gravity value; corresponds to a geoid shift), as observed at the Earth's surface by co-located gravimeter and optical clock instruments. The groundwater body is positioned 5 m below the instrument site and taken as a planar disk with 1 m thickness and varying radius $s_0$ (courtesy L. Timmen, Institut für Erdmessung, Leibniz Universität Hannover).*

In a second scenario, let us assume a surface-bound measurement point, which serves for monitoring the vertical stability of the site, e.g. a tide gauge, a fundamental geodetic observatory (ITRS station), or a station of a geodynamic monitoring network. In order to be independent from a single measurement technique and to increase the redundancy and reliability, ideally several different methods should be employed, where the instruments should complement each other to show an agreement or to reveal a contradiction in the case of failure. Taking the Zugspitze mountain in southern Germany as an example case, the mountain area is affected by a diminishing permafrost area and a snow cover that varies dramatically with the seasons of a year and also between the years, which is to be monitored by a geodetic observatory on the top of the mountain, equipped with co-located GNSS, gravimeter, and optical clock instruments (see Fig. 2.4). The increasing instability of the rock material causes a vertical ground subsidence of about 1 cm/year at the mountaintop site. Moreover, the snow cover changes lead to mainly annual (but also long-period) signals, associated with gravity variations of 0.5 μms⁻², which can be applied as a reduction to the gravity observations with an arguable uncertainty of 0.05 μms⁻² (10%). Furthermore, as the vertical movement of 1 cm per year of the observatory at the peak corresponds to a gravity change of only about 0.03 μms⁻² per year, this means that the gravimeter recording will not allow a significant detection of the change after one year of measurements. On the other hand, the continuous GNSS (GPS) measurements at the 3000 m high site have to be processed relative to an antenna in Garmisch-Partenkirchen in the valley (height about 750 m); therefore, the GNSS results will suffer from the different atmospheric conditions and thus the discovery of the 1 cm height change is unlikely. However, a clock with a long-term instability below $10^{-18}$ can observe the variation in the potential field of 0.1 m²s⁻² (corresponding to 1 cm vertical movement of the equipotential surface in one year), as it is almost unaffected by the local mass variations from the snow cover and atmospheric changes.

The publication by Bondarescu *et al.* (2015) already covers the deployment of clocks in conjunction with gravimeters for monitoring geodynamic processes. They discuss two applications in detail. First, an inflating magma chamber is simulated and modelled for the Etna volcano, where the vertical uplift of the instrument (ground displacement) within the potential field by some centimeters is the dominating effect on the clock observations, being accompanied by a much smaller effect from the mass redistribution in the underground. Secondly, the globally acting tides due to the Newtonian attraction of the Moon and Sun cause continuously ongoing deformations of the Earth and its geopotential field, which can be monitored by geometric techniques such as GNSS, by gravimeters, and possibly by clocks in the future. As these measurement techniques are affected differently by the Earth's elasticity parameters (Love numbers), Bondarescu *et al.* (2015) discuss the calibration of existing Earth tide models as well as the separation of the Love numbers *h* (vertical deformation) and *k* (deformation potential) by gravimeter and clock measurements. While vertical tidal displacements from present models are confirmed to better than 1 mm by space-geodetic techniques such as VLBI and GNSS (i.e. about one part per thousand; Agnew 2007, Yuan and Chao 2012, Krásná *et al.* 2013, Torge and Müller 2012), the deformation potential and associated Love number *k* are mainly based on geophysical Earth models with largely unknown accuracy (Seitz *et al.* 2012) and thus not verified by direct observations, which could be a target for future accurate clock networks.

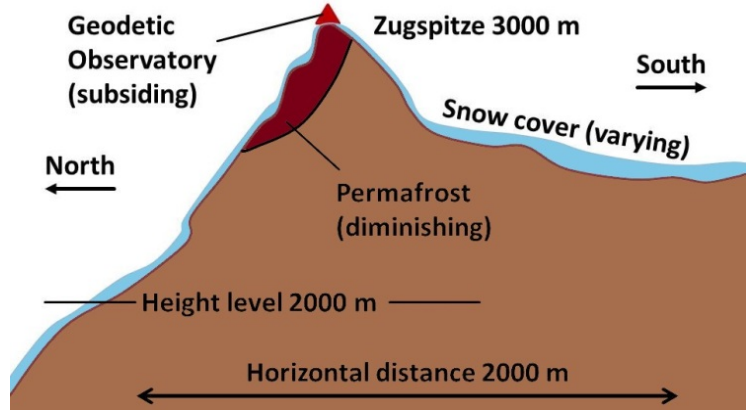

*Figure 2.4: Situation around the Zugspitze mountain in the German Alps. The area is affected by an increasing destabilization of the mountain in connection with the diminishing permafrost area and by mass variations from seasonal and long-term snow cover changes. This mainly leads to a subsidence of the mountaintop of about one centimetre per year and further seasonal and other periodical effects, which are to be monitored by a geodetic observatory on the top of the mountain, equipped with co-located GNSS, gravimeter, and optical clock instruments* (courtesy *L. Timmen, Institut für Erdmessung, Leibniz Universität Hannover*).

In summary, geodynamic investigations could benefit from a combination of geometric measurements with physical observations from gravimeters and clocks, because these techniques are affected differently by geometry changes and gravity field variations associated with corresponding mass changes. Especially clocks provide a new tool to directly observe potential changes, which are quite insensitive to small local mass changes in contrast to gravity measurements (related to the gradient of the potential) or even gradiometry (related to the 2nd derivatives or curvature of the potential). As a consequence, clocks are likely not to be deployed as stand-alone instruments, but should be understood as a complementary tool in the realm of applied geophysics and geometric techniques.

## 3. Relativistic Redshift Measurements - To which level can we trust Einstein's GR?

Chronometric levelling experiments rely on the predictions of Einstein's general relativity (GR), in particular motional time dilation shifts and the gravitational redshift. The validity of GR is frequently questioned as quantum theory and general relativity, the two pillars of modern physics, are incompatible (Will 2014). Thus, in order to exploit these relativistic effects, their validity under the given conditions has to be trusted. In the following, we briefly review tests of motional time dilation, which needs to be considered for clocks on the rotating Earth, and the gravitational redshift, which has been tested to a much lower level, both representing the basis for geodetic applications of clocks. A more thorough treatment of clocks and relativity is found in (Mai 2013, Petit *et al.* 2014).

Einstein's general relativity theory predicts that an ideal clock at rest will run at a slower rate, when under the influence of a gravitational potential, compared to a clock outside of it. For clocks placed at different sites with local gravitational potential $V_1$ and $V_2$ this redshift is expressed in terms of the difference in ticking rate, i.e. the difference of the clock frequencies[1],

$$\frac{\Delta f}{f} = \frac{f_1 - f_2}{f} = -\frac{\Delta V}{c^2}, \quad \text{where} \quad \Delta V = V_1 - V_2. \quad (3.1)$$

The existence of the gravitational redshift in the gravitational field of Earth was first verified in 1960 using nuclear Mößbauer spectroscopy. With a gamma ray source and absorber at an effective height difference of 45 m on Earth's surface a fractional frequency shift between the two sources of 5 x $10^{-15}$ was resolved with an uncertainty of 10 % (Pound and Rebka 1960). Today, when comparing modern atomic Cs fountains, which operate with a fractional frequency uncertainty below $10^{-15}$ (BIPM) and are stationed at different locations on Earth, the gravitational redshift and $2^{nd}$ order Doppler shifts are accounted for on an everyday basis. For the establishment of the international time scale (TAI) a gravitational redshift correction of about $1 \times 10^{-16}$ per metre height difference with respect to a virtual clock at zero gravity potential surface is applied to the atomic clock contributions (Pavlis and Weiss 2003, Calonico *et al.* 2007, Guinot 2011). Since 2003, this has provided a definition of an international time scale independent of the geoid (Soffel *et al.* 2003), still relying on the knowledge of local geodetic heights, see section 2.4. Today's realization of TAI with an uncertainty of $10^{-15}$ requires determining the geodetic height at the contributing clock on the order of one metre.

---

[1] Note that we are using the geodetic convention for the definition of the gravitational potential in equation (3.1), where water flows from low to high gravity potentials, as used in section 2.

The capability of modern optical clocks to resolve height differences of tens of cm in the gravitational field on Earth's surface was demonstrated first in 2010, where two $Al^+$ quantum logic clocks were compared against each other while elevating one of them on an optical table to a height difference of 30 cm (Chou *et al.* 2010b). With continuing improvement, the accuracy and the frequency stability of optical clocks approaches today a sensitivity to geodetic height differences at the level of a centimetre. This means, however, that integrating optical clocks into the international time scales TAI and UTC to improve their uncertainty will require a better understanding of the geopotential across the continents and careful consideration of higher relativistic effects (Denker *et al.* 2017, Petit *et al.* 2014).

As clocks are improving further and further, also the reversed scheme of using clocks to test the prediction of GR becomes an important application. One of the foundations of the theory of general relativity is Einstein Equivalence Principle (EEP), which can be split in three sub-principles: the Local Lorentz Invariance (LLI), the Local Position Invariance (LPI) and the Universality of Free Fall (UFF). In the search for unifying theories and the puzzle of Dark Energy and Dark Matter (NASA), several alternative theories of gravitation predict violations of the EEP. The precise measurement of gravitational redshifts come in here as an important tool to test the validity of general relativity and search for violations of the LPI.

On Earth, tests of the universality of the gravitational redshift can be performed in lab experiments with optical clocks of different species using the annual orbit of our planet around the sun. Due to the ellipticity of Earth's orbit, the annual variation of the gravitational redshift in the gravitational potential of the sun is $\Delta V/c^2 \approx \pm 3.3 \times 10^{-10}$ (Fortier *et al.* 2007). A violation of LPI would manifest itself in a modulation of the frequency ratios between different clocks. The supposed differential effect is parameterized as $\Delta f/f = (\beta_1 - \beta_2) \Delta V/c^2$, where $\beta_1$ and $\beta_2$ are two clock species dependent parameters. Long term clock comparisons over several years have been performed using Cs and Rb fountains, hydrogen masers and optical clocks (Fortier *et al.* 2007, Blatt *et al.* 2008, Tobar *et al.* 2013), giving the most stringent limit on $\beta_1 - \beta_2$ with $\beta_H - \beta_{Cs} = 4.8 \times 10^{-6}$ and $\beta_H - \beta_{Rb} < 10^{-5}$ (Tobar *et al.* 2013). The best absolute test of the gravity redshift to date was performed with an atomic clock in space, in the frame of the Gravity Probe-A (GP-A) mission in 1976, during which the frequency of a hydrogen maser launched on a rocket to 10000 km altitude was compared to a hydrogen maser on ground (Vessot and Levine 1976). With this experiment, the predicted gravitational red shift was verified with an uncertainty of $1.4 \times 10^{-4}$.

Already the recent launch of Galileo satellites 5 and 6 could enable further tests of the relativistic redshift. Recent work proposes to use the stable on-board clocks to perform a test of the LPI. Considering realistic noise and systematic effects, and thanks to an (accidental) highly eccentric orbit, it is possible to improve on the GP-A limit to an uncertainty around $(3-4) \times 10^{-5}$ after one year of integration of Galileo 5 and 6 data (Delva *et al.* 2015). An improvement to $10^{-5}$ is expected, when using the on-board atomic hydrogen maser clock on board of the RadioAstron satellite with highly eccentric orbit (Litvinov *et al.* 2017). An even more stringent limit can come from the planned space mission ACES with an ultra-stable Cs clock in orbit, which aims for a 35-fold improvement of the GP-A results (Heß *et al.* 2011). Proposals for future optical clocks in space target at gravitational redshift tests at the level of $10^{-7}$ (Altschul *et al.* 2015).

While the present boundaries on relative gravitational redshifts have so far been limited by the performance of clock measurements at accessible gravity differences and are still at the $10^{-4}$-level, tests of motional time dilation have been boosted by the use of relativistic atomic beams in accelerators. Using precision laser spectroscopy on accelerated atomic beams is a highly sensitive testbed of time dilation due to the motion of an emitting and absorbing quantum object (Gwinner 2005). Using $Li^+$ ions as clocks at relativistic speed, the lasers' Doppler shifted resonance frequency was measured with a relative accuracy of $< 4 \times 10^{-9}$ (Botermann *et al.* 2014) and a constraint on the

LLI violating parameter of $|\alpha| \leq 2 \times 10^{-8}$ was deduced in this experiment. Only recently, these limits have been overcome by optical clock measurements. A new test based on a fibre-network based frequency comparison of four optical lattice clocks using Sr atoms, two located in Paris, France, one in Braunschweig, Germany, and one in Teddington, UK, improved upon this best previous constraint on the LLI violating parameter α by a factor of around two (Delva *et al.* 2017).

If clocks are to be used in the future for geodetic applications on Earth, these validated tests of GR will be the boundary of trust of geodetic clock measurements. The best absolute test of the gravitational redshift at the level of $10^{-4}$ presents the strictest boundary: it allows to measure changes in the gravitational potential on Earth's surface due to tides (with decimetre amplitudes in terms of height, see section 2) with a trusted validity of GR corresponding to a few 0.01 millimetres and less, but when geodetic height differences of more than a 1 km are to be determined by direct clock comparison, the tested validity of GR already corresponds to an uncertainty of 10 cm. This means the remote clock comparison cannot distinguish between the gravitational signal and a possible deviation from Einstein's predictions. To push beyond this limit, verified clock tests at known gravity potential differences are needed. Also, space missions like ACES, targeting redshift tests at $10^{-6}$ and below, will remedy the situation and push the limits of verified GR tests to 1 mm at 1 km height difference. In general, the continuous improvement of optical clocks, in particular the improved technological readiness for robust and reliable operation, will enable to place better and better constraints on the fundamental principles of nature or to search for new physics with Earth bound and space clocks.

## 4. Atomic Clocks and Frequency Standards

In this section, we provide an overview over the basic concepts and techniques of operating optical frequency standards, aimed at the interested geodesist. Typical characteristics of clocks, and in particular those relevant for geodetic applications, are discussed.

### 4.1 Introduction to frequency standards

Atoms with their discrete energy levels are ideal references for time and frequency, since all atoms of the same isotope are identical. It is the main goal of atomic frequency standards to provide a signal corresponding to the unperturbed transition frequency between two such long-lived energy levels, which is defined by natural constants. The output of a frequency standard is a microwave or optical frequency. The standard is operated as a clock by counting the number of cycles that have passed since a well-defined starting time (Grebing *et al.*, 2016; Le Targat *et al.*, 2013). Despite this conceptual difference, "frequency standard" and "clock" are terms that are sometimes used synonymously. In the following, we will discuss the principle of (optical) frequency standards, their characterization, and some of their critical components.

Figure 4.1(a) shows the schematic of an optical frequency standard. It consists of a narrow-linewidth clock laser that periodically probes a single atom or an ensemble of atoms. The signal from the atoms is proportional to the excitation probability of the atom(s) during this probe time (figure 4.1(b)). It is used to stabilize the frequency of the clock laser to the atomic resonance. Some of the clock laser light is split off and represents the output of the frequency standard. An optical frequency comb is used to compare the frequency standard to other standards (typically operating at different transition frequencies), or to count the optical cycles. The two main characteristics of a frequency standard are its (in)stability and its systematic uncertainty, colloquially termed "accuracy". Instability quantifies the statistical uncertainty of the frequency standard's output frequency $f$, whereas accuracy refers to (usually an estimate of) the uncertainty of the deviation of the frequency standard's output frequency $f$ from the unperturbed reference frequency $f_0$ of the atom. These statistical and systematic uncertainties are also known as Type A (denoted by $u_A$) and Type B (denoted by $u_B$) uncertainties (BIPM et al., 2008), respectively.

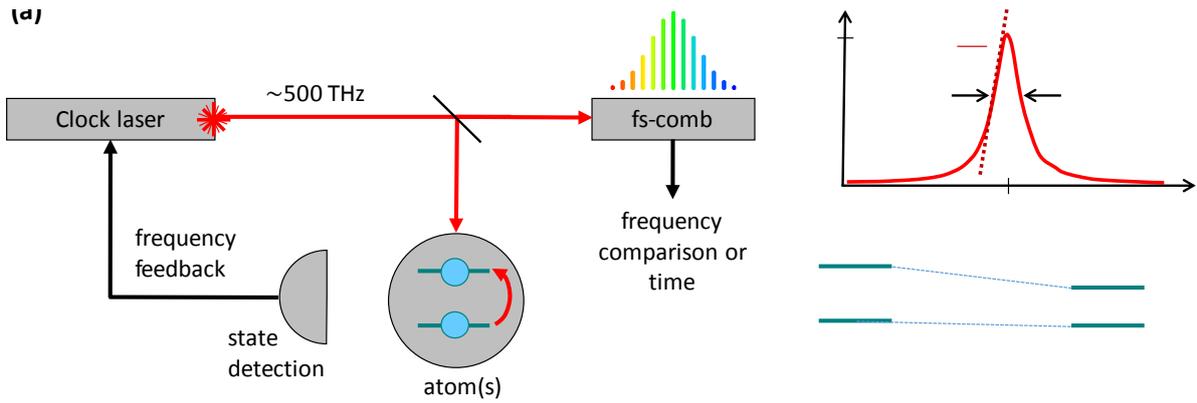

*Figure 4.1: (a) Schematic of an optical frequency standard. A laser with high frequency stability and small linewidth is used to probe an atomic transition. The transition probability of the atomic reference is detected, an error signal is generated and fed back to the laser to steer its frequency onto the atomic resonance. Part of the laser's radiation is split off and its oscillations are either counted or compared to another clock. (b) Excitation signal $S(f)$ of an atomic resonance with center frequency $f_0$ and width $\Delta f$ as a function of frequency $f$ during interrogation. It is used to steer the clock laser onto the atomic transition frequency. (c) The unperturbed energy levels are shifted by external fields and other effects to their measured values.*

## 4.2 Accuracy and systematic shifts

The unperturbed frequency of the reference transition is defined for a free and isolated atom at rest (with respect to a chosen reference frame) in the absence of any external fields or radiation. A close approximation of this ideal situation has been achieved through the development of laser cooling techniques (Chu, 1998; Cohen-Tannoudji, 1998; Phillips, 1998) and by trapping the atoms either in an optical lattice (Katori *et al.*, 2003) or in an ion trap (Bergquist *et al.*, 1987) under ultra-high vacuum conditions. This way almost perfect isolation from the environment can be achieved, while having exquisite control over all quantum mechanical degrees of freedom. The accuracy of a frequency standard is determined by estimating all possible effects (and their uncertainty) that could lead to a shift of the reference transition. The output frequency of the actual standard in the laboratory is corrected by the magnitude of the shifts, whereas the root mean square (RMS) of the sum of the uncertainties of all shifts (assuming they are independent) is its estimated uncertainty $u_B$. The magnitude and uncertainty of all shifts are typically listed in the form of an estimated error budget. In general, the most promising atomic candidates as references for frequency standards are those for which the number of different shift contributions, their magnitudes, and – in particular – their uncertainties are small. One of the important shifts is the 2$^{nd}$ order Doppler or relativistic time dilation shift

$$\Delta f_D/f_0 = -\frac{E_{kin}}{mc^2}, \qquad (4.1)$$

where $E_{kin}$ is the kinetic energy, $m$ the mass of the atom, and $c$ the speed of light. It arises from residual thermal motion or driven micromotion (in ion traps) and is a consequence of the time dilation in special relativity. Laser cooling to temperatures of 1 mK and below suppresses this shift that scales inversely with the mass and is thus more relevant for light clock atoms. Most other shifts depend on an external field that results in a shift proportional to an atomic property (Itano, 2000). Examples are the linear and quadratic Zeeman effects that result in a differential shift of the clock levels induced by magnetic fields. The linear Zeeman effect is typically cancelled by averaging two clock transitions that experience an almost equal shift in magnitude but with opposite sign, or by probing a transition without magnetic field sensitivity. From the difference of two transitions with a known sensitivity, the magnetic field can be extracted to calculate the differential quadratic Zeeman shift of the clock transition from known atomic parameters with high accuracy. Another example is the black-body radiation (BBR) shift, which arises from thermal radiation interacting with the polarizability of the clock levels, also known as the ac Stark effect (Gallagher and Cooke, 1979; Itano *et al.*, 1982). Neglecting higher order effects, it can be written as (Pal'chikov *et al.*, 2003; Porsev and Derevianko, 2006):

$$\hbar \Delta f_{BBR} = -\Delta \alpha_S \langle E^2(T) \rangle / 2, \tag{4.2}$$

where $\Delta \alpha_S$ is the differential scalar polarizability of the clock levels, and $\langle E^2(T) \rangle = \left(831.9 \frac{V}{m}\right)^2 \left(\frac{T}{300\,K}\right)^4$ is the square of the electric field of the ambient thermal radiation of a black body at temperature $T$ (see figure 4.1(c)). At room temperature this is currently one of the dominant shifts for most optical frequency standards, ranging between 10's and 100's of mHz (Ludlow *et al.*, 2015). Its uncertainty arises from imprecise knowledge of $\Delta \alpha_S$ (uncertainty $\delta \Delta \alpha_S$), obtained from measurements (Dubé *et al.*, 2014; Huntemann *et al.*, 2016; Middelmann *et al.*, 2012; Sherman *et al.*, 2012) or calculations (Safronova et al., 2011, 2013), and the effective temperature (uncertainty $\delta T$) seen by the atoms

$$u_B(\Delta f_{BBR}/f_0) = \sqrt{\left(\frac{\delta \Delta \alpha_S}{\Delta \alpha_S}\right)^2 + \left(4 \frac{\delta T}{T}\right)^2}. \tag{4.3}$$

The effective temperature is estimated through detailed modelling of the environment of the atoms, typically combined with measurements (Bloom *et al.*, 2014; Doležal *et al.*, 2015; Madej *et al.*, 2012). While the atomic property $\Delta \alpha_S$ may not be known with high precision, we can assume it to be constant between measurements. When evaluating differences between e.g. frequency ratio measurements involving the same standards, the uncertainty of all fixed atomic parameters drops out, resulting in an improved accuracy of the *change* in a frequency ratio. This type of evaluation, which we call "reproducibility", can lead to improved height resolution in chronometric levelling.

A straightforward procedure to evaluate shifts is to change a parameter (e.g. laser power of the lattice laser in lattice clocks) and measure the resulting change in frequency. Through extrapolation the unperturbed frequency can be obtained. If the change in frequency cannot be made very large, averaging to the desired level of uncertainty can be tedious and take a long time if the targeted accuracy is high (Westergaard *et al.*, 2011). Other systematic frequency shifts can be probed via "leverage". This either means that the shift can be made very large, or that the physical property resulting in the shift can be measured precisely by some other means. An example of the latter is the 2nd order Doppler shift from residual temperature. Assuming a thermal motional state, the mean kinetic energy can be probed via sideband spectroscopy within seconds (Ludlow *et al.*, 2015; Monroe *et al.*, 1995), and without the need to observe the clock frequency shift as a function of residual thermal energy of the atoms. Figure 4.2(left) gives an overview over the relative systematic frequency uncertainties of the past and current state-of-the-art for caesium microwave and optical clocks based on different species. The tremendous progress in optical clocks has been made by detailed measurements of atomic and environmental properties, culminating in the systematic shifts and their uncertainties of the currently most accurate lattice and ion clocks as shown in Table 4.1. However, it should be mentioned that this table provides an *estimate* of the frequency uncertainty, based on measurements and calculated properties. For the most advanced clocks this has not yet been verified through repeated frequency comparisons with other clocks, a mandatory requirement for using them for chronometric levelling. This is one of the motivations for developing transportable clocks that can be verified or calibrated against a laboratory or other transportable reference clock before moving to a remote site for frequency comparison.

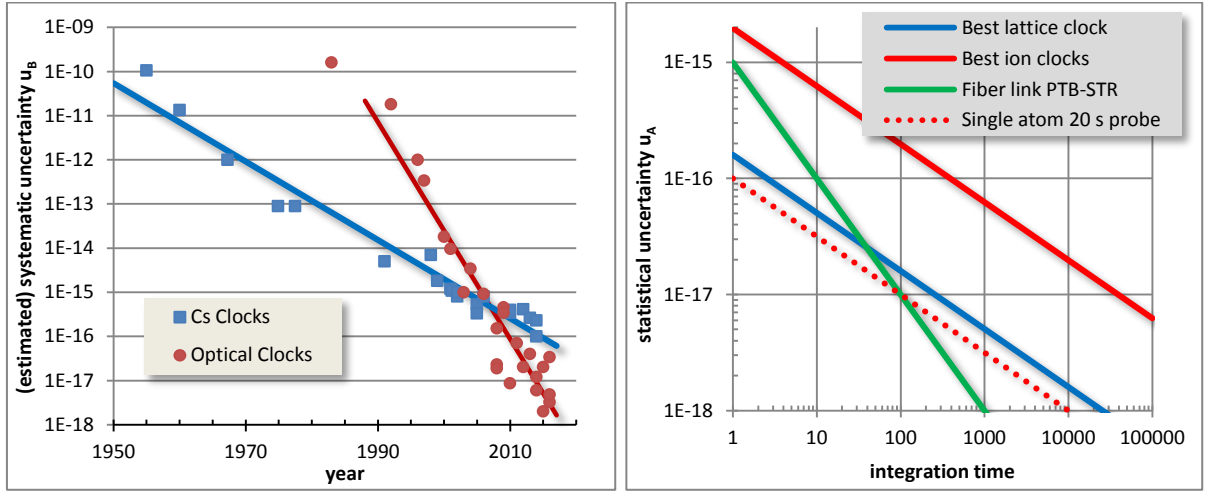

*Figure 4.2: Clock performance overview. Left: Historical development of clock inaccuracies for caesium microwave clocks and some of the most accurate optical frequency standards. Right: Typical instability (Allan deviation) for single-ion (Chou et al., 2010a) and single-ensemble neutral atom lattice clocks (Al-Masoudi et al., 2015; Nicholson et al., 2012) together with the instability given by MDEV (Allan and Barnes 1981) of a 1400 km long fibre link between PTB and Strasbourg (Raupach et al., 2015). The dashed line is the quantum projection noise limit for a single atom probed for 20 s.*

| Shift | Physical origin | Sr ($\times 10^{-18}$) | Yb$^+$ ($\times 10^{-18}$) |
|---|---|---|---|
| Lattice light shift | Light shifts that do not cancel at the magic wavelength | -1.3(1.1) | - |
| BBR shift | BBR coupling to the differential polarizability of clock levels | -4867.4(1.4) | -70.5(1.8) |
| Second-order Zeeman | dc and ac magnetic fields coupling to internal levels | -51.7(0.3) | -40.4(0.6) |
| Second-order Doppler | Time dilation shift due to non-zero motion | 0.0(0.1) | -3.7(2.1) |
| Probe light related shift | Non-cancelled ac Stark shift from probe light | | 0.0(1.1) |
| Quadratic dc Stark | Differential shift from dc electric fields | 0.0(0.1) | -1.2(0.6) |
| Quadrupole shift | Electric field gradient coupling to electric quadrupole moment of excited clock state | - | 0.0(0.3) |
| Density | Density-dependent interaction between atoms | -3.5(0.4) | - |
| Background gas collisions | Phase-changing collisions with residual background gas | 0.0(0.6) | 0.0(0.5) |
| Servo error | Residual non-linear drifts of clock interrogation laser | -0.5(0.4) | 0.0(0.5) |
| AOM phase chirp | Residual thermal and electrical phase chirps from after turning AOM on | 0.6(0.4) | |

*Table 4.1: Dominant relative systematic shifts, their physical origin, magnitude and uncertainty in the most accurate Sr lattice clock (Nicholson et al., 2012, Nicholson et al., 2015) and Yb$^+$ single ion clock (Huntemann et al., 2016).*

To summarize, the inaccuracy of a clock is the uncertainty with which we can determine the unperturbed transition of the atomic reference, whereas the reproducibility neglects uncertainties from shifts that are believed to be constant, such as atomic properties. For the determination of frequency differences as is the case for e.g. chronometric levelling, reproducibility will thus provide an improved resolution, which, however, needs to be verified by a side-by-side calibration with a reference clock.

### 4.3 Frequency instability and statistical uncertainty

The instability of a frequency standard determines the averaging or integration time required to achieve a certain frequency resolution. In the context of chronometric levelling, the instability also determines the resolution with which temporal changes in the gravity potential (see section 2) can be determined. It can be derived from simple considerations (Riehle 2004; Schmidt and Leroux, 2015). The relative frequency uncertainty in a single measurement near the maximum slope of the excitation signal $\left(\frac{dS}{df} \approx \frac{S_0}{\Delta f}\right)$ is given by the noise of the signal ($\delta S$), divided by its slope: $\left(\frac{\delta f}{f_0}\right)_1 = \frac{\delta S}{f_0(dS/df)} \approx \frac{\delta S}{S_0 Q}$, where $S_0 = N$ is the maximum signal amplitude for $N$ atoms and $Q = f_0/\Delta f$ is the quality factor of the transition (see figure 4.1(b)). Quantum mechanics dictates that the outcome of a single measurement on a single particle is binary: either the atom is found to be in the ground or in the excited state. The discreteness of the measurement results in so-called quantum projection noise (QPN) and is given by $\delta S = \sqrt{Np(1-p)}$ (Itano *et al.*, 1993), where $N$ is the number of atoms and $p$ is the excitation probability ($p \approx 0.5$ in our scenario). The observed linewidth scales with the inverse of the probe time $T$ as $\Delta f = \kappa/\pi T$, where $\kappa$ is a numerical factor on the order of 1 that depends on the probe scheme (Rabi, Ramsey, Hyper-Ramsey, …). Assuming Ramsey interrogation ($\kappa = 1$), we obtain a frequency uncertainty of $\left(\frac{\delta f}{f}\right)_1 = \frac{1}{2\pi f_0 T \sqrt{N}}$. Repeating the measurement $M$ times reduces the uncertainty by $1/\sqrt{M}$. Assuming a cycle time of $T_C$, $M = \tau/T_C$ measurements can be performed during an integration time $\tau$, resulting in a relative frequency uncertainty of (Ludlow et al., 2015)

$$\frac{\delta f}{f} = \frac{1}{2\pi f_0 T} \sqrt{\frac{T_c}{N\tau}} \ . \tag{4.4}$$

For white frequency noise, the statistical uncertainty, $u_A$, is equivalent to the Allan deviation $\sigma_y(\tau)$ (Allan, 1966; Riehle, 2004; Riley, 2008). This expression illustrates that the smallest instability is achieved for high clock frequencies $f_0$, long interrogation times $T$, and a large number of atoms $N$. The higher frequency of optical clocks, together with smaller systematic shifts, are the main reasons for the success of optical clocks. In an ideal situation, the cycle time equals the interrogation time, $T = T_c$, resulting in

$$\sigma_y(\tau) = \frac{1}{2\pi f_0} \sqrt{\frac{1}{NT\tau}} \ . \tag{4.5}$$

The limit to the interrogation time is ultimately given by the lifetime of the excited clock state. Figure 4.2(right) shows typical instabilities of single-ion and neutral atom lattice clocks together with the quantum projection noise limit for a single atom probe for 20 s. Furthermore, the typical instability of a 1400 km long length-stabilized optical fibre link is shown (for more details see section 6.2). In contrast to the clock instabilities, the fibre link averages down with $1/\tau$, since it is based on stabilizing the optical phase. The figure illustrates that the instability of the fibre link is not a limitation beyond averaging times of 100 s.

**4.4 Clock laser stabilization**

Probe times beyond a few seconds are currently not achievable, since they are limited by the phase coherence of the clock laser (Kessler *et al.*, 2014). Phase coherence needs to be maintained during interrogation and frequency stability until the servo loop has stabilized the laser's frequency to the atomic transition (Leroux *et al.*, 2017; Santarelli *et al.*, 1998). The stability of clock lasers is derived from narrow optical resonance features to which the laser is stabilized using an electronic feedback loop. The most successful approaches include stable optical cavities and spectral hole burning. Spectral hole burning is based on rare-earth doped crystals at cryogenic temperatures in which optical pumping with a laser pulse generates narrow resonances (spectral holes in the population), which in turn can be used to stabilize the laser's frequency (Chen *et al.*, 2011; Thorpe *et al.*, 2011). A short-term instability of $10^{-15}/\sqrt{\tau/s}$ has been achieved, limited by the resonance's sensitivity to electric and magnetic fields, vibrations, pressure and temperature fluctuations (Cook *et al.*, 2015).

The stability of optical resonances in high-finesse optical cavities with an optical path length $L$ is derived from their passive length stability according to $\Delta f/f = -\Delta L/L$. For a relative frequency instability of $10^{-16}$, the length of a 50 cm long cavity needs to be stable to less than 1/100 of the size of a proton. This is achievable only in vibration-isolated cavities with mirror spacers made from ultra-low expansion material (e.g. ULE$^{TM}$) inside several layers of thermal shielding and active temperature stabilization in a vacuum with pressure fluctuations lower than $10^{-10}$ mbar. Brownian motion of the spacer, the mirrors, and the optical coatings fundamentally limit the achievable instability (Kessler *et al.* 2012, Numata *et al.*, 2004). The fractional frequency shift due to thermal noise can be reduced by using long cavities made from titanium-doped glass (e.g. ULE$^{TM}$) and fused silica mirrors (Amairi *et*

*al.*, 2013; Keller *et al.*, 2013; Häfner *et al.*, 2015; Jiang *et al.*, 2011; Young *et al.*, 1999). This way, an instability below $10^{-16}$ from 1 to 1000 s has been achieved (Häfner *et al.*, 2015). By choosing materials with a high mechanical quality factor, thermal noise can be further suppressed. The best lasers today are based on cryogenic single-crystal silicon optical cavities and achieve a phase coherence of around 16 s and a flicker floor-limited Allan deviation of $5 \times 10^{-17}$ between 0.8 s and 10 s (T. Kessler *et al.*, 2012a; Matei *et al.*, 2017). By employing crystalline coatings, these values can be further improved (Cole et al., 2013, 2016).

For single ion clocks with limited signal-to-noise ratio, but very fast cycle times ($T_c \approx T$), several probe cycles are required to stabilize the laser's frequency to the atomic reference (Leroux *et al.*, 2017; Peik *et al.*, 2006; Riis and Sinclair, 2004). In contrast, neutral atom lattice clocks that operate with 1000s of atoms can stabilize the laser to the atomic reference after a single interrogation cycle. However, since lattice clocks require significant time for loading and preparing the atoms ($T_c > T$), they are limited by the so-called Dick effect (Dick, 1987; Dick et al., 1990; Quessada et al., 2003; Santarelli et al., 1998). It deteriorates the instability, since the phase deviations of the laser during the preparation time of the atoms (dead time) goes unnoticed. In other words, the frequency standard cannot correct for spectral components in the laser noise corresponding to the inverse dead time. This represents a limitation in stability that can be suppressed through correlated interrogation (Bize *et al.,* 2000; Lodewyck *et al.,* 2010) or by using two references that are operated in an interleaved fashion (Schioppo *et al.*, 2017) as discussed in further detail in section 5.1.

## 4.5 Counting optical cycles: the optical frequency comb

Operating a frequency standard as a clock also requires counting the optical cycles at several hundred THz. Since no electronic device can count these high frequencies, optical frequency combs are employed to divide the optical cycles by a known large integer number. The invention of stabilized optical combs was instrumental in the advance of optical frequency metrology and optical clocks (Diddams, 2010; Hall, 2006, 2017; Hänsch, 2006). They are based on pulsed lasers with repetition rate $f_r$ that emit a comb-like spectrum as shown in figure 4.3. The frequency of the $m^{th}$ comb tooth (when counting from zero) is given by the comb equation $f_m = mf_r + f_{CEO}$ (Jones *et al.*, 2000; Telle *et al.*, 1999; Udem *et al.*, 1999). While the repetition frequency is easily determined by measuring the pulsed output intensity of the laser with a photodiode, the so-called carrier envelope offset frequency, $f_{CEO}$ (see figure 4.3), is typically obtained in a nonlinear interferometer. By frequency-doubling part of the lower-frequency spectrum of the comb and beating it with the high-frequency part of an octave-spanning comb, gives $f_{CEO}$ according to $2f_n - f_{2n} = 2nf_r + 2f_{CEO} - (2nf_r - f_{CEO}) = f_{CEO}$. The comb can be stabilized to the laser output of a frequency standard by stabilizing its beatnote $f_{beat}$ with the nearest comb tooth by feeding back on the repetition rate. Through stabilization of $f_{CEO}$ to $f_r$, a phase-locked gear box between the frequency standard and the two radio frequencies is established, thus realizing an optical clock with a frequency of

$$f_{clock} = mf_r + f_{CEO} \pm f_{beat} \ .  \quad (4.6)$$

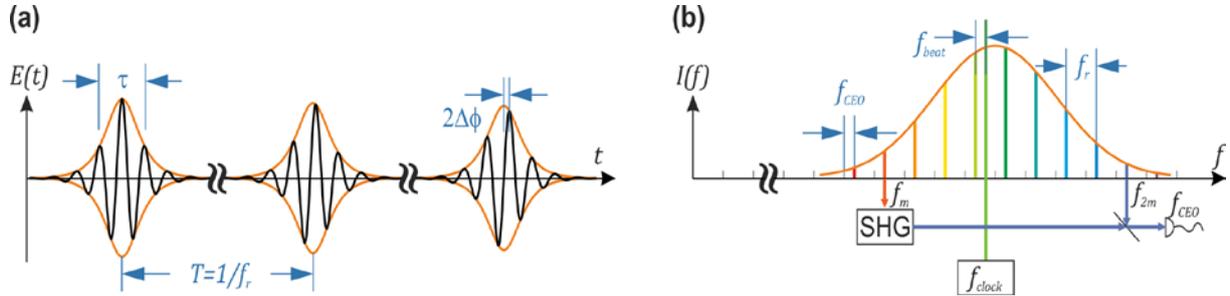

*Figure 4.3: Principle of an optical frequency comb. (a) The output of a frequency comb consists of pulses with duration τ, separated by a time $T = 1/f_r$. The phase difference between carrier (solid line) and pulse envelope (dashed line) advances by Δϕ from pulse to pulse. (b) The spectrum of a frequency comb consists of equally spaced comb teeth, separated by $f_r$ with an offset of $f_{CEO} = f_r \Delta\phi/2\pi$ when extrapolated to zero frequency. The light from a clock laser is overlapped with part of the comb spectrum to observe a beat signal $f_{beat}$ with the nearest comb line. The beat between the high-frequency part of the comb spectrum with the second harmonic (SHG) of the low-frequency part provides $f_{CEO}$.*

This relation enables an absolute frequency measurement of the optical frequency by counting the frequencies $f_r$ and $f_{CEO}$ with Cs- or GPS-referenced counters. Such measurements are typically limited by the accuracy of the Cs fountain clocks to a few parts in $10^{16}$ (Campbell *et al.*, 2008; Falke *et al.*, 2014; Huntemann *et al.*, 2014; Kim *et al.*, 2017; Oskay *et al.*, 2006; Pizzocaro *et al.*, 2017; Tamm *et al.*, 2014, Le Targat *et al.*, 2013). Higher accuracy can be obtained by measuring optical frequency ratios between optical standards. This is accomplished by measuring the beat note between the output of a second standard with the nearest comb tooth of a frequency comb referenced to the first standard, reaching inaccuracies of $5 \times 10^{-17}$ for clocks based on different species (Godun *et al.*, 2014; Nemitz *et al.*, 2016; Rosenband *et al.*, 2008; Takamoto *et al.*, 2015; Yamanaka *et al.*, 2015). Frequency conversion to other wavelength regimes e.g. to 1.5 $\mu$m for remote frequency comparison with optical fibre links as discussed in section 6.2 is achieved by stabilizing the output of a laser operating near the target wavelength with one of the stabilized comb teeth. Stabilization of the entire comb in all optical-optical comparisons can be avoided by employing a transfer scheme in which the noise of the comb is eliminated electronically (Scharnhorst *et al.*, 2015; Stenger *et al.*, 2002; Telle *et al.*, 2002). This way, frequency stability transfer between different spectral regimes at the level of $4 \times 10^{-18}$ has been achieved (Nicolodi *et al.*, 2014).

## 5. State-of-the-Art of Optical Frequency Standards

### 5.1 Neutral Atoms in Optical Lattices

Optical lattice clocks with their high number of simultaneously interrogated atoms are promising excellent signal to noise ratio as the quantum projection noise (see equation (4.5)) is averaged over an ensemble of few hundreds to several thousands of atoms. In particular, for the application of optical clock in geodesy, high clock stability is a very convenient quality since it reduces the averaging time required to reach a desired level of uncertainty. It also allows to monitor time-variable effects like tides, which have to be reduced to obtain a static solution. Because the targeted height resolution of geodetic applications in the centimetre region, statistical clock uncertainties $u_A$ of few $10^{-18}$ have to be reached. If temporal resolution of gravity potential variations is desired, time scales of few hours to a day become relevant in which such a statistical uncertainty has to be reached. With the typical dependence of $u_A$ on the averaging time $\tau$ in clocks of $u_A \propto \tau^{-1/2}$ an instability of less than $3 \times 10^{-16}$ is required in one second.

The concept of lattice clocks can be realized with several different atomic species that provide sufficiently narrow clock transitions and have a level structure that allows to operate an optical trap i.e. the lattice at a wavelength (so-called 'magic' wavelength) where the polarizabilities of upper and lower clock state are equal (Katori *et al.* 1999 and figure 5.1). These are typically group-II like atoms; most often Sr (Nicholson *et al.* 2015; Tanabe *et al.* 2015; Lin *et al.* 2015; Lodewyck *et al.* 2016; Grebing *et al.* 2016; Takano *et al.* 2016; Hachisu *et al.* 2017), Yb (Takamoto *et al.* 2015; Nemitz *et al.* 2016; Schioppo *et al.* 2017; Kim *et al.* 2017), or Hg (Yamanaka *et al.* 2015; Tyumenev *et al.* 2016) are used, but other species like Mg or Cd are investigated as well (Kulosa *et al.* 2015). Interrogation of the atoms in an optical lattice operated at the 'magic' wavelength provides the tool to observe the clock transition without first order light shifts by the trap. Line broadening by atomic motion is also inhibited by the quantized level structure in the trap (Lemonde and Wolf 2005).

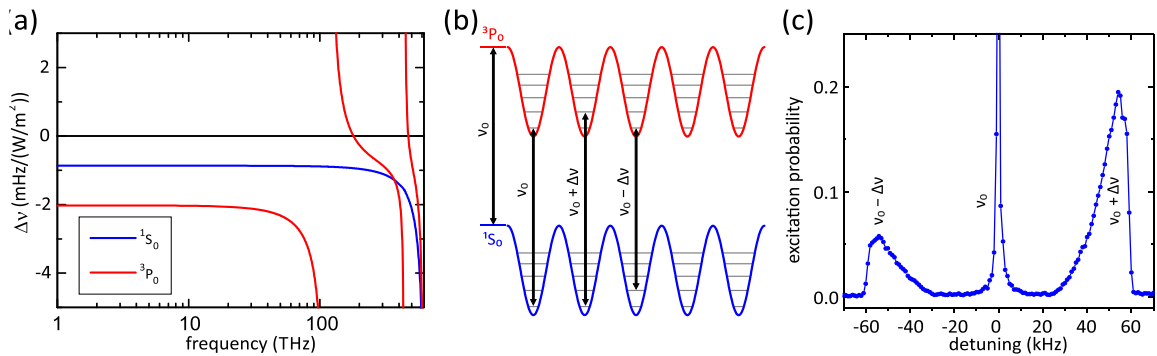

*Figure 5.1: Optical lattices. (a) Typical light shift of an alkaline earth atom (here Sr) of lower ($^1S_0$) and upper ($^3P_0$) clock state. At the crossing points of the curves, 'magic' optical traps can be operated. (b) Schematic of the interrogation in a lattice at the magic wavelength: The transition frequency $\nu_0$ is the same as without trap. The atomic motion in the trap is quantized leading to sidebands shifted by $\pm\Delta\nu$. (c) Experimental observation of the clock transition and its motional sidebands in an optical lattice. The sideband transitions are broadened due to inhomogeneity of the trapping potential.*

As the preparation of the atomic sample in an optical lattice by laser cooling is causing considerable dead times, in which the atomic reference transition cannot be interrogated by the clock laser, the stability of optical lattice clocks suffered for a long time from aliasing of laser noise known as the Dick effect (see section 4.4). Furthermore, the state-dependent interaction of the atoms in the lattice light field with the atoms and interactions between atoms themselves and their environment had to be investigated in detail before fractional uncertainties $10^{-17}$ could be achieved.

With the advent of laser sources with outstanding coherence properties as discussed in section 4, the duty cycle of the interrogation in the clock cycles has improved and the influence of the Dick effect has been reduced such that clock instabilities of few $10^{-16}/\sqrt{\tau/s}$ have been demonstrated (Nicholson *et al.* 2015; Al-Masoudi *et al.* 2015; Hinkley *et al.* 2013). The long coherence time of the interrogation laser paired with the preservation of atomic coherence in the lattice allows for long interrogation times (Campbell *et al.* 2017). Consequently, the observed transitions have sub-Hertz linewidth (figure 5.2), which pushes the stability limitations imposed by the QPN to even lower instability. Beyond the aspect of reduced averaging time, the outstanding spectral resolution allows to resolve inhomogeneous influences on the transition frequency of the reference transition caused by e.g. interactions between atoms (Martin *et al.* 2013) or with the trapping light field (Campbell *et al.* 2017). This may not only provide better understanding of the underlying processes but can lead to smaller clock uncertainties due spectral resolution of atoms with perturbed transition frequencies from unperturbed ones.

The approach of improving the duty cycle of the clock interrogation has consequently been followed in the NIST Yb lattice clock. There, instead of focusing on improved laser coherence, a virtually dead time free interrogation scheme was implemented by interleaving the interrogation of two samples of atoms in independent physics packages (Schioppo *et al.* 2017). With this approach, the currently smallest clock instability of $6\times10^{-17}/\sqrt{\tau}$ was demonstrated.

Alternatively, stability limitations by the Dick effect can be mitigated by synchronized probing of two clocks with one interrogation laser (Bize *et al.*, 2000; Lodewyck *et al.*, 2010, Campbell *et al.* 2017). Though the Dick effect related noise is not removed, it is identical in both clocks. The frequency ratio of both systems can thus be observed with high stability (Ushijima *et al.* 2015). This approach however realizes its full potential only for clocks operating at the same (or very similar) frequencies (Nemitz *et al.* 2016) and in situations, where interrogation light with identical spectral properties can be provided to both systems. The latter requirement so far inhibits long-distance clock comparisons but has led to improved stability for clocks separated by few 10 km (Takano *et al.* 2016).

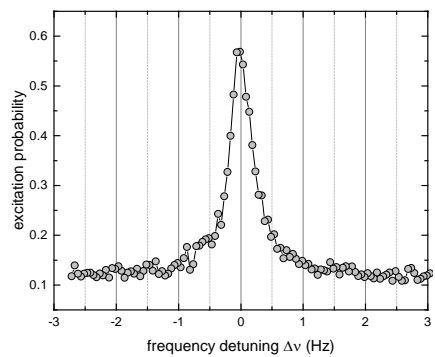

*Figure 5.2: Narrow linewidth scan over the 429 THz clock transition in $^{87}$Sr (PTB). The interrogation time by the clock laser is 2.6 s, every point is recorded in a single interrogation cycle.*

Not only the stability of optical lattice clocks has been greatly improved in the past years but also the accuracy of the lattice clocks has made a large leap forward. The progress is mainly enabled by detailed investigations of the in Sr and Yb lattice clocks largest perturbation caused by coupling between the thermal radiation of the environment (BBR) and the atoms. The atomic polarizabilities parametrizing this coupling have been accurately measured now (Sherman *et al.* 2012; Beloy *et al.* 2012; Middelmann *et al.* 2012; Nicholson *et al.* 2015) and sophisticated temperature control systems have been implemented in the apparatuses, both for operation at room temperature (Beloy *et al.* 2014; Bloom *et al.* 2014) as well as for cryogenic environments (Ushijima *et al.* 2015). These measures are accompanied by revised approaches to control the light shifts induced by the optical lattice: At uncertainties of $10^{-17}$ and below, light shifts originating from higher order multipoles than electric dipole transitions or even from two photon transitions can be expected to become relevant (Taichenachev *et al.* 2006; Barber *et al.* 2008; Westergaard *et al.* 2011; Ovsiannikov *et al.* 2013; Le Targat *et al.* 2013, Katori *et al.* 2015, Porsev *et al.* 2017). Since these couplings exhibit different scaling with the lattice light intensity than the first order electric dipole related light shift, a universal cancellation of the light shift for all lattice depths is no longer possible. Furthermore, the light shifts become dependent on the motional state of the atoms in the optical lattice, which leads to the necessity of precise control of the atomic motion or temperature. Therefore, methods have been devised that do not necessarily aim at a perfect cancellation of the lattice light shift but try to maintain an acceptably small shift over a reasonably large range of experimental conditions (Ovsiannikov *et al.* 2016; Katori *et al.* 2015, Brown *et al.* 2017).

Strontium lattice clocks currently have the lead among all frequency standards with an uncertainty of only $2 \times 10^{-18}$ (Nicholson *et al.* 2015). This implies the possibility of chronometric levelling with uncertainties of 2 cm, which rivals the performance of classical geodetic methods (see section 2). It is also noteworthy that the uncertainty of Sr lattice clocks has been validated on a very high level. Besides long-distance comparisons (see section 6), highly accurate side-by-side measurements have been performed that have tested the agreement of two Sr lattice clocks at $4.4 \times 10^{-18}$ (Ushijima *et al.* 2015). This result remains however so far a singular example, surpassing the next best clock validation, which was done between to Al$^+$ ion clocks (Chou *et al.* 2010a), by a factor of two. Other validations (Lisdat *et al.* 2016; Bloom *et al.* 2014) of Sr lattice clocks reach mid-$10^{-17}$ uncertainties. On the long run, optical ratio measurements (Rosenband *et al.* 2008; Akamatsu *et al.* 2014; Yamanaka *et al.* 2015; Nemitz *et al.* 2016; Grotti *et al.* 2017) between clocks of different types will provide a matrix-like data field that will allow judging the quality of the clock uncertainty (Margolis and Gill 2015). This is an essential step for applications of optical clocks in metrology and geodesy since the rapid development of clocks must be supported to build a sound foundation for applications. As this process is developing, it can be expected that optical lattice clocks will soon be powerful instruments for geodesy providing (sub-) centimetre height resolution within hours of data integration. It should however be recalled (section 3) that the validation of general relativity must keep track with these developments, if chronometric levelling over >100 m height differences shall be applied.

## 5.2 Trapped Single Ions

Optical clocks based on trapped ions are currently based on a single reference atom. While this puts constraints to the achievable signal-to-noise ratio when interrogating a single atom (see equation (4.5)), single ions trapped in a Paul trap (Ghosh, 1995; Paul, 1990) offer highest accuracy for selected species that have advantageous atomic properties, such as a small polarizability and electric quadrupole moment (Dehmelt, 1981; Ludlow *et al.*, 2015). Additionally, the single atomic ion is strongly confined and located to a region of a few nm, almost ideally shielded from its environment. All relevant shifts can be characterized using "leverage". This means that all systematic frequency shifts can be measured without averaging to the desired clock uncertainty as described in section 4. Furthermore, ion trap systems can be very compact since ion traps have a volume of a few cm$^3$, are loaded from sources with a volume of only a few mm$^3$ and require lasers with only mW powers levels. Ion traps offer long storage times of hours to months, making them robust and reliable candidates for user-friendly clocks operating in the field (Nisbet-Jones *et al.*, 2016). The continuous trapping enables high duty cycles in the clock interrogation, reducing the Dick effect.

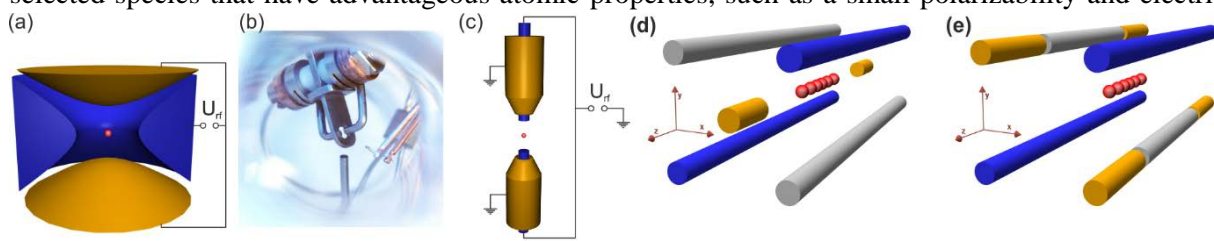

*Figure 5.3: Examples of ion traps. (a) Schematic of a spherical ion trap. (b) Picture of a spherical ring trap such as used in (Tamm et al., 2000), courtesy Chr. Tamm. (c) Endcap trap, providing more optical access. (d) Linear trap with axial dc electrodes. (e) Linear ion trap with split axial dc electrodes. Orange, blue, and grey electrodes are at dc, rf, and ground potential, respectively.*

For optical clocks, Paul traps being based on electric fields are preferred over Penning traps (Dehmelt, 1990), which require a strong magnetic field with exceptional stability to avoid fluctuations in the Zeeman shifts on the clock transition. In a spherical Paul trap, a single ion is confined in the node of an electric field oscillating at tens of MHz trap drive frequency $\Omega_{rf}$. In a linear Paul trap (Wineland *et al.*, 1998) radial confinement is provided by an oscillating 2D rf quadrupole electric field, while axial trapping is achieved by applying an additional 3D dc quadrupole (see figure 5.3). In properly engineered linear traps (Herschbach *et al.*, 2012; Pyka *et al.*, 2014), the rf field exhibits a nodal line of vanishing electric field in axial direction which allows trapping of strings of ions as discussed in more detail in section 5.3.

Since the trap drive frequency is much larger than the oscillation frequency (secular frequency) of the ions in the trap (typically on the order of one MHz), the time scales separate and one can apply the pseudo-potential approximation (Dehmelt, 1981). In this approximation, the trapping fields provide a 3D harmonic confinement with secular frequencies $\omega_{x,y,z}$ along three orthogonal directions. The most dominant shifts in ion clocks arise from electric and magnetic fields and 2$^{nd}$ order Doppler (time dilation) shifts from residual motion that have been discussed already in section 4. In the following, we will concentrate on shifts that are unique to trapped ions. Since the ion is always located at zero dc electric field, clock frequency shifts arise exclusively from oscillating fields via the ac Stark effect, and from electric field gradients. The latter can interact with an electric quadrupole of the electron distribution of the ion, resulting in the electric quadrupole shift (Itano 2000). For typical ions, this shift can be on the order of 10 Hz. It can be eliminated by averaging over several clock transitions (Dubé *et al.*, 2005, p. 20), similar to the linear Zeeman effect, or by applying a magnetic field along three mutually orthogonal directions (Itano, 2000), which can also be viewed as averaging over three different transitions. Effects from secular motion can be made small via laser cooling. The residual small oscillations from the trap drive field is called intrinsic micromotion and is unavoidable, since it is part of the trapping mechanism. Stray electric fields or asymmetries in the rf field can cause excess micromotion, which can be compensated to reduce Doppler-induced frequency shifts (Berkeland *et al.* 1998; Keller *et al.*, 2015). At the same time, the finite size of the ion's wave packet (which is on the order of 10 nm in the ground state for 1 MHz trap frequency and grows with the ion's kinetic energy) samples the non-zero electric field of the rf trapping potential. This results in a differential ac Stark effect on the clock transition. For some ion species (e.g. Ca$^+$ and Sr$^+$) this ac Stark shift has the opposite sign as the 2$^{nd}$ order Doppler shift, such that they can be made to cancel at a particular "magic" trap drive frequency (Madej *et al.*, 2012). Clock transitions that are much narrower than available laser source need power broadening to be excited by the clock laser. This can induce additional ac Stark shifts through interaction of the clock laser with the differential polarizability of the clock levels, similar to black-body radiation (Webster et al., 2002). Special interrogation schemes such as Hyper-Ramsey (Hobson *et al.*, 2016; Huntemann *et al.*, 2012a; Yudin *et al.*, 2010; Zanon-Willette *et al.*, 2016) have been implemented to null this shift.

Several ion species have been developed into optical frequency standards with an inaccuracy below $10^{-16}$ that surpass the best cesium clocks. The $^1S_0$-$^3P_0$ clock transition in Al$^+$ has been proposed as a frequency reference in the 1990's by Dehmelt and co-workers (Yu *et al.*, 1992). It is essentially free of the electric quadrupole shift (Beloy *et al.*, 2017), has a small linear and quadratic Zeeman effect and exhibits the smallest sensitivity to black-body radiation of all investigated atomic species (Rosenband *et al.*, 2006; Safronova *et al.*, 2011). However, the transition for laser cooling and detection in Al$^+$ is at 167 nm, which is not yet accessible by lasers (Wang *et al.*, 2016). Therefore, a co-trapped well-controllable logic ion (such as Be$^+$, Mg$^+$ or Ca$^+$) takes on these tasks using a technique called quantum logic spectroscopy (Schmidt *et al*., 2005). This way, an Al$^+$ optical clock with an estimated inaccuracy of $2.3 \times 10^{-17}$ has been demonstrated and compared to a single-ion Hg$^+$ clock with an inaccuracy of $1.9 \times 10^{-17}$ (Rosenband *et al.*, 2008). The frequency ratio between these two different clocks has been determined with a total uncertainty (statistical and systematic) of $5.2 \times 10^{-17}$, which was the most accurately measured physical quantity for many years. By employing $^{25}$Mg$^+$ instead of $^{9}$Be$^+$ as the logic ion species, sympathetic cooling improved (Wübbena *et al.*, 2012) and a second generation Al$^+$ clock with an inaccuracy of $8 \times 10^{-18}$ could be demonstrated (Chou *et al.*, 2010a), limited by residual secular and micromotion. A comparison between the two Al$^+$ clocks has been performed with an instability of $2.8 \times 10^{-15}/\sqrt{\tau/s}$ and the two frequencies agreed to within $(-1.8 \pm 0.7) \times 10^{-17}$, consistent with their combined evaluated inaccuracy of $2.5 \times 10^{-17}$ (Chou *et al.*, 2010a). As discussed in section 4, the maximum probe time (and thus the instability) is restricted by the coherence time of the laser. This limitation can be overcome through correlated probing of two systems, in which the common mode phase noise of the laser drops out (Bize *et al.*, 2000; Lodewyck *et al.*, 2010). Such a measurement has been implemented with two Al$^+$ ions trapped in the same trap and probed by the same probe laser (Chou *et al.,* 2011). This way, a relative coherence time of $T_c = 9.7^{+6.9}_{-3.1}$ s was demonstrated, limited by the excited state lifetime of $20.6 \pm 1.4$ s (Rosenband *et al*., 2007). This scheme can also be extended to frequency comparisons between ion and lattice clocks to take advantage of the high stability of a lattice clock to correct phase excursions of the laser (Hume and Leibrandt, 2016)

The $^2$S$_{1/2}$-$^2$D$_{5/2}$ transition in $^{40}$Ca$^+$ (Chwalla et al., 2009; Huang et al., 2012, 2014; Matsubara et al., 2012) and $^{88}$Sr$^+$ (Barwood et al., 2014; Dubé et al., 2013 and 2017, Madej et al., 2004, 2012; Margolis et al., 2004) have also been investigated by several groups as frequency standards. Both species have similar properties and dominant systematic frequency shifts, which include a large (10 kHz/µT) linear Zeeman effect, a vanishingly small quadratic Zeeman shift, and significant BBR and quadrupole shifts. As a consequence of their negative differential polarizability, 2$^{nd}$ order Doppler and ac Stark shifts cancel at a "magic" trap drive frequency (Huang *et al.*, 2016; Madej *et al.*, 2012). By averaging several transitions, quadrupole and linear Zeeman shifts are cancelled. Together with numerical simulations that enable an assessment of the BBR environment, Ca$^+$ has been evaluated to an inaccuracy of $5.1 \times 10^{-17}$ (Huang *et al.*, 2016) and Sr$^+$ to $1.2 \times 10^{-17}$ (Dubé *et al.*, 2014) inaccuracy and $3 \times 10^{-17}$ reproducibility (Barwood *et al.*, 2014). Frequency comparisons below $10^{-16}$ have been performed by comparing two Ca$^+$ standards that agreed to within $(3.2 \pm 5.5) \times 10^{-17}$ (Huang *et al.,* 2016) and similarly for two Sr$^+$ standards which agreed to within $(0.9 \pm 4) \times 10^{-17}$ (Barwood *et al*., 2014).

Currently the most accurate operational optical clock, with an inaccuracy of $3.2 \times 10^{-18}$, is based on the octupole (E3) transition in $^{171}$Yb$^+$ (Huntemann et al., 2016) with a nano-Hertz linewidth. It features an exceptionally small electric quadrupole shift and small magnetic field shifts. As a consequence of its large mass, 2$^{nd}$ order Doppler shifts are also reduced. Heating of the ion during the clock interrogation increases the uncertainty of these shifts which together with the evaluation of the BBR environment and the knowledge about the differential polarizability, are the major contributors to the clock inaccuracy (Doležal *et al.,* 2015; Huntemann *et al.*, 2016). Power broadening of the narrow clock transition by the clock laser results in significant ac Stark shifts that are nulled by a special pulse sequence called Hyper-Ramsey (Hobson *et al.*, 2016; Huntemann *et al.*, 2012a; Yudin et al., 2010; Zanon-Willette *et al.*, 2016). $^{171}$Yb$^+$ offers a second frequency reference on the $^2$S$_{1/2}$-$^2$D$_{3/2}$ quadrupole transition that has been evaluated to $1.1 \times 10^{-16}$ (Tamm *et al.,* 2014) and $6.6 \times 10^{-16}$ (Godun *et al.*, 2014). This transition has a relatively short excited state lifetime of 52 ms and exhibits a significant quadrupole, 2$^{nd}$ order Zeeman and BBR shift. Cancellation of quadrupole and magnetic field shifts is accomplished by averaging several transitions.

In summary, ion clocks are highly accurate frequency standards that provide cm resolution in height difference measurements that can be made compact and reliable for field use in geodesy.

The best demonstrated stability of a single ion clock is $2 \times 10^{-15}/\sqrt{\tau/s}$ (Chou *et al.*, 2010a), limited by the probe time given by the coherence time of the laser. Therefore, improved laser sources, prestabilization of the laser light using high-stability frequency references (e.g. neutral atom lattice clocks or atomic beam references) would allow instabilities competitive with the best neutral atom lattice clocks of $10^{-16}/\sqrt{\tau/s}$ for Al$^+$ or even better for Yb$^+$ E3. Alternatively, higher signal-to-noise ratio can be obtained by employing multi-ion clocks as discussed in the next section.

**5.3 Multi-Ion Frequency Standards**

As discussed in the previous section, the low signal-to-noise ratio of the single ion interrogation limits the obtainable short-term stability. With state-of-the-art optical ion clocks, integration times of 10´s of days to weeks are necessary to reach a fractional frequency resolution of $1 \times 10^{-18}$, being the current target for geodetic missions where a height resolution of 1 cm is required. Therefore, improving the short-term stability of ion frequency standards is a vital issue.

One possible path to improve the stability of single ion clocks is by interrogating narrower atomic transitions and thus improving the quality factor $Q = f_0/\Delta f$ of the resolved atomic line. This, however, requires a superior stability of the optical clock laser and comes along with additional technical challenges: For Fourier limited spectroscopy of mHz wide atomic transitions a short-term linewidth of the clock laser at the mHz level is required. Due to the limited quantum information obtained in single ion spectroscopy, long integration times are necessary to lock the laser onto the atomic signal. For example, resolving and locking a laser to a mHz wide clock transition by standard quantum jump spectroscopy, will require a clock laser frequency stability in the low $10^{-17}$ range over several minutes, in order to reach the quantum projection noise limit of the single ion (see equation (4.5)). A large effort is made in the frequency metrology community to push the stability limits of laser oscillators, discussed in section 4, and lasers locked to cryogenic cavities at the thermal noise limit of $\sigma = 4 \times 10^{-17}$ with a phase coherence time of about 16 s have recently been demonstrated (Matei *et al.* 2017). Still, extending the clock interrogation to tens of seconds will have to go hand in hand with improving the control of the ion's motion on this timescale, which is deteriorated by excess heating of the ion due to electric field noise, micromotion and background collisions.

Also, faster averaging, phase sensitive locking schemes are investigated (Borregaard and Sørensen 2013; Rosenband and Leibrandt 2013, Kessler *et al.* 2014), where multiple ensembles are employed to successively extend the clock laser coherence time in a cascade of measurements with increasing interrogation time. First proof-of-principle measurements of a microwave hyper-fine transition in a large cloud of Yb$^+$ ions have shown a factor of $\sqrt{3}$ improvement in stability compared to single projective measurements at the same averaging time (Shiga *et al.* 2014). The quantum projection noise limit can be overcome in the interrogation of multiple atoms by introducing quantum correlations, such as e.g. in spin-squeezed states, see (Ludlow *et al.* 2015) for a more comprehensive overview. In order to benefit from any of these techniques, next-generation ion frequency standards will need to be operated with multiple (>10) ions.

Already classical clock interrogation of multi-ion ensembles will lead to a major leap in obtainable instabilities of frequency standards. As the stability of the frequency measurement averages down with the square root of both atom number *N* and integration time $\tau$ (see equation (4.5)), increasing the number of ions to *N* can shorten the integration time to 1/*N* times compared to the single ion performance. Together with improved laser stability and possibly new detection methods, increasing the number of ions within a well-controlled ensemble of ions will substantially reduce integration times and provide new ultra-stable reference systems for determining variations in gravitational potentials.

A fundamental question is, whether a larger sample of ions can be controlled sufficiently well to reach fractional frequency inaccuracies or long-term stabilities as low as $10^{-18}$ and below. So far frequency standards based on microwave transitions in a cloud of buffer gas cooled ions have been investigated for a frequency standard with a long-term stability of $10^{-15}$ (Prestage and Weaver 2007). A ring of ions in an anharmonic linear Paul trap has been proposed (Champenois *et al.* 2010) for an optical frequency standard with improved short-term stability, conserving the long-term stability at the level of $10^{-15}$.

For clock spectroscopy at the accuracy level of $10^{-18}$, a multitude of systematic effects come into play. Laser-cooled ions trapped in a common potential form Coulomb crystals (Dubin and O'Neil

1999), in which the ions take on equilibrium positions at zero mean electric field. However, the large static electric field gradients of neighbouring ions interact with the atomic electron shell via the quadrupole moment (Itano 2000). Unless atomic clock states with electron angular momentum of $J \leq$ ½ and thus vanishing electronic quadrupole moment $\theta$ are chosen, these shifts are typically on the order of 10 Hz. Also, static field gradients from the ion trap confinement itself or residual patch potentials lead to frequency shifts on this order.

In linear rf ion traps, micromotion (Berkeland *et al.* 1998, Keller *et al.* 2015) causes Stark shifts and $2^{nd}$ order Doppler shifts at the level of several Hz, as soon as ions in 2D or 3D crystals leave the rf nodal line, defining the weak trapping axis. Even linear strings of ions experience residual on-axis rf fields due to imperfect electrode geometries and finite size effects leading to significant frequency shifts in extended crystals (Chou *et al.* 2010a, Herschbach *et al.* 2012). Finally, motional shifts due to excess ion heating (Brownnutt *et al.* 2015), the stability of crystal configurations and a complex spectrum of motional modes are expected to place a limit on the controllable ion number. In particular, heating of the secular motion of ions due to their interaction with fluctuating patch potentials or electric field noise scales proportional to the ion number (Turchette *et al.* 2000). For large ensembles of ions this can lead to significant time dilation.

For the application of optical clocks as sensors for geodetic heights, the present systematic frequency shifts need to be well controlled and stable in time, allowing reproducible frequency measurements. Spatially inhomogeneous frequency shifts across the ion Coulomb crystal are required to be within the single ion linewidth of the interrogated clock transition.

One approach to scale up the number of ions for optical frequency standards is to choose atomic candidates with clock states with very low, close to zero quadrupole moment. Already the $Al^+$ quantum logic clock, discussed in section 5.2, utilizes the feature of an atomic clock transition with vanishing electric quadrupole moment of the bare clock states to accommodate a second ion in the ion trap for sympathetic cooling and clock read-out. Just recently, quantum-logic clock read-out with multiple clock ions was proposed (Schulte *et al.* 2016).

The first proposal to scale up the number of ions for optical clock spectroscopy was made in 2012 (Herschbach *et al.* 2012), with the idea to avoid and reduce systematic shifts by choice of ion candidates with intrinsic low sensitivity on field gradients, and engineering ion traps that provide an environment with lowest systematic shifts and a high level of control of the ions' motion (Herschbach 2012, Pyka *et al.* 2014). In Herschbach *et al.* 2012 a scalable ion trap design was introduced, which provides a linear array of multiple ion traps. On the order of 10 clock ions can be stored in each trapping segment simultaneously, while still providing a defined and controlled environment for each ion ensemble. This way, optical frequency standards operating some 10´s to 100 ions can be realized, with the ability to reduce the averaging time of clock measurements by up to two orders of magnitude.

Candidates for this approach are ions with clock transitions between states of electron angular momentum $J \leq$ ½, such as the intercombination lines $^1S_0$ and $^3P_0$ of two-electron systems, where $\theta \approx 0$. The $Al^+$ ion, which is investigated at several major research labs in a quantum logic approach and discussed in section 5.2, fulfils this requirement. Direct detection of excitation on the $^1S_0$ to $^3P_0$ clock transition is possible in the $In^+$ ion, where relativistic effects sufficiently broaden the $^1S_0$ to $^3P_1$ intercombination line. This easily scalable detection scheme and the possibility for direct laser cooling on the same transition (Peik *et al.* 1995, Peik *et al.* 1999), make $In^+$ a ready candidate for a multi-ion approach. Clock spectroscopy in $^{115}In^+$ on the $^1S_0$ to $^3P_0$ transition was demonstrated (von Zanthier *et al.* 2000, Wang *et al.* 2007) and is presently investigated in Tokyo and Braunschweig for multi-ion clock operation (Pyka *et al.* 2014, Ohtsubo *et al.* 2017). Recently, higher order contributions to the quadrupole shift due to hyperfine coupling of the electrons to the electromagnetic moments of the nucleus have been considered for these two-ion species. A fractional frequency shift $\Delta\nu/\nu < 4\times10^{-19}$ for eight $In^+$ ions and $\Delta\nu/\nu < 2\times10^{-20}$ for an $Al^+$ ion clock was derived (Beloy *et al.* 2017). Besides atomic states with $J \leq$ ½, also the octupole transition to the highly shielded F state of the $Yb^+$ ion with low quadrupole moment can be a possible candidate for a multi-ion approach (Huntemann *et al.* 2012).

A major technological challenge is the design of multi-ion traps in which systematic frequency shifts are controlled. Scalable ion traps for frequency metrology have been built at PTB over the past 5 years. A prototype trap with integrated electronics is shown in figure 5.4 together with a laser-cooled ion Coulomb crystal stored therein. In such type of traps, micromotion and heating rate measurements with fractional clock frequency shifts of $10^{-20}$ have been demonstrated for single ions (Pyka *et al.* 2014, Keller *et al.* 2016). Based on this design, fully non-magnetic, precision laser-engineered ion traps made of AlN ceramic wafers with Ti/Au coated electrodes have been developed. These types of traps allow for multi-ion clock operation at the level of $10^{-18}$ over a spatial extent of several millimetres. Due to the low RF losses and high heat conductivity of the AlN stack, black-body shifts due to the trap environment are on the order or below $10^{-19}$ for $Yb^+$, $In^+$, and $Al^+$ ions (Dolezal *et al.* 2015, Keller *et al.* 2016). In these AlN traps, recently, also 3D control of time dilation shifts due to micromotion below $10^{-18}$ was demonstrated over the full ion Coulomb crystal of 400 µm size (Keller *et al.* 2017).

Another multi-ion approach aims at the cancellation and active compensation of shifts, together with averaging of frequencies to synthesize effective atomic transitions, which are insensitive to external fields (Arnold *et al.* 2015). This approach applies to atoms with negative differential scalar polarizability, for which micromotion shifts can be eliminated to a sufficient degree (Dubé *et al.* 2013). Averaging over multiple hyperfine levels, when $I \geq J$, cancels the quadrupole shift and creates an effective $m = 0 \rightarrow m = 0$ transition (Barrett 2015, Arnold *et al.* 2016a). In this approach, a single ion Coulomb crystal consisting of a large number of ions (100s to 1000) is envisioned to provide an ion frequency standard with an order of magnitude improved stability (Arnold *et al.* 2015). Spectroscopy of the lutetium ion clock transition was recently reported (Arnold *et al.* 2016b). In this context, an active optical frequency standard was proposed using a Coulomb crystal composed of ions with negative differential polarizability on the clock transition (Kazakov *et al.* 2017). Here, ions with narrow clock transitions are coupled to an optical cavity and play the role of the gain medium. As an example, an active clock lasing on the $^3D_2 \rightarrow {}^1S_0$ transition in $^{176}Lu^+$ can serve as a novel optical frequency standard.

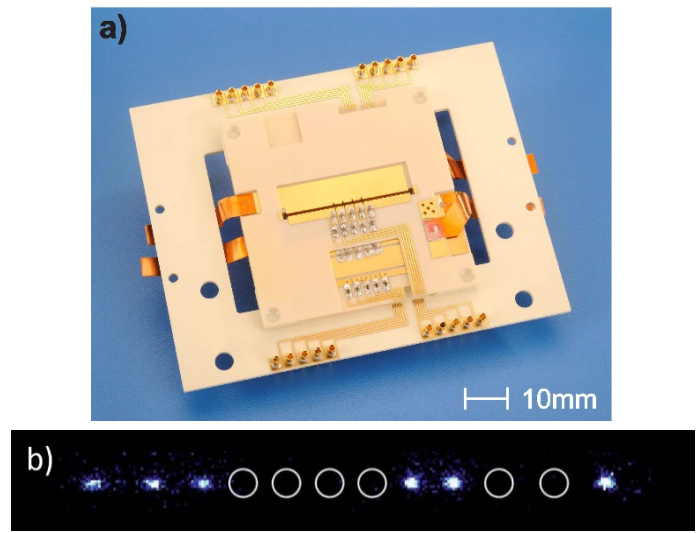

*Figure 5.4(a): Prototype ion trap of a chip based, laser cut and structured ion trap design for scalable multi-ion operation. The chips have integrated filter electronics. 5.4(b): Photo of a $Yb^+/In^+$ Coulomb crystal with fluorescing $^{172}Yb^+$ ions and sympathetically cooled $^{115}In^+$ ions (indicated with circles). Photos are taken from (Pyka et al. 2014).*

While single ion clocks can already provide excellent long-term stable frequency references, multi-ion clocks are a promising path to realize ultra-stable optical frequency standards based on trapped ions. They have the capability of providing compact frequency standards with a $10^{-18}$ frequency resolution in short averaging times of a few hours and represent a basis for the future development of ion clocks with advanced interrogation protocols or quantum-enhanced operation below the quantum projection noise limit. With the use of multiple ions for precision spectroscopy, also clock operation using specially engineered substates, which are protected against systematic shifts can be envisioned (Roos *et al.* 2006, Ludlow *et al.* 2015). Yet, multi-ion clocks are still at the verge of technological realization. In addition, for field missions outside the lab, when technical demanding, cutting edge clock lasers are not available, the development of optical clocks operating larger ensembles of ions promises the realization of ultra-stable frequency standards with short enough measurement times to even resolve the daily tidal effects on Earth's surface (see section 2).

## 6. Clock Comparisons and Frequency transfer

Optical clock comparisons have already been used in several pathfinder experiments testing fundamental physics (Blatt *et al.* 2008; Rosenband *et al.* 2008; Le Targat *et al.* 2013; Huntemann *et al.* 2014; Godun *et al.* 2014; Delva *et al.* 2017), or demonstrating chronometric levelling (Lisdat *et al.* 2016; Takano *et al.* 2016). They also support the characterization and validation of state-of-the art optical clocks, testing the reproducibility of optical clock frequency ratios. Most important for the scope of this paper, all geodetic applications as discussed in section 2 rely on the clock frequency comparison of remote clocks.

Frequency comparisons between different clocks are traditionally performed by side-by-side comparison. To span any geographical distances, a natural way is to bring portable atomic clocks from one site to another. This approach was pursued for optical systems almost from their inception, for example with a specially developed optical frequency standard based on Ca atoms, with fractional uncertainty of $10^{-12}$, that was transported from the PTB in Germany to the USA (Kersten *et al.* 1999).

An alternative path is to transport only the *information* that an atomic clock provides, i.e. its frequency or absolute timing signals (Grosche, 2015). Frequency and time transfer is intimately related to the highly evolved methods for establishing an international time scale, so far using satellite techniques (Kirchner *et al*. 1993, Petit *et al*. 2015). Frequency transfer via satellites achieves a fractional frequency uncertainty of around $10^{-15}$, in special cases close to $10^{-16}$ (Petit *et al*. 2015a, Droste *et al*. 2015), i.e. sufficient for most microwave frequency standards. Satellite techniques are advantageous in bridging long-distances, giving global access and flexibility of location. In recent years, methods based on optical telecommunication fibre networks developed rapidly: early attempts already achieved a fractional uncertainty of $10^{-14}$ over a short distance of 3 km (de Beauvoir *et al*. 1998). More recently, a fractional frequency uncertainty close to $10^{-20}$ was obtained over a loop of 1400 km rented telecom fibre in optimal cases, when transferring an optical frequency (Raupach *et al*. 2015). This is currently *four orders of magnitude* more accurate than the best operational satellite techniques.

Finally, the combination of both techniques, i.e. portable clocks *and* accurate frequency transfer, enables chronometric levelling experiments (as explained in section 2) between locations that are of geodetic interest but happen not to be equipped with an optical clock validated to within a few $10^{-18}$. For geodetic applications, portable clock systems may even be operated simultaneously at several locations *along* the route of an existing fibre link and be connected by frequency extraction (Grosche 2014) to another optical clock that is typically found at the end of such a link. Together with the future development of optical satellite links, more flexible choice of sites, mountainous areas, islands or other relevant locations may become accessible, where no institute with highly developed infrastructure will be found.

In the following, we briefly review frequency transfer techniques using satellites (section 6.1), including some basic concepts applicable to all transfer mediums, and then turn to techniques using optical fibre as a transfer medium (section 6.2). Here we concentrate on the method with the lowest frequency uncertainty, i.e. optical frequency transfer (Grosche *et al*. 2009). For recent brief overviews of frequency transfer see also (Riehle 2017) and (Piester *et al*. 2011) and the references therein. The current development of portable atomic clocks is discussed in section 6.3.

### 6.1 Satellite Links

Satellite links based on radio frequency or microwave signals predominantly involve either the passive reception of signals broadcast from Global Navigation Satellite Systems (GNSS), especially GPS, or the active, bi-directional exchange of signals via communications satellites, i.e. Two-Way Satellite

Time and Frequency Transfer (TWSTFT). Detailed reviews of these two well-established approaches their variants and state-of-art in year 2002 are given by (Bauch and Telle 2002, pages 828-834) and (Levine 2002) and references therein, in addition to many books. An excellent recent summary of both techniques is provided by (Bauch 2015).

The achievable fractional frequency uncertainty of both GNSS and TWSTFT improved by less than one order of magnitude between 2002 and 2015, hovering near $10^{-15}$ for measurements exceeding one day (Huang *et al.* 2016a). In this section, we outline the basic principles and concentrate on ideas and developments that penetrate towards $1\times10^{-16}$ fractional frequency uncertainty.

Satellite links using *optical* signals usually employ two-way techniques. One system in operation for several decades is "Time Transfer by Laser Link" (T2L2) based on short optical pulses (Samain *et al.*, 2014, and references therein). Its design is similar to the earlier "Laser Synchronisation from Stationary Orbit" (LASSO)-system (Fridelance P. and Veillet C. 1995). T2L2 is embedded on the satellite Jason-2 and was designed to allow synchronization of remote clocks with an instability of 1 ps over 1000s, i.e. $10^{-15}$ fractional frequency uncertainty. Encouraging earth-bound tests of free space optical frequency transfer have been performed over distances of several km (horizontally) both for a continuous optical carrier ((Djerroud *et al.* 2010), with a projected frequency uncertainty below $10^{-17}$), and for femtosecond-pulse transfer (Deschênes *et al.* 2016, and references therein). The latter system *demonstrated* a fractional frequency uncertainty well below $10^{-18}$ across a turbulent 4-km atmospheric path, and points the way towards an "optically based global navigation satellite system" (Deschênes *et al.* 2016).

6.1.1    Passive reception of microwave signals: Global Navigation Satellite System Links

In the 1970s, a system of "miniature interferometer terminals for Earth surveying" was suggested, where "each receiver detects radio signals from low-power transmitters on Earth-orbiting satellites" (Counselman and Shapiro, 1979). This idea was inspired by Very-Long-Baseline Interferometry (VLBI), where radio signals from distant stars are observed at several positions on Earth. The Global Positioning System (GPS), now available for more than three decades, maintains at least 24 satellites whose positions are precisely known. These broadcast time-tagged, modulated microwave signals (with additional information about the satellite clock, position etc.). If at least four satellites are visible simultaneously, the receiver can determine its location and time by processing the signal arrival times. Since the 1980s, GNSS, and its most common representative, GPS, has therefore been used to realise time and frequency links; different techniques are reviewed e.g. in (Bauch 2015).

GNSS-carrier phase (CP) uses the phase of the microwave carrier in addition to the radio frequency code, to achieve higher time resolution. Using precise point positioning (PPP) methods, phase and code measurements are combined using precise satellite orbit and clock information, and modelling of the one-way signal propagation, provided by the International GPS Service (IGS). Such systems now achieve a time instability of about 100 ps over one day, and 300 ps over one month, corresponding to a fractional frequency resolution of about $1\times10^{-15}$ at one day averaging and close to $1\times10^{-16}$ for 30-day averaging (Petit *et al.* 2015, Droste *et al.* 2015). Using GPS PPP, two $^{171}$Yb$^+$ ion optical clocks, one at NPL in London (UK), and one at PTB in Braunschweig (Germany) have recently been compared, observing a fractional frequency difference of $1.3\times10^{-15}$ with a combined uncertainty of $1.2\times10^{-15}$ for a total measurement time of 67 hours (Leute *et al.* 2016). Further, recent improvements in data processing techniques, e.g. improved "integer ambiguity resolution", now open the path to sub-$10^{-16}$ frequency resolution (Petit *et al.* 2015a).

6.1.2    Two-Way Satellite Time and Frequency Transfer

Two-way (TW) techniques were developed for point-to-point time and frequency transfer with lowest uncertainty (Kirchner *et al.* 1993), They are used for radio frequency, microwave and optical signals, propagating in any medium, including free-space and optical fibres (see section 6.2).

The basic idea is shown in figure 6.1 for exchange of timing signals: two endpoints 1 and 2 each have their own time-scale $T_1$ and $T_2$ with some unknown offset between them $\Delta T := T_1 - T_2$; both operate transmitting and receiving stations, and each performs a local time-interval measurement, $\Delta T_1$ and $\Delta T_2$, respectively, between *sending* out their own pulse-per second (PPS) generated by their local clock, and *receiving* the PPS from the remote station. Thus, we measure at station 1: $\Delta T_1 = \tau_{21} + T_1 - T_2$ and at station 2: $\Delta T_2 = \tau_{12} + T_2 - T_1$.

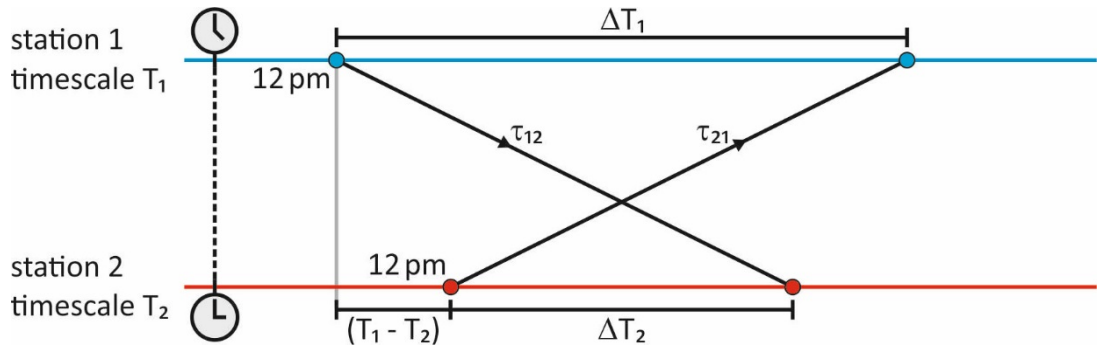

*Figure 6.1: Basic scheme of two-way time transfer*

If the overall signal delay is *completely symmetric*, i.e. the same in both directions: $\tau_{21} = \tau_{12} := \tau_0$, the (current) difference of the two time-scales $\Delta T$ is simply given by

$\Delta T := T_1 - T_2 = (\Delta T_1 - \Delta T_2)/2$. More generally: $T_1 - T_2 = (\Delta T_1 - \Delta T_2)/2 + (\tau_{12} - \tau_{21})/2$, i.e. the second term $(\tau_{12} - \tau_{21})/2$ summarizes all asymmetries. Only these delay asymmetries affect the accuracy of time transfer.

Importantly, for *frequency transfer,* constant delays do not contribute to the frequency uncertainty; so, in the case of TW-transfer, only changes of the asymmetric components of delays are relevant.

In a satellite transfer system using microwave signals, as shown in figure 6.2, there are several contributions to the delay asymmetry, following (Kirchner *et al.* 1993):

(a) differences in signal delay (to and from the satellite) in the two directions (station 1→ station 2, and 2→ 1),

(b) differences in satellite transponder delays in the two directions,

(c) path non-reciprocity due to the rotation of Earth (Sagnac effect)

(d) delay differences in the transmit and receive parts of the earth stations 1 and 2 (station delays).

Thus, e.g. contribution (b) should cancel if the same transponder is used in both directions. However, the same transponder is typically not available for both directions and periodic variations up to 2 ns/day (equivalent to frequency variations exceeding $10^{-14}$) have been observed in transcontinental links (Piester *et al.* 2008) - different thermal gradients as the satellite rotates daily with respect to the sun are believed to be responsible. Contribution c) can be calculated to within 1 ps if the positions of the ground stations are known within ~ 30 m (Piester *et al.* 2008); (d) may be measured prior to operation, e.g. by collocating the receive/transmit stations (Kirchner *et al.* 1993). Contribution (a) is estimated to be < 100 ps ((Kirchner *et al.* 1993) and references therein) for simultaneous transmission in the "Ku-Band".

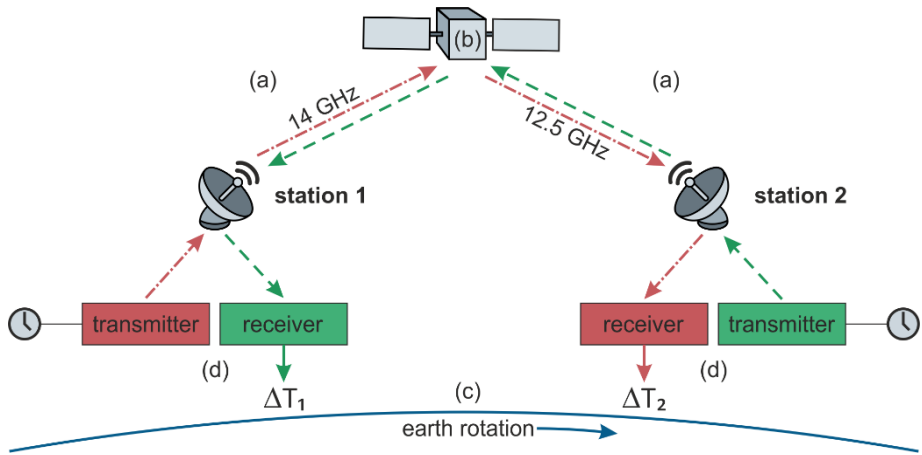

*Figure 6.2: TW- satellite transfer using microwave signals: dominant uncertainty contributions (a-d) as described in the text above.*

TW-satellite links normally exchange microwave signals in the "Ku-Band" (transmit ~14 GHz, receive ~12.5 GHz) between dedicated ground stations via geostationary telecommunication satellites, that relay the signals (nominal satellite translation frequency ~1.5 GHz). The signals are coded with pseudo-random noise (prn): pulses, or "ticks" from a local clock drive a prn-code generator that typically runs at a frequency of order 10 MHz. Each transmitting station has a dedicated prn-code, and for a code of length 10000 "chips" and a chip rate of 2.5 MChips/s (so that one chip lasts 400 ns), 250 codes per second are transmitted. The receiver itself generates copies of the code it expects to receive, and detects the timing of the received code by correlation between the received signal and local copies, as shown in figure 6.3.

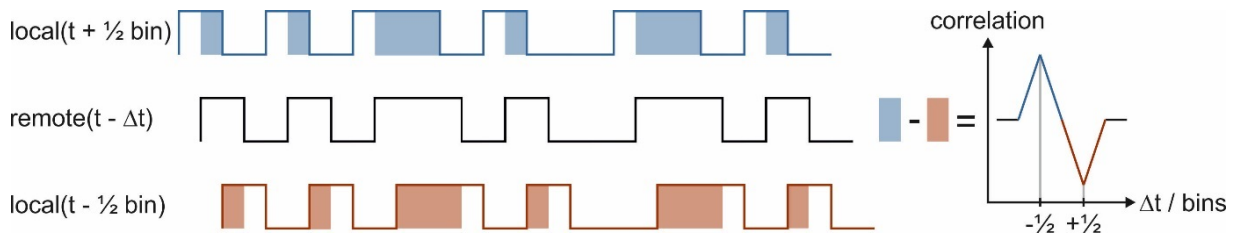

*Figure 6.3: Timing resolution in the early/late copy cross-correlation process. The receiver generates two delayed copies of the expected code signal, displaced by ± ½ bin (or chip), and compares each copy to the received signal; Δt is adjusted such that the output of the correlator difference is zero. At a chip rate of 2.5MCh, ½ bin corresponds to 200 ns. More details are summarised e.g. in (Levine 2002, p.1153).*

For a given carrier-to-noise ratio, the timing instability (and hence the frequency uncertainty) is expected to be inversely proportional to the chip rate, as the correlation process becomes more sensitive for higher chip rates (see figure 6.3).

Already in 1993, a short term *timing* instability of $\sigma_x(\tau) = 8\times10^{-10}\ (\tau/s)^{-1/2}$ was reported between 1 s and 100 s; this would result in a frequency instability expressed as Modified Allan Deviation (MDEV) (Allan and Barnes 1981) mod $\sigma_y(\tau) = 1.4\times10^{-9}\ (\tau/s)^{-3/2}$, i.e. reaching $1.4\times10^{-12}$ at 100 s. The theoretical prediction of the frequency instability expressed as Allan Deviation (ADEV) for higher chip rates is shown in figure 6.4: a frequency resolution of $10^{-10}$ at 1 s, and $1\times10^{-16}$ at $10^6$ seconds may be achievable for 20 MCh/s. However, increasing the chip rate increases the bandwidth required and hence the rental cost.

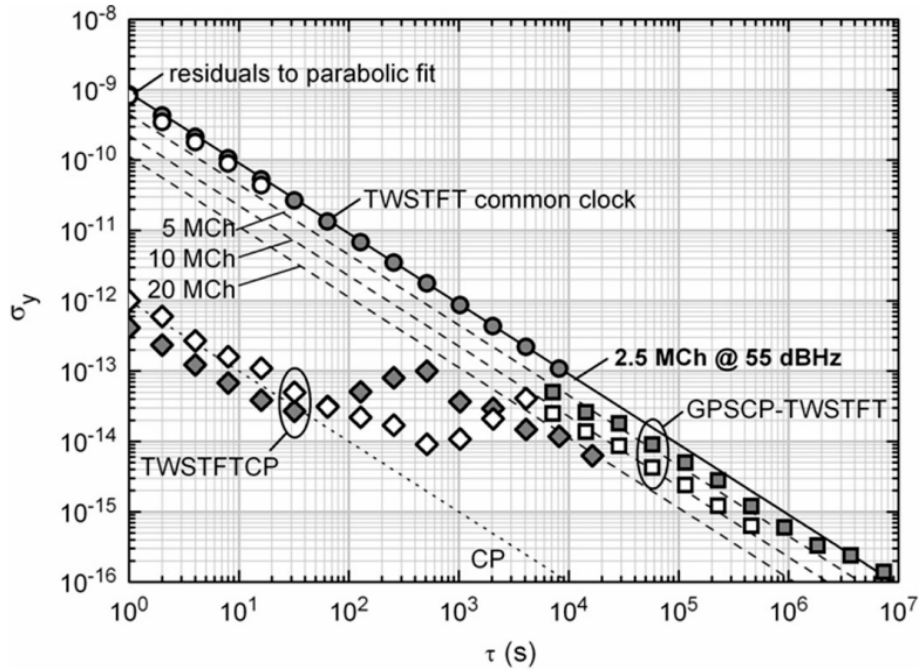

*Figure 6.4: Improved TWSTFT-systems, summarised with GPS-CP: the figure is reproduced from (Piester et al. 2008a, figure 2). It shows measured data taken with (i) a TWSTFT code-system (common clock 2.5 MCh) (ii) a TWSTFT-carrier-phase system (CP), see also (Fonville et al. 2004), and (iii) theoretical predictions for code-systems with chip-rates up to 20 MCh/s. Also shown is (iv) the difference between two links based on GPS-carrier phase and TWSTFT code.*

To keep the required bandwidth small but still improve the short-term frequency instability by up to a factor of 1000, the carrier phase (CP) of the microwave carrier (~ 10 GHz) has been used; this is referred to as "TWSTFT-CP", or "TW-CP". Early experiments reported in 2004 already achieved a frequency instability just below $10^{-12}$ at 1 s, (Fonville *et al.* 2004), but then levelled off near $10^{-14}$ for longer measurement times. More recent TW-CP experiments reached a frequency instability (MDEV) close to $10^{-13}$ at 1 s and a few $10^{-16}$ beyond 2000 s (Fujieda *et al.* 2014). Petit *et al.* (2015), however, question whether it will be practical to operate TW-CP with commercial geostationary satellites, and expect present techniques to remain limited close to $1 \times 10^{-16}$, even at averaging times of at least several days.

In addition to cost considerations, having to share commercial satellite transponders impedes improvements (towards lower time and frequency transfer uncertainty), as the signal strength, the usable bandwidth and the experiment durations are constrained (Petit *et al.* 2015, Huang *et al.* 2016a). Having to relay via telecommunication satellites also "limits the choice of locations for ground terminal installations and the operational distance to ~ 10000 km" (Piester *et al.* 2011). A dedicated space payload, such as envisioned for the Atomic Clock Ensemble in Space (ACES) mission (Heß *et al.* 2011), could be an attractive solution: here a dedicated microwave link is planned, aiming for a timing instability of about 7 ps at one day, and 20 ps at ten-day averaging. This corresponds to fractional frequency instabilities below $1 \times 10^{-16}$ (1 day) and about $2 \times 10^{-17}$ (ten days); recent ground tests have been encouraging (Schäfer and Feldmann 2016).

## 6.2 Fibre Links: transferring precision signals via optical telecommunication fibres

"Fibre links", i.e. frequency and time transfer connections using *guided* light in optical telecommunication single-mode fibre, have rapidly developed in the last two decades. Fibre optic long-distance connections are ubiquitous and "underpin the global communications infrastructure - with current commercial systems transmitting data rates exceeding 10 Terabit/s over a single fibre core" (Maher *et al.* 2016). Dense wavelength division multiplexing (DWDM) for telecommunication specifies ~72 channels spaced 100 GHz apart per fibre: compared to communication demands, the bandwidth required by frequency signals is negligible. Furthermore, a single long-distance fibre cable contains 12 to 96 fibre cores, the deployed infrastructure is continually expanding and can be rented for communication or other purposes.

Compared to free-space (or rf-cable) links, three further advantages of optical fibre are important:

a) signals experience an attenuation of only 0.2 dB/km, i.e. 20 dB per 100 km, at wavelengths in the C-band (~1530-1560 nm) for standard backbone telecom fibre, such as SMF28$^{TM}$.
b) the telecom industry has developed many components, such as beam-splitters, optical filters, or signal modulators in a very compact, reliable designs and at comparatively low cost.
c) an optical fibre buried underground is comparatively well shielded and experiences very small fluctuations of refractive index, compared to paths through the atmosphere near ground.

Standard telecom fibres are not polarisation maintaining: polarisation maintaining (PM) fibres are sold for almost any wavelength, but are much more expensive. Within laboratories or over short distances, fibre links can be established very easily, using similar (but usually simplified) techniques (Rosenband *et al.* 2008) to those presented in this chapter.

Some basic ideas in time/frequency transfer via fibre links using "intensity modulation and direct detection" (IM-DD), i.e. transmitting radio frequency signals modulated onto an optical carrier, are reviewed in (Sliwczynski *et al.* 2010); that review concentrates on methods analogous to satellite techniques. Additionally, already from the 1980s, reference frequencies (and later time-stamps) were distributed via rf signals over a few km to several sites, in support of the deep space network, see (Calhoun *et al.* 2007). Similarly, several international groups used network data packets themselves to transfer frequency or timing information (Ebenhag *et al.* 2011), simply "listening to the existing data frames" (chapter. 5, page 33 in Ebenhag 2013). Based on the Ethernet protocol, combined with the "precision time protocol" (PTP), a frequency and timing distribution network for data acquisition at accelerator sites of CERN has been established with the "White Rabbit" system (Serrano *et al.* 2009).

Ground-breaking work for IM-DD frequency transfer, achieving an instability near $10^{-17}$ (at $10^5$ s) already in 2006 using 100 MHz rf-modulation, and impressive developments reaching to a few $10^{-19}$ (at $10^5$ s) in 2010 using microwave modulation at 9 GHz over a distance of 86 km, are documented in (Narbonneau *et al.* 2006, Lopez *et al.* 2008, Lopez *et al.* 2010a) and led to several applications. The key idea to improve frequency transfer by direct modulation in optical fibre links – akin to that shown in section 6.1.2 for satellite links – was to move to ever higher modulation frequencies. IM-DD techniques continue to be studied, often achieving highly reliable systems, but akin to satellite rf techniques, they mostly achieve instabilities of order $10^{-13}/(\tau/s)$ at best.

For high accuracy applications, the logical step is to move to the optical carrier frequency itself (near 200 THz for lowest attenuation in optical telecommunication fibre); this approach achieves excellent short-term instabilities matching the performance of optical clocks (i.e. ~ $10^{-16}$ or so at 10s) and very low uncertainties, close to $10^{-19}$. This corresponds to a height resolution of about 1mm when performing chronometric levelling experiments. We thus concentrate only on optical carrier frequency transfer in this review, because of its relevance to chronometric levelling experiments performed with optical clocks. Many other important tasks that require less accuracy are more easily served by rf-fibre links. Previous reviews of fibre optical frequency and time transfer can be found in (Ye *et al.* 2003; Foreman *et al.* 2007a, Lopez *et al.* 2015).

6.2.1 One-way frequency dissemination by fibre

Analogous to satellite links, "fibre links", i.e. frequency or time links that use fibre optic transmission, are either "one-way" systems, where a signal is transmitted through the optical fibre to a passive receiver (akin to GNSS frequency links), or "two-way" systems, where signals are exchanged between two locations to determine the delay fluctuations and compensate them (akin to two-way satellite links).

For optical frequency transfer, the relevant changes in the signal delay $\tau_0$ are dominated by environmental temperature fluctuations, which change the refractive index $n$; in the case of modulated signals, $n$ is replaced by the group refractive index $n_g$, see e.g. (Sliwczynski *et al.* 2010). For a fibre of length $L$, we calculate the momentary fractional frequency offset $\Delta\nu(t)/\nu$ due to changes of the optical path length $nL$:

$$\Delta \nu (t)/\nu \sim \lambda_0/c_0 \times 1/(2\pi)\, d\phi(t)/dt = \lambda_0/c_0 \times d/dt\,(nL/\lambda_0) \sim L/c_0 \times d/dt\,(n)$$
$$= \tau_0/n \times dn/dT \times dT(t)/dt. \tag{6.1}$$

Here $1/n\, dn/dT \sim 7$ ppm/K or close to $10^{-5}$ per Kelvin (Wray and Neu 1969). Early experimental studies gave a similar temperature dependence for the *group* refractive index $n_g$ (Lutes and Diener 1989), consistently reported later (Narbonneau *et al.* 2006, Calhoun *et al.* 2007); the corresponding signal delay change per kilometre and Kelvin, of 35 – 40 ps/(km·K), was also confirmed by (Sliwczynski *et al.* 2010).

Underground fibres are very well shielded: from the above, a temperature drift of 1 mK/s will result in a frequency offset of only $\Delta\nu/\nu = 3.5\times10^{-14}$ per km fibre. Uni-directional propagation over 3 km yielded a fractional frequency shift $\Delta\nu/\nu < 10^{-14}$ (de Beauvoir *et al.* 1998); similarly, on German fibre links fractional frequency instabilities near $10^{-14}$ are routinely observed for distances up to 1400 km (Raupach *et al.* 2015); the frequency offsets are typically below $10^{-13}$.

One-way frequency dissemination by optical fibre is thus entirely sufficient for most applications, such as remote calibration of measuring instruments. This is true for simple, passive reception, and further improvements may be achievable by applying corrections, e.g. using systems with dual wavelength transmission (akin to GPS approaches; (Ebenhag 2013)); for one-way timing dissemination, similarly, Sliwczynski *et al.* (2010) cite slow delay changes of about 1 ns per month on a 6 km underground fibre duct for modulated signals (~ a mean frequency offset of $4\times10^{-16}$ over the 30 day integration time; short-term changes may be much larger).

### 6.2.2 Bi-directional links – large scale optical fibre interferometers

Aiming to achieve lower uncertainty, below $\sim10^{-15}$ optical frequency links may be operated in a bi-directional way, to detect and correct one-way delay changes. One idea to implement this was presented as early as 1994 by (Ma *et al.* 1994, Ye *et al.* 2003). The most common set-up developed later (Foreman *et al.* 2007a; Grosche *et al.* 2007; Foreman *et al.* 2007; Newbury *et al.* 2007; Grosche *et al.* 2009) is shown in a schematic way in figure 6.5: we incorporate the fibre optic link in the measurement arm of a long-distance unbalanced optical interferometer, and stabilize the optical phase (modulo a fixed frequency offset) of the transmitted carrier signal by applying a correction signal to an actuator (here an acousto-optic modulator, AOM1) *within* this arm. Analogous designs had previously enabled phase-stabilized transfer of radio frequency signals modulated onto the optical carrier (Narbonneau *et al.* 2006) and later supported the stabilized transfer of pulsed signals from a frequency comb (Marra *et al.* 2012).

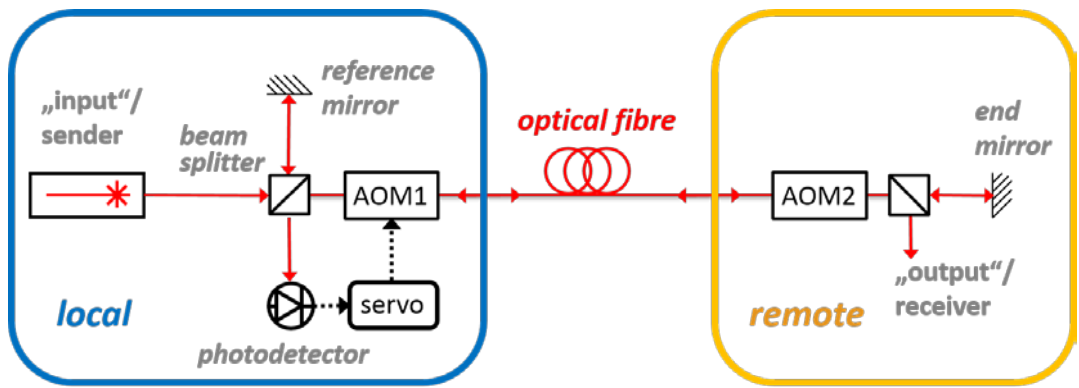

*Figure 6.5: Interferometer for detecting and compensating fibre noise between local and remote end points of an optical fibre link. An ultrastable laser (sender) is injected at the local end, where reference light split off at the first beam splitter travels via a reference mirror to the photodetector. Signal light travels through a frequency shifter (acousto-optic modulator; AOM1) via the optical fibre to the remote end. Part of the light is reflected by the end mirror, after being shifted by a fixed frequency by a second frequency shifter (AOM2), and returns through the entire path back to the local end. Here it is re-united with reference light on the photo-detector. The "servo"-box detects the phase and frequency offset between reference light and returned light, and stabilises this (i.e. keeps it constant) by generating a suitable correction frequency applied to AOM1.*

The first long-distance measurements using such a system transferring an optical carrier over 86 km deployed fibre and extended by fibre spools to 211 km, carried out jointly by French and German researchers in Paris in 2006 (Grosche *et al.* 2007), demonstrated the high potential of this method, with similar demonstrations quickly following over a total of 32 km fibre distance (Foreman *et al.* 2007a) and over 76 km deployed fibre – extended to 251 km by fibre spools- (Newbury *et al.* 2007). While only spanning relatively short distances compared to satellite links, the instability and uncertainty compared favourably with satellite methods, and fully supported the best optical clock performance for several years to come, see figure 6.6.

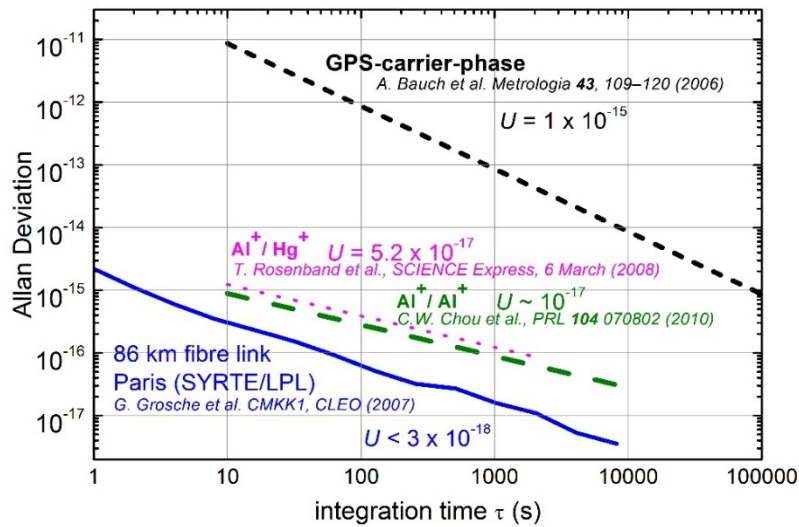

*Figure 6.6: Overview of the instability and uncertainty of global (GPS-carrier phase, guide to the eye) and regional (86 km fibre, data) methods of frequency transfer, differing by 3-5 orders of magnitude (in 2006/2007) and performance of the best reported optical frequency standards (in 2008 and 2010).*

Uncertainty contributions of this novel approach are presented in (Williams *et al.* 2008) – and include (but are not limited to) the ones listed there: (i) laser phase noise, (ii) interferometer phase noise, (iii) limited fibre noise suppression due to the time delay in the fibre, (iv) phase noise of rf components, (v) fibre non-linearities, (vi) polarisation effects, and (vii) detection noise (see Foreman *et al.* 2007).

Most of these contributions can be reduced by optimised design; we briefly consider two examples:

(i) Laser phase noise of the "link" (or "source") laser is converted by the delay of the link into self-heterodyne noise, falsely appearing as fibre noise.
To suppress this effect, a narrow-linewidth ultrastable "link" laser is used. One option is to stabilise the "link" laser to an optical clock laser via a fs frequency comb (Grosche *et al.* 2008). Here the optical beats between lasers and fs-comb are first combined to the so-called "transfer beat" that suppresses fs-comb noise (Telle *et al.* 2002). An actuator signal is derived from the transfer beat and stabilises the link laser to the clock laser, resulting in a relative linewidth of less than one hertz, as demonstrated by using *two* fs-combs in parallel to implement and test this method (Grosche *et al.* 2008). The self-heterodyne noise thus becomes negligible (Williams *et al.* 2008).
Alternatively, the link laser is stabilized directly to an ultrastable optical resonator operating at 1.54 µm (Newbury *et al.* 2007). These now significantly outperform systems in the visible (Kessler *et al.* 2012) (see section 4.4). Thus, more recently, ultrastable 1.54 µm-light is often transferred by the same fs-comb method (Grosche *et al.* 2008) to an optical clock frequency, to interrogate the clock transition, see section 4.1. This process allows to relate the link laser frequency to the optical clock frequency, as required for remote frequency comparisons.

(ii) "Interferometer phase noise", refers to residual noise entering via uncommon paths, and disturbances affecting the reference arm; this was found to limit the achievable instability (Williams *et al.* 2008); similarly, optical paths required to *characterise* the "remote" output by comparison to the "local input" may limit the measurement. Special set-ups were developed to minimise these influences.

Thus (over a distance of 146 km) $1\times10^{-19}$ uncertainty was soon demonstrated (Grosche *et al.* 2009), and no asymmetry or non-reciprocity of delay changes in the optical fibre were found down to this level.

### 6.2.3 Amplification concepts

In contrast to free-space propagation through vacuum, where the signal strength detectable in a fixed aperture decreases with the distance $R$ as $1/R^2$, the optical signal power transmitted through a fibre falls off *exponentially* with fibre length $L$; the signal is reduced by a factor 100 (20 dB) per 100 km. Hence, short fibre links (within a laboratory, or on a campus, with $L$ less than a few km) are often straightforward to implement in terms of signal power, but long distances become literally exponentially more difficult.

Fortunately, fibre optic infrastructure allows the insertion of amplifiers, which is necessary for $L > 100 – 200$ km. To compensate phase fluctuations, signal light must travel the same physical path in both directions (see figure 6.5); standard unidirectional telecom devices, such as amplifiers incorporating an optical isolator, are not suitable. The attenuation between accessible huts along a link, where amplifiers can be placed, is ~ 20 dB. This exceeds the maximum gain, of around ~15 dB, that a broad-band erbium-doped fibre amplifier can achieve in bi-directional operation without amplitude fluctuations or even lasing. An excess loss of ~5 dB per 100 km fibre thus builds up for longer distances that prevents reliable, phase-stable operation.

Overcoming this by dedicated development of suitable *bi-directional* amplification methods over the last eight years has enabled improvements and extension of fibre links, now spanning up to 1840 km (Droste *et al.* 2013). Today, in optimal cases the long-term instability and uncertainty reaches $< 10^{-19}$ (Chiodo *et al.* 2015), or even $< 2\times10^{-20}$ over 1400 km (Raupach *et al.* 2015). This is more precise than initial reports over much shorter distances, i.e. 25 m (Ma *et al.* 1994) or ~ 100 km about 10 years ago.

A discussion of the many concepts and developments in signal amplification, signal regeneration and remote control (of gain modules installed in huts along the fibre link) is outside the scope of this review. We refer the reader to the following publications and the references therein, as a start, assuming textbook knowledge of telecommunication amplifier physics. Without completeness: (i) bi-directional erbium doped fibre amplifiers (bi-EDFA) – (Newbury *et al.* 2007; Grosche *et al.* 2009; Predehl *et al.* 2012; Sliwczynski *et al.* 2012; Calonico *et al.* 2014; Lopez *et al.* 2015); ii) repeater lasers/ signal regeneration – (Foreman *et al.* 2007b; Lopez *et al.* 2010; Lopez *et al.* 2012; Lopez *et al.*, 2015; Chiodo *et al.* 2015); iii) Raman amplification – (Clivati *et al.* 2013; Clivati *et al.* 2015) and iv) fibre Brillouin amplification – (Ferreira *et al.* 1994; Terra *et al.* 2010; Raupach *et al.* 2014; Raupach *et al.* 2015; Soudi *et al.* 2016).

### 6.2.4 How to test a link – and perform accurate frequency comparisons of clocks

Frequency and time transfer links are commonly validated and characterised using either one of two approaches: a) characterising the constituent parts of the link, and estimating the overall uncertainty from the uncertainty contributions for each part, or b) summarily, by comparing the frequency (or time) difference reported by the link under test (LUT) versus a reference.

For approach (b), such a reference is usually another frequency (or time) link. In case of satellite links, the LUT and the reference are often of similar expected performance. As fibre links achieve much lower frequency uncertainty than satellite links, over distances up to ~1000 km, they have recently been used as such a reference ("ground truth") to characterise satellite links in Europe (Petit *et al.* 2015a, Droste *et al.* 2015) down to the level of $1\times10^{-16}$.

Similarly, where the frequency difference between two distant optical clocks is *a priori* known, we can test a frequency link between them down to the uncertainty of the clocks, or to the statistical uncertainty of the link (whichever is higher). This approach is useful for satellite links but it gives only a weak bound (or confirmation) for optical fibre links.

Furthermore, to perform chronometric levelling, we require the reverse, namely to measure the frequency difference between two distant optical clocks (that have a known side-by-side frequency difference) to determine the gravitational redshift between them. Therefore, fibre links are usually characterized prior to using them for frequency comparisons.

Link characterization is most often based on forming a loop, transporting the frequency information to a far-away lab on one fibre and back to the origin on a second fibre (Jiang *et al.* 2008; Grosche *et al.* 2009), with a variant using only one optical fibre but different frequencies (Calonico *et al.* 2014). Such a measurement precedes the application, and only once characterised, is the loop "split": the end mirror (figure 6.5) is moved to the far-away lab, to disseminate the optical frequency to this remote point.

After the 146 km loop PTB-Hanover-PTB had been tested down to the $1 \times 10^{-19}$ level (Grosche *et al.* 2009), an ultra-stable laser in the near-infrared was remotely characterised in Hanover (via 73 km fibre) with sub Hz-resolution (Pape *et al.* 2010)- and subsequently, the transition of the Mg optical clock (in the same lab in Hanover) was remotely measured versus a primary frequency reference at PTB, Braunschweig on the same link (Friebe *et. al.* 2011). Similar frequency measurements were performed, e.g. in Japan, where the Sr lattice clock frequency was measured over 120 km of fibre (Hong *et al.* 2009).

To confirm correct operation of the stabilized link *during a remote optical frequency measurement*, two stabilized links have been used, one disseminating the optical frequency to the remote lab, and a second one to return the frequency information to the local lab, where it is compared to the (virtual) loop input. Errors (or signal loss) on either the outgoing ("up-link") or the returning link ("down-link") will then show in the sum. A dual 920 km link was implemented in this way (Predehl *et al.* 2012) and subsequently used to measure the transition frequency in atomic hydrogen remotely in the lab in MPQ Garching, near Munich (Matveev *et al.* 2013). A similar approach was used already much earlier for rf-fibre links in France (Narbonneau *et al.* 2006), which supported remote measurements of transition in the $OsO_4$-molecule over a distance of 43 km (Daussy *et al.* 2005), and remote frequency synthesis (Argerence *et al.* 2015).

Cascading an "up-link" and a "down-link" requires a full link stabilisation system at both local and remote end, and – depending on the level of automation – some fine-tuning and supervision by dedicated link researchers at both ends. French and German groups have cooperated over more than six years to connect optical clocks at LNE-SYRTE in Paris with those in PTB, Braunschweig: each side built a 700 km scale national "up-link" from their respective institutes to the meeting point in the University of Strasbourg. There two repeater laser stations (Lopez et al. 2015) start the two national "down-links" and also compare the signals from both sides.

This 1400 km long, first international stabilised optical frequency link enabled frequency comparisons between LNE-SYRTE in Paris and PTB in Braunschweig, spanning a geographical distance of 700 km. Agreement of the clock frequencies limited by the uncertainty budget of the clocks themselves was found: for primary frequency standards (Guena *et al.* 2017) at a few $10^{-16}$, and for two Sr optical lattice clocks (Lisdat *et al.* 2016) at $5 \times 10^{-17}$.

The 700 km baseline frequency link connecting optical clocks in Paris and Braunschweig has thus enabled the first sub-1m proof of principle experiment of chronometric levelling over a larger geographical distance. In the same year, in Japan, frequency transfer over a short distance link (30 km) connecting two optical Sr clocks with a combined uncertainty of a $\sim 6 \times 10^{-18}$ was reported, to verify chronometric levelling at the 5 cm level (Takano *et al.* 2016).

A growing *European* network incorporates a 340 km baseline link between London and Paris: first data have been used already from comparisons between NPL-LNE/SYRTE and LNE/SYRTE-PTB to test special relativity by a detailed analysis of the diurnal dependence of optical clock frequencies (Delva *et al.* 2017). Work now continues to improve monitoring, reliability and automation of this connection and compare ever more precise optical clocks.

6.2.5   Current trends and efforts towards an optical frequency network

Compared to satellite techniques, which span very large distances quite easily and -in principle- allow access almost anywhere on Earth (because the satellites are already flying), a disadvantage of interferometrically stabilised point-to-point optical fibre links is the effort involved in establishing a connection between any two locations, which also increases with the distance involved.

Thus, an "important question is how to distribute reference frequencies to many users simultaneously in a cost-effective way" (page 1 in (Grosche 2014)). It turns out that simply adding suitable *detection units* can convert a point-to-point link into a "multi-point" system which disseminates a reference frequency to multiple users (each using their own, independent detection unit), as shown in figure 6.7 (Grosche 2010).

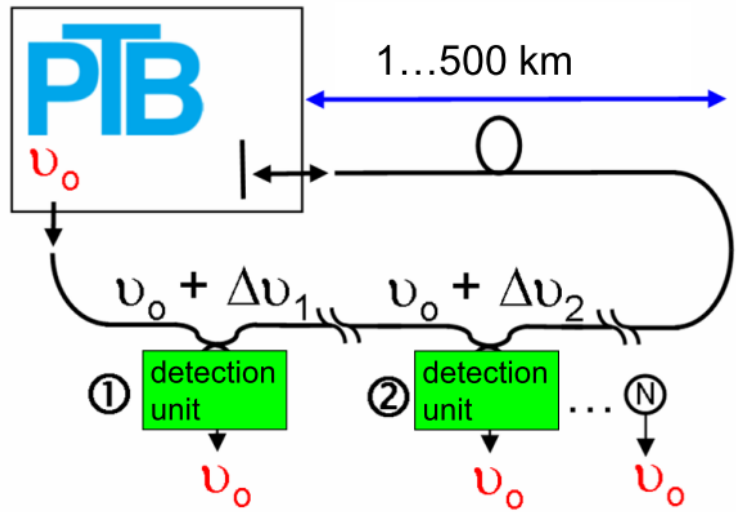

*Figure 6.7: Simplified schematic of multi-point distribution along a phase-stabilised fibre loop, beginning and ending e.g. at a national metrology institute like PTB, Braunschweig. At each user site, a detection unit extracts and suitably mixes signals, such that the reference frequency $v_0$ is delivered to the user despite frequency offsets along the link due to environmental noise.*

Suitable tapping of the transmitted signals – which can be an optical frequency, a modulated signal, timing information, or a combination of these, on either optical fibre or some other transmission path (Grosche 2010, Grosche 2014) – provides the same precision as that available at the remote end. The patented idea was quickly implemented, tested and analysed on several types of systems, with some examples given here: optical frequency distribution (Bai *et al.* 2013; Bercy *et al.* 2014; Bercy *et al.* 2016; Grosche 2014); rf-distribution (Gao *et al.* 2012; Krehlik *et al.* 2013; Li *et al.* 2016), and microwave distribution using an fs-comb output (Zhang *et al.* 2015). Note that other approaches are possible, in particular for branched topologies, see e.g. (Schediwy *et al.* 2013; Wu *et al.* 2016).

The set-up as shown in figure 6.7 allows monitored and automatic operation, as the so-called end mirror ("remote") can now be *permanently co-located* with the "local" end, thus forming a loop. The link can be continuously assessed, while operating only one fibre stabilization, and no specialized knowledge is required by the user who just receives the output of his detection unit.

In France, work on extending this method supports the construction of a national optical frequency metrology fibre network ("REFIMEVE+"), where "a reference optical signal generated at SYRTE will be distributed to about 20 academic and institutional users using the national academic network RENATER." (page7, Bercy *et al.* 2016). Potential applications, beyond frequency metrology and optical clock comparisons, are antenna arrays in radio astronomy, such as the square kilometre array (Dewdney *et al.* 2009).

As a future vision, we may see a combination of satellite methods (to link the continents), possibly using optical techniques based on femto-second pulse transfer (Deschênes *et al.* 2016), and national optical frequency metrology fibre networks, connecting space agencies, university laboratories, specialized laboratories in industry, telecommunication nodes and national metrology institutes.

Table 6.1 summarizes the different types of frequency transfer links discussed in this chapter, with their state-of-the-art capabilities, and suitability for geodetic experiments. State-of-the art frequency, frequency instability and the maximum demonstrated link distance with that performance are listed, as well as the geographical flexibility, availability/costs and an assessment, whether the method supports chronometric levelling with sub-10 cm uncertainty.

| link type | experiments since | state-of-the-art capabilities (2017) a fractional frequency uncertainty of $10^{-19}$ corresponds to a height uncertainty of ~ 1mm | | | geo-graphical flexibility | running costs/ availability | supports chronometric levelling < 10 cm |
|---|---|---|---|---|---|---|---|
| | | fractional frequency uncertainty | fractional frequency instability (ADEV) | link distance demonstrated with this performance | | | |
| **satellite: passive reception of GNSS-CP with PPP** | 1980s | ~$1\times10^{-16}$ (Petit 2015) | >$10^{-15}$ @ 1 d >$10^{-16}$ @ 30 d | 290 km geographical distance | worldwide coverage, just GNSS receiver | - / always | no |
| **-sSatellite: TWSTFT** | 1993 | ~ $10^{-15}$ (Huang 2016) | ~ $10^{-15}$ @ 1d | 2080 km | requires installing special satellite ground terminals | rental cost for satellite bandwidth | no |
| **satellite: TWSTFT-CP** | 2004 | < $10^{-15}$ (Fujieda 2014) | few $10^{-16}$ @ > 2000 s | 9000 km | as above | as above | no |
| ***ACES MWL (planned)*** | < 2009 | *projected: < $2\times10^{-17}$ (Schäfer 2016)* | *projected: < $10^{-16}$ @ 1 d ~$1\times10^{-17}$ @ 10d* | *projected: global* | *satellite ground terminals* | *18 months to 3 years mission on International space station (ISS)* | *projected performance nearly sufficient* |
| ***aatellite: Time Transfer by Laser Link (T2L2)*** | < 1997 | *specified: ~ $10^{-15}$ (Samain 2014)* | *specified: $10^{-15}$ at 1000s* | *specified: intercontinental* | *optical ground terminals* | *dedicated satellite / passing satellite, favourable weather condition* | *no* |
| **free space: femtosecond pulse transfer** | < 2014 | direct free-space link: < $10^{-18}$ (Deschênes 2016) | < $10^{-18}$ @ >1000s | 4 km | line of sight | telescope/ favourable weather condition | yes for 4 km; longer, if satellite links are feasible with this technique |
| **optical fibre links: IM-DD with noise compensation** | 1980s | < $4\times10^{-19}$ (Lopez 2010a) | ~ $10^{-15}$ @ 1s ~ $4\times10^{-19}$ @ $10^5$ s | 86 km fibre link | access to optical fibres | fibre rental | yes, over relatively short distance |
| **optical fibre links: one way frequency transfer** | < 1994 | - (Sliwczynski 2010) | $4\times10^{-16}$ @ 30 d | 6 km fibre link | access to optical fibres | fibre rental | no |
| **optical fibre links: bidirectional/ interferometric** | 1994 | ~$1\times10^{-20}$ (Raupach 2015) | $2\times10^{-15}/\tau$ (s) ~$2\times10^{-20}$ @ 1d | 1400 km fibre link | access to optical fibres, ultrastable laser | fibre rental | yes |

*Table 6.1 State-of-the-art capabilities of different frequency transfer techniques*

Chronometric levelling, i.e. using frequency transfer links and optical clocks to determine gravity potential differences, will become increasingly useful to geodesy as lower and lower uncertainties are reached, and over larger distances. As described in section 2, two geodetic ("classical") approaches to derive static (i.e. time-independent) gravity potential differences exist, the geometric levelling and the GNSS/geoid approach. The practical results of both approaches are currently inconsistent at the decimetre level across Europe. Thus, chronometric levelling will provide useful input and an independent reference for classical techniques from an uncertainty of about 10 cm (or $1\times10^{-17}$) and below.

Note that the most accurate frequency transfer techniques to date can only be verified by loop experiments, where the frequency is brought back to the origin (see section 6.2.4). Still, with the best stationary optical clocks at the level of a few $10^{-18}$, and the best frequency transfer techniques at a level of $10^{-20}$ (optical fibre-based frequency transfer) and short free-space links readily demonstrating below $10^{-18}$ chronometric levelling looks very promising even for determining static gravity potential differences. Additional, time-dependent information is available, as chronometric levelling measurements can be made rapidly (within a few hours) and also repeatedly, once a frequency link connection between two optical clocks is established.

### 6.3 Portable Atomic Clocks

Optical fibre links are currently the only alternative for top quality comparisons of optical clocks. However, due to the required rigid infrastructure the options for validating clocks or geodetic measurements are so far limited to dedicated laboratories, the development of portable (optical) clocks opens the door to higher flexibility. For the validation of clocks, a state-of-the-art frequency standard can be brought to an institute where a reference system is required. For geodetic applications, such a portable system can be operated along the route of existing fibre links and there be connected by frequency extraction (Grosche 2014) to another optical clock that is typically found at the end of such a link. Together with future satellite links, by the more flexible choice of sites, mountainous areas, islands or other relevant locations can become accessible, where no institute with highly developed infrastructure will be found.

The development of frequency standards and clocks for applications outside dedicated metrology laboratories have been pursued for decades. Microwave frequency standards like hydrogen masers or caesium beam clocks are commercial products and thus commonly available. The comparative simplicity of these systems makes them easily applicable, however on the cost of performance compared to state of the art laboratory frequency standards.

Though masers provide with $10^{-15}$ and below the best fractional instability of all commercial devices on timescales of several thousand seconds to days, their unpredictable long-term frequency drift makes their application difficult when a reliable absolute frequency is required. For relative measurements with small instability, they are nevertheless very efficient and were the heart of outstanding experiments like gravity probe A, see section 3. Commercial Cs clocks reach a mid-$10^{-13}$ uncertainty. For applications like chronometric levelling, the height resolution related to the frequency accuracy is too low to be of interest.

The transportable caesium fountain clock developed by LNE-SYRTE (Bize *et al.* 2004; Guéna *et al.* 2012) pushed the limits on fractional frequency uncertainty to below $6\times10^{-16}$ (Abgrall *et al.* 2012). This microwave clock uses laser cooled atoms and is thus increasing the complexity of the apparatus. This was albeit the necessary step to achieve the uncertainties in this regime that were necessary for high precision spectroscopic application such as the determination of the hydrogen 1S – 2S two-photon transition frequency (Niering *et al.* 2000; Fischer *et al.* 2004; Parthey *et al.* 2011) or the $^{40}$Ca$^+$ clock transition frequency (Chwalla *et al.* 2009). Aiming at chronometric levelling with clocks on a competitive level to established techniques, the clock accuracy is still too low and ways to apply optical clocks must be pursued that have demonstrated the relevant frequency resolution of $10^{-17}$ and below.

Realizing a full transportable optical clock is however a difficult task since highly advanced pieces of experimental equipment must be combined: traps for ions or neutral atoms including the necessary laser sources for cooling and state preparation on site, and the highly stable interrogation laser with its high finesse reference resonator on the other. The latter requires new approaches for the mounting of the optical resonator. Since variations in the mounting forces of the optical resonator usually lead to length variations of the cavity and thus frequency instabilities of the laser, the cavity commonly lies on its support and is not rigidly clamped (Chen *et al.* 2006; Nazarova *et al.* 2006; Webster *et al.* 2007;

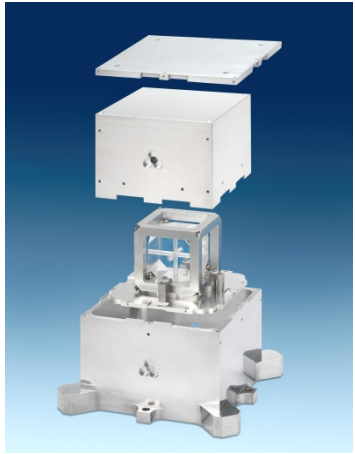

*Figure 6.8: A transportable reference resonator, which can be used to pre-stabilize the interrogation laser of an optical clock. The resonator (glass cube) is mounted in a metal cage such that the resonator length is insensitive to elastic deformation by accelerations (Webster and Gill. 2011). The cage is surrounded by two heat shields for thermal decoupling of the resonator from the environment. The resonator was supplied by National Physical Laboratory and integrated Physikalisch-Technische Bundesanstalt within the EMRP project JRP IND14.*

Millo *et al.* 2009; Kessler *et al.* 2012a; Häfner *et al.* 2015; Matei *et al.* 2017). Despite the advantage that by the proper choice of support points a suppression of cavity length changes induced by residual vibrations can be achieved, these supports are unsuitable for mobile optical clocks as the cavity has to remain in place during transport. Therefore, other approaches have been developed (figure 6.8) that combine the advantages of rigid mounting and vibration noise suppression (Webster and Gill 2011; Vogt *et al.* 2011; Leibrandt *et al.* 2011; Argence *et al.* 2012; Chen *et al.* 2014; Parker *et al.* 2014; Didier *et al.* 2016; Świerad *et al.* 2016; Koller *et al.* 2017), which have successfully demonstrated transportation tests (Leibrandt *et al.* 2011a; Vogt *et al.* 2011; Poli *et al.* 2014).

For reasons of compactness and reduction of inertial forces, often cavities of only about 5 cm length have been chosen (Webster *et al.* 2011; Leibrandt *et al.* 2011; Parker *et al.* 2014). As the thermal noise limit for the frequency instability of the resonator is decreasing for the usual cavity parameters with cavity length, one cannot expect a performance similar to laboratory lasers with longer resonators. In consequence, the stability of a transportable optical clock will be reduced. Since fractional laser instabilities of down to $10^{-15}$ have been achieved with these lasers, they still can perform well for clocks. With designs using longer cavities, frequency instabilities approaching $5\times10^{-16}$ and below have been achieved (Argence *et al.* 2012; Vogt *et al.* 2016; Koller *et al.* 2017), promising better clock instability. In combination with new materials for the mirror coatings exhibiting smaller thermal noise (Cole *et al.* 2013; Cole *et al.* 2016), further improved coherence properties of transportable lasers can be expected.

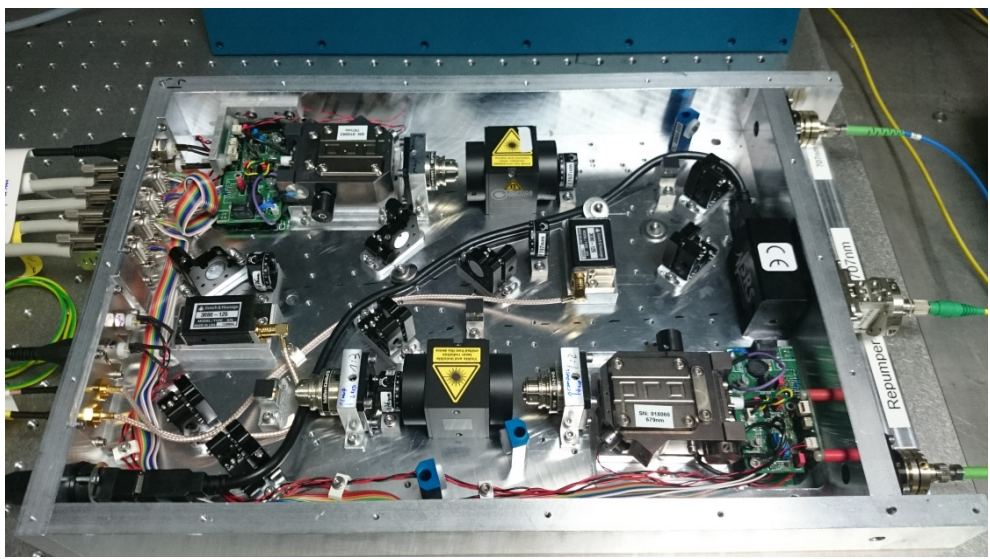

*Figure 6.9: A laser system including two laser heads on a compact breadboard with acousto-optical modulators for frequency tuning and intensity modulation, an optical shutter and coupling into three optical fibres. Similar assemblies are used in the transportable optical clocks reported by Koller et al. 2017 and Origlia et al. 2016.*

As the interrogation or clock laser is only one – though admittedly a very important – part of a complete optical clock, a functional transportable system requires more developments. These may require very different solutions of experimental details depending on the desired application of the optical clock.

On the background of a potential follow-up space mission of ACES (Heß *et al.* 2011), efforts have been made to increase the technological readiness of optical lattice clocks. The activities of the "Space Optical Clock" consortium (SOC) are coordinated by the University of Düsseldorf (Bongs *et al.* 2015). Originally following the development of both Sr (Poli *et al.* 2014) and Yb lattice clocks (Mura *et al.*

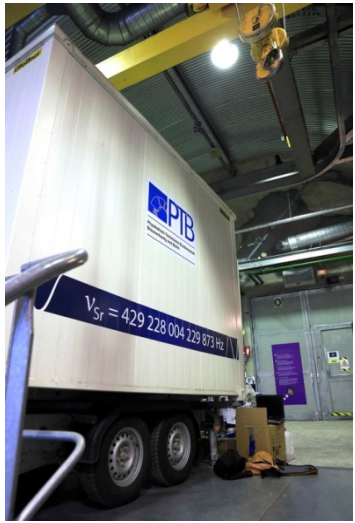

*Figure 6.10: A Sr transportable lattice clock during its first measurement campaign in the Laboratoire Souterrain de Modane (Grotti et al. 2017).*

2013) it focusses now on a very compact Sr system (Bongs *et al.* 2015; Origlia *et al.* 2016). So far, spectroscopy of the 698 nm clock transition on $^{88}$Sr has been demonstrated with an apparatus that houses all laser systems (figure 6.9) and vacuum components necessary for the clock excluding electronics and the clock laser in a volume of less than 1 m$^3$. This marks already a substantial achievement in compactification in view of the complexity of a lattice clock apparatus. The characterization of this optical clock is ongoing to reach the target fractional uncertainty of few $10^{-17}$, accompanied by further measures to reduce the mass, volume, and power consumption of the apparatus.

A more compact transportable clock based on a $^{40}$Ca$^+$ has been realized (Cao *et al.* 2017). Apart from electronics, the apparatus requires a volume of 0.54 m$^3$ only. The apparatus contains a clock laser with a frequency instability of $2\times10^{-15}$ in 1 s. The uncertainty budget has already been evaluated, presently reaching $7.7\times10^{-17}$. The instability of the system was found to be $2.3\times10^{-14}/\tau^{1/2}$, being very close to the QPN limited instability of $2.0\times10^{-14}/\tau^{1/2}$ expected for the operation conditions. This apparatus is impressively demonstrating the advantage of comparative simplicity of laser systems and vacuum system some single ion clocks can offer, which may lead to high compactness of the clock apparatus. This may be an important prerequisite for applications or even a commercialization. It comes however at the cost of clock instability related with the principle of single ion optical clocks. While $10^{-16}$ fractional frequency resolution is rapidly achieved within less than 15 hours, $10^{-17}$ resolution corresponding to a height resolution of 10 cm will already require averaging times of about two months given these experimental parameters. The route to shortening these is however clear: As the clock instability is practically QPN-limited, longer probe times i.e. an interrogation laser with improved coherence properties will enable shorter measurement times. With the present state of development, a reduction of laser instability of at least a factor of four seems to be realistic. If this leads to equivalently longer probe times, the averaging times could be shortened about fourfold.

Shorter measurements have been demonstrated with a transportable Sr optical lattice clock (Koller *et al.* 2017). With an interrogation laser of mid-$10^{-16}$ fractional frequency instability, a clock instability of $1.3\times10^{-15}/\tau^{1/2}$ has been demonstrated by comparison against a stationary Sr lattice clock with a state of the art instability of few $10^{-16}$ in a second (Al-Masoudi *et al.* 2015; Grebing *et al.* 2016). This instability contrasts with the case of the Ca$^+$ clock, which is not limited by the QPN but via the Dick effect (Dick 1988). Still, a fractional statistical uncertainty of $10^{-17}$ will be achieved after less than 5 hours of averaging time. This clock instability opens the door to monitor time dependent variations of the gravity potential (see section 2, Voigt *et al.* 2016), in particular if a high stability reference clock can be used.

The transportable lattice clock is compact enough to be installed in car trailer in which it can be operated. Secondly, its uncertainty of $7.4 \times 10^{-17}$ has been validated against the PTB Sr laboratory clock demonstrating agreement within the combined clock uncertainties (Koller *et al.* 2017). Further lowering of the systematic clock uncertainty is likely since the largest contributions are typically well controlled and do not e.g. require technical modifications to be reduced. These are important ingredients for actual campaigns outside metrology laboratories and a transportable lattice clock, shown in figure 6.10, was indeed used for out-of-the-laboratory experiments with negligible metrological infrastructure (Grotti *et al.* 2017).

The combination with optical fibre links, that are as discussed in section 6.2 currently the only alternative for top quality comparisons of non-transportable optical clocks, offers excellent perspectives for geodetic applications. A portable system can be operated along the route of existing fibre links and there be connected by frequency extraction (Grosche 2014) to another optical clock that is typically found at the end of such a link. The higher flexibility in the choice of measurement locations that is offered by a transportable clock grants access to mountainous areas, islands or other relevant locations, where no institute with highly developed infrastructure will be found. Portable optical clocks may also prove to be very useful for precision measurements in other research fields because a state-of-the-art frequency standard can be brought to an institute where a reference is required. This applies not only to the validation of optical clocks themselves but also for measurements that require particular properties of a clock like e.g. high frequency stability, high or low sensitivity of its frequency to possible variations of fundamental constants, or a clock frequency that is especially well suited for the planned experiment.

Further miniaturization of laser systems (figure 6.9) and other components will benefit different applications, in particular those connected to the commercialization of optical frequency standards. Examples of applications are robust replacements of high stability frequency standards like masers, optical clocks as ready-to-use tools in geodetic reference stations or tide gauges, or for research in space. These ask for different optimization strategies like good performance at reasonable price, accuracy at the level of laboratory optical clocks and beyond, or extreme robustness and compactness paired with very high performance, respectively. Still, these development branches will mutually benefit each other.

### 6.4 First Geodetic Campaigns with Clocks

At present, the information exchange between geodesy and frequency metrology is mostly directed towards the metrology side. Geodetic determinations of geopotentials serve for the evaluation of the optical clocks and their uncertainty budgets in order to gain confidence in the new generation clocks within the international metrology community and beyond. In the future, this will however change and the clocks will provide the reference for geodetic heights and the geopotential as well as for changes of these quantities. As discussed in section 2.3 and 2.6, the GNSS/geoid approach gives absolute gravity potential values, presently with an uncertainty of about 2 cm in terms of heights, however, requiring sufficient high resolution and quality terrestrial (gravity and terrain) data around the sites of interest, which may not exist in remote areas. On the other side, the geometric levelling approach can deliver potential differences with millimetre uncertainties over shorter distances, but is susceptible to systematic errors at the decimetre level over large distances. Results from both approaches show an agreement at the few centimetre level over a few 100 km distance, while being presently inconsistent at the decimetre level across Europe (see Denker *et al.* 2017, Kenyeres *et al.* 2010, Gruber *et al.* 2011). For this reason, the more or less direct observation of gravity potential differences through optical clock comparisons (with targeted fractional accuracies of $10^{-18}$, corresponding to 1 cm in height) is eagerly awaited as a means for resolving the existing discrepancies between different geodetic techniques and remedying the geodetic height determination problem over large distances. In this connection, the present geodetic results are by far sufficient for the existing global network of microwave clocks and the associated international timescales, operating in the $10^{-16}$ regime. However, for the integration of optical clocks to international timescales, also the international height systems need to be improved. Once optical clocks are validated by side-by-side clock comparisons at the level of $10^{-18}$, their long-distance comparison will eventually contribute to the improvement of geodetic height systems.

Experimentally, several comparisons between remote optical clocks have been performed, which require geodetic information on the differential relativistic redshift between the involved systems that cannot be trivially derived. The dissemination of the clock frequencies involves, e.g. satellites (Hachisu *et al.* 2014), but more often optical fibre links, discussed in section 6.2, are employed (Yamaguchi *et al.* 2011, Lisdat *et al.* 2016, Takano *et al.* 2016). The distances between the clocks varied from a few 10 km to about 9000 km, thus spanning distances that are relevant in geodesy.

In all these experiments clocks of the same type where used. Thus, the frequency ratio between both clocks is expected to be unity if they are operated under the same conditions and locations in spacetime, and consequently the observed deviation of the frequency ratio from unity can be interpreted as an effect of the potential difference between both systems. Although the respective clocks were very carefully evaluated in their uncertainties, no independent validation of their performance was available. As mentioned before, this is a critical point since the optical clocks were operated at the limit of their performance. For this reason, the related publications interpret their measurements as validations of their clocks rather than measurements of their potential difference. Depending on the separation between the involved clocks, the gravity potential induced frequency shift was derived from dedicated levelling campaigns for smaller separations of the clocks (Takano *et al.* 2016) or a combination of geoid modelling and GNSS results (Lisdat *et al.* 2016).

Takano *et al.* (2016) report an agreement of two Sr lattice clocks separated by 15 km at the level of $6\times10^{-18}$, which would correspond to a height resolution of 5 cm. This is to be compared to the accuracy of the levelling of 6 mm. The two Sr lattice clocks used by Lisdat *et al.* (2016) where separated by 690 km. The potential difference was known equivalent to 3.6 cm in height, while the clock uncertainty corresponds to 46 cm. These experiments convincingly demonstrate the potential of clocks in geodesy to resolve small height differences also over long distances, and hence clocks may provide the future reference for geodetic heights and the geopotential as well as for changes of these quantities.

To be certain that the observed frequency difference correctly reflects the relativistic frequency shift and is not biased by undetected clock errors, additional information about the clocks is required. The application of a transportable clock is one way to circumvent this issue since it can be locally calibrated against the clock at the reference position and then operated remotely. One example for this approach was reported by Fateev *et al.* (2017) and Kopeikin *et al.* (2016), where hydrogen masers were used as transportable clocks. Unfortunately, masers are known to experience unpredictable frequency changes, which limit the achievable accuracy when this type of clock is used. Grotti *et al.* (2017) used a transportable optical lattice clock in a proof of concept experiment to measure a height difference of about 1000 m. Though the accuracy of this clock comparison – corresponding to about 18 m in terms of height difference - did not reach the resolution demonstrated by Takano *et al.* (2016) or Lisdat *et al.* (2016), it was the first demonstration of a full chronometric levelling experiment with an optical clock, because it included both a remote and a local measurement.

Beyond this, optical clock networks and corresponding link technologies are currently established or are under discussion, from which distinct advantages are expected for the dissemination of time, geodesy, astronomy and basic research; for a review see (Riehle 2017). Also under investigation is the installation of optical master clocks in space, where spatial and temporal variations of Earth's gravitational field are smoothed out, such that the space clocks could serve as a reference for ground-based optical clock networks (e.g. Gill *et al.* 2008, Bongs *et al.* 2015, Schuldt *et al.* 2016). Such clock networks could also help to establish a new global height reference system. Moreover, clock measurements for spaceborne gravity field recovery missions have been discussed in (Mayrhofer and Pail 2012) and (Müller *et al.* 2017). Naturally, transportable optical clocks will typically follow their laboratory counterparts. With the latter targeting the $10^{-19}$ range of fractional uncertainty, transportable clocks with low $10^{-18}$ uncertainties are expected in the next decade. This would render them at latest a highly useful tool in geodesy.

## 7. Conclusions

In this work, we have discussed recent progress with optical atomic clocks based on trapped ions or neutral atoms. For both types of clocks estimated uncertainties of a few $10^{-18}$ have been reported in several laboratories now. While neutral lattice clocks have a superior short-time stability operating several 1000s of atoms and thus can work at extremely short integration times, ion clocks have a high potential to achieve lowest inaccuracies with well-controlled single atomic particles. Here, the recently proposed multi-ion clocks can be a promising and combining path to profit best from both worlds. We have presented the state-of-the-art of methods to compare optical clocks over long distances, relying on optical fibre networks, satellites or portable frequency standards. These techniques have advanced rapidly in the past years, e.g. an optical frequency transfer over more than 1000 km distance via optical fibres with a fractional frequency uncertainty close to $10^{-20}$ was demonstrated. For the more flexible use of clocks outside dedicated laboratories, advanced microwave or optical satellite links still need to be developed. A first important step towards intercontinental clock comparison at the level of $10^{-17}$ will be enabled by the ACES mission, which will provide a high-stability microwave link together with a microwave clock acting as a flywheel on the international space station (ISS).

Due to the relativistic redshift effect, optical atomic clocks are sensitive to the (otherwise not directly observable) Earth's gravity potential at an entirely new level of detail, which offers completely new perspectives in geodesy, e.g. the establishment of a unified International Height Reference System, contributions to the mapping of the geoid, and the measurement of time variations of Earth's gravity field. Presently, very long wavelength gravity field structures can be determined accurately from satellite data, but only with a spatial resolution of a few 100 km, and hence the (neglected) short wavelength signals are still at the several decimetre-level for the geoid. The combination of satellite and terrestrial data (mainly gravity measurements) gives the geoid with an uncertainty of about 2 cm in a best-case scenario, but this situation does not exist in remote areas such as parts of Africa, South America, Asia, etc. Together with precise GNSS positions, this leads to the GNSS/geoid approach, which gives absolute gravity potential values, at best also with an uncertainty of about 2 cm in terms of heights, but with some perspective for further improvements. On the other hand, the geometric levelling approach can deliver potential differences with millimetre uncertainties over shorter distances, but is susceptible to systematic errors at the decimetre-level over large distances. Consequently, over long distances across national borders, the GNSS/geoid approach should be a better approach than geometric levelling. This is also confirmed by practical results, showing a good agreement at the centimetre level over short distances, while the two approaches are presently inconsistent at the decimetre level across Europe. A further complication with classical geometric levelling is that the existing national height systems are based on different tide gauges related to different level surfaces, e.g. resulting in inconsistencies of more than 0.5 m across Europe (due to the dynamic ocean topography (DOT) and possible vertical crust movements as discussed in section 2.2). For this reason, the more or less direct observation of gravity potential differences through optical clock comparisons (with targeted fractional accuracies of $10^{-18}$, corresponding to 1 cm in height) is eagerly awaited as a means for resolving the existing discrepancies between different geodetic techniques and remedying the geodetic height determination problem over large distances. Further on, optical clocks might become a geodetic tool for monitoring geodynamics overcoming the drawback of superimposed and dominating local effects in gravimetry to a large extent as described in section 2.7.

Several clock comparisons at the level of $10^{-15}$ to $10^{-18}$ between remote optical clocks have already been performed, which require geodetic information on the differential relativistic redshift between the involved systems that cannot be trivially derived. Also, recently, a first proof-of-principle, full chronometric levelling experiment with a portable optical clock was carried out in the French/Italian alps, see section 6.4. At present, the performance of the optical clocks still needs to be improved and the reliable operation at $10^{-18}$ be verified by repeated side-by-side clock comparisons, but in the future long-distance clock comparisons may resolve the vertical datum problem and provide the reference for geodetic heights and the geopotential as well as for changes of these quantities.

Another interesting use of clock measurements is the mapping of the (mean) DOT, which globally reaches maximum values of about ±2 m. This deviation of the mean sea level from a best fitting equipotential surface (geoid) is due to winds, atmospheric pressure, buoyancy and ocean currents. For this purpose, robust portable optical clocks and optical satellite links, which are currently under development, are needed to work on ships and to reach remote places.

Moving optical clocks out of the laboratories is also an essential step for two other developments: the commercialization of these instruments and the positioning of optical clocks in space, similar to the microwave clock ACES. Both topics are very relevant for the use of optical clocks in geodesy, as reference stations require reliable apparatuses and moderate user surveillance. A spaceborne clock with appropriate frequency links may serve as an optical master clock, i.e. a stable and universal height reference that is commonly accessible from the ground. As spatial and temporal variations of Earth's gravitational field are smoothed out, such a space clock could provide a reference for ground based optical clock networks.

Finally, the impact of optical clocks on geodetic applications will naturally also depend on the size (transportability) and price of optical clocks (and link technology) in the future; in general, the perspectives for miniaturisation and cost reduction are good and it is possible that developments may occur similar to those which have taken place for GNSS receivers, which started as 100 kg boxes but are now the size of a wristwatch, along with corresponding reductions in price.

**Acknowledgements**

The authors would like to thank Ludger Timmen for valuable input to this work, Jonas Keller and Thomas Waterholter for reading of this manuscript and helpful comments, Thomas Waterholter, Erik Benkler, Nimrod Hausser and Katharina Dudde for assistance with preparing the manuscript. Support has been received from the projects EMPIR 15SIB03 OC18 and EMPIR 15SIB05 OFTEN. These projects have received funding from the EMPIR programme co-financed by the Participating States and from the European Union's Horizon 2020 research and innovation programme. We acknowledge support by the German Research Foundation (DFG) through grant ME 3648/1-1 and ME 3648/3-1, and within CRC 1128 geo-Q (projects A03, A04, C04), CRC 1227 DQ-mat (projects B02, B03), RTG 1729, and the Deutsche Akademische Austauschdienst (DAAD).

# List of Acronyms

| | |
|---|---|
| **AC** | Alternating current |
| **ACES** | Atomic Clock Ensemble in Space |
| **ADEV** | Allan Deviation |
| **AOM** | Acousto-optic modulator |
| **BBR** | Black body radiation |
| **bi-EDFA** | Bi-directional erbium doped fibre amplifier |
| **BIPM** | Bureau International des Poids et Mesures (International Bureau of Weights and Measures) |
| **CEO** | Carrier envelope offset |
| **CERN** | Conseil Européen pour la Recherche Nucléaire (European Organization for Nuclear Research) |
| **CP** | Carrier phase |
| **DC** | Direct current |
| **DOT** | Dynamic ocean topography |
| **DWDM** | Dense wavelength division multiplexing |
| **EEP** | Einstein Equivalence Principle |
| **EMRP** | European Metrology Research Programme |
| **EVRF** | European Vertical Reference Frame |
| **GBVP** | geodetic boundary value problem |
| **GCRS** | Geocentric Celestial Reference System |
| **GGOS** | Global Geodetic Observation System |
| **GNSS** | Global Navigation Satellite Systems |
| **GOCE** | Gravity Field and Steady-State Ocean Circulation Explorer |
| **GP-A** | Gravity Probe-A |
| **GPS** | Global Positioning System |
| **Gpu** | geopotential unit |
| **GR** | General Relativity |
| **GRACE** | Gravity Recovery and Climate Experiment |
| **GRS80** | Geodetic Reference System 1980 |
| **IAG** | International Association of Geodesy |
| **IAU** | International Astronomical Union |
| **IERS** | International Earth Rotation Service |
| **IGS** | International GPS Service |
| **IHRS** | International Height Reference System |
| **IM-DD** | Intensity modulation and direct detection |
| **ITRF** | International Terrestrial Reference Frame |
| **ITRS** | International Terrestrial Reference System |
| **Jason** | Journées Altimétriques Satellitaires pour l'Océanographie |
| **LASSO** | Laser Synchronization from Stationary Orbit |
| **LLI** | Local Lorentz Invariance |
| **LNE-SYRTE** | Laboratoire national de métrologie et d'essais - Système de Références Temps-Espace (French national metrology institute for time and frequency) |
| **LPI** | Local Position Invariance |
| **LSC** | least-square collocation |
| **LUT** | Link under test |
| **MDEV** | Modified Allan Deviation |
| **MSL** | Mean Sea Level |
| **NASA** | National Aeronautics and Space Administration |
| **NGS** | National Geodetic Survey |
| **NIST** | National Institute of Standards and Technology |
| **NOAA** | National Oceanic and Atmospheric Administration |
| **NPL** | National Physical Laboratory (British national metrology institute) |
| **PM** | Polarisation maintaining |
| **PPP** | Precise point positioning |
| **PPS** | Pulse-per-second |
| **PRN** | Pseudo-random noise |
| **PTB** | Physikalisch-Technische Bundesanstalt (German national metrology institute) |
| **PTP** | Precision Time Protocol |
| **QPN** | Quantum projection noise |
| **RCR** | remove-compute-response |
| **REFIMEVE+** | Reseau Fibré Metrologique à Vocation Européenne+ |
| **RENATER** | Réseau national de télécommunications pour la technologie, l'enseignement et la recherche (French national research and education network) |
| **RF** | Radio frequency |
| **RMS** | Root mean square |
| **SI** | International System of Units |
| **SOC** | Space optical clock |
| **SSH** | sea surface height |
| **T2L2** | Time Transfer by Laser Link |
| **TAI** | Temps Atomique International (International Atomic Time) |
| **TCG** | Geocentric Coordinate Time |
| **TT** | Terrestrial Time |
| **TW** | Two-way |
| **TWSTFT** | Two-way Satellite Time and Frequency Transfer |
| **UFF** | Universality of Free Fall |
| **UTC** | Coordinated Universal Time |
| **VLBI** | Very-long Baseline Interferometry |